\documentclass[12pt]{article}
\usepackage{epsf}
\usepackage{amsmath}
\def\hybrid{\topmargin 0pt      \oddsidemargin 0pt
	\headheight 0pt \headsep 0pt
	\textheight 9in         
	\textwidth 6.25in       
	\marginparwidth .875in
	\parskip 5pt plus 1pt   \jot = 1.5ex}

\catcode`\@=11
\def\marginnote#1{}
\newcount\hour
\newcount\minute
\newtoks\amorpm
\hour=\time\divide\hour by60
\minute=\time{\multiply\hour by60 \global\advance\minute by-\hour}
\edef\standardtime{{\ifnum\hour<12 \global\amorpm={am}%
	\else\global\amorpm={pm}\advance\hour by-12 \fi
	\ifnum\hour=0 \hour=12 \fi
	\number\hour:\ifnum\minute<10 0\fi\number\minute\the\amorpm}}
\edef\militarytime{\number\hour:\ifnum\minute<10 0\fi\number\minute}

\def\draftlabel#1{{\@bsphack\if@filesw {\let\thepage\relax
   \xdef\@gtempa{\write\@auxout{\string
      \newlabel{#1}{{\@currentlabel}{\thepage}}}}}\@gtempa
   \if@nobreak \ifvmode\nobreak\fi\fi\fi\@esphack}
	\gdef\@eqnlabel{#1}}
\def\@eqnlabel{}
\def\@vacuum{}
\def\draftmarginnote#1{\marginpar{\raggedright\scriptsize\tt#1}}

\def\draft{\oddsidemargin -.5truein
	\def\@oddfoot{\sl preliminary draft \hfil
	\rm\thepage\hfil\sl\today\quad\militarytime}
	\let\@evenfoot\@oddfoot \overfullrule 3pt
	\let\label=\draftlabel
\let\marginnote=\draftmarginnote
	\let\marginnote=\draftmarginnote
   \def\@eqnnum{(\theequation)\rlap{\kern\marginparsep\tt\@eqnlabel}%
\global\let\@eqnlabel\@vacuum}  }

\def\numberbysection{\@addtoreset{equation}{section}
\def\theequation{\thesection.\arabic{equation}}}

\def\underline#1{\relax\ifmmode\@@underline#1\else
	$\@@underline{\hbox{#1}}$\relax\fi}

\def\titlepage{\@restonecolfalse\if@twocolumn\@restonecoltrue\onecolumn
     \else \newpage \fi \thispagestyle{empty}\c@page\z@
	\def\thefootnote{\fnsymbol{footnote}} }

\def\endtitlepage{\if@restonecol\twocolumn \else  \fi
	\def\thefootnote{\arabic{footnote}}
	\setcounter{footnote}{0}}  
\catcode`@=12
\relax

\def\beq{\begin{equation}}
\def\eeq{\end{equation}}
\def\bea{\begin{eqnarray}}
\def\eea{\end{eqnarray}}
\def\nn{\nonumber}

\relax
\hybrid

\begin{document}

\begin{titlepage}
\begin{center}
February~2012 \hfill . \\[.5in]
{\large\bf Parafermionic chiral algebra $Z_{3}$ with the dimension of the principal parafermion fields $\psi(z)$, $\psi^{+}(z)$, $\Delta_{\psi}=8/3$. }
\\[.5in] 
{\bf Vladimir S.~Dotsenko}\\[.2in]
{\it LPTHE, CNRS, Universit{\'e} Pierre et Marie Curie, Paris VI, UMR 7589\\
               4 place Jussieu,75252 Paris Cedex 05, France.}\\[.2in]
               
 \end{center}
 
\underline{Abstract.}

We analyze, and prove, the associativity of the new $Z_{3}$ 
parafermionic chiral algebra which has been announced some time ago,
with principal parafermionic fields having the conformal dimension
$\Delta_{\psi}=8/3$. In doing so we have developed a new method 
for analyzing the associativity of a given chiral algebra of parafermionic
type, the method which might be of a more general significance
than a particular conformal field theory studied in detail in this paper.
Still, even in the context of our particular chiral algebra, of $Z_{3}$
parafermions with $\Delta_{\psi}=8/3$, the new method allowed us
to give a proof of associativity which we consider to be complete.

\end{titlepage}

\newpage

\numberwithin{equation}{section}

\section{Introduction.}

In the paper [1] the solution has been announced for the associativity constraints of the $Z_{3}$ parafermionic algebra with dimension of the principal parafermionic fields $\psi(z)$, $\psi^{+}(z)$,  $\Delta_{\psi}=8/3$. In this paper we shall give the details of the calculations leading to this solution. 

In addition, we shall also present, but in the next publication, two other (non-minimal) solutions for the $Z_{3}$ parafermionic algebra, still with $\Delta_{\psi}=8/3$, but with a bigger content of chiral fields: with two additional bosonic fields [6].

We begin by reproducing the results announced in [1].

In addition to the principal fields $\psi(z)$, $\psi^{+}(z)$, with dimensions $\Delta_{\psi}=8/3$, the parafermionic algebra contains extra 
parafermionic fields $\tilde{\psi}(z)$ and $\tilde{\psi}^{+}(z)$, with dimensions $\Delta_{\tilde{\psi}}=\Delta_{\psi}+2$, and also one bosonic field $B(z)$, with dimension $\Delta_{B}=4$. Their operator product expansions (OPE) are of the form :
\bea
\psi(z')\psi(z)=\frac{1}{(z'-z)^{\Delta_{\psi}}}\{\lambda\psi^{+}(z)+(z'-z)\lambda
\beta^{(1)}_{\psi\psi,\psi^{+}}\partial\psi^{+}(z)\nn\\+(z'-z)^{2}[\lambda\beta^{(11)}
_{\psi\psi,\psi^{+}}\partial^{2}\psi^{+}(z)+\lambda\beta^{(2)}_{\psi\psi,\psi^{+}}
L_{-2}\psi^{+}(z)+\zeta\tilde{\psi}(z)]+...\} \label{eq1.1}
\eea
\bea
\psi(z')\psi^{+}(z)=\frac{1}{(z'-z)^{2\Delta_{\psi}}}
\{1+(z'-z)^{2}\frac{2\Delta_{\psi}}{c}T(z)
+(z'-z)^{3}\frac{\Delta_{\psi}}{c}\partial T(z)\nn\\
+(z'-z)^{4}[\beta^{(112)}_{\psi\psi^{+},I}\partial^{2}T(z)+\beta^{(22)}_{\psi\psi^{+},
I}\Lambda(z)+\gamma B(z)]\nn\\
+(z'-z)^{5}[\beta^{(1112)}_{\psi\psi^{+},I}\partial^{3}T(z)+\beta^{(122)}_{\psi\psi^{+},I}\partial\Lambda(z)+\gamma
\beta^{(1)}_{\psi\psi^{+},B}\partial B(z)]+...\} \label{eq1.2}
\eea
\bea
\psi(z')\tilde{\psi}(z)=\frac{1}{(z'-z)^{\Delta_{\psi}+2}}\{\zeta\psi^{+}(z)+
(z'-z)\zeta\beta^{(1)}_{\psi\tilde{\psi},\psi^{+}}\partial\psi^{+}(z)\nn\\+
(z'-z)^{2}
[\zeta\beta^{(11)}_{\psi\tilde{\psi},\psi^{+}}\partial^{2}\psi^{+}(z)+
\zeta\beta^{(2)}_{\psi\tilde{\psi},\psi^{+}} L_{-2}\psi^{+}(z)\nn\\+\eta
\tilde{\psi}^{+}(z)]+...\} \label{eq1.3}
\eea
\beq
\psi(z')\tilde{\psi}^{+}(z)=\frac{1}{(z'-z)^{2\Delta_{\psi}-2}}\{\mu B(z)+(z'-z)\mu
\beta^{(1)}_{\psi\tilde{\psi}^{+},B}\partial B(z)+...\} \label{eq1.4}
\eeq
\bea
\psi(z')B(z)=\frac{1}{(z'-z)^{4}}\{\gamma\psi(z)+(z'-z)\gamma\beta^{(1)}_{\psi B,
\psi}\partial\psi(z)+(z'-z)^{2}[\gamma\beta^{(11)}_{\psi B,\psi}\partial^{2}
\psi(z)\nn\\+\gamma\beta^{(2)}_{\psi B,\psi}L_{-2}\psi(z)+\mu\tilde{\psi}(z)]
+...\} \label{eq1.5}
\eea

In the equations above, $\wedge(z)=L_{-2}T(z)$, $T(z)$ is the stress-energy tensor. The coefficients $\{\beta\}$ are all fixed by the conformal invariance; all the necessary values of the $\beta$ coefficients are listed in the Appendix A.

In addition to the expansions presented above, with first terms given explicitly, one has also to consider the following products and their decompositions:
\beq
\tilde{\psi}\times\tilde{\psi}\sim\eta[\psi^{+}]+\tilde{\lambda}[\tilde{\psi}^{+}]
\label{eq1.6}
\eeq
\beq
\tilde{\psi}\times\tilde{\psi}^{+}\sim[I]+\tilde{\gamma}[B] \label{eq1.7}
\eeq
\beq
\tilde{\psi}\times B\sim\mu[\psi]+\tilde{\gamma}[\tilde{\psi}] \label{eq1.8}
\eeq
\beq
B\times B\sim[I]+b[B] \label{eq1.9}
\eeq
Here $I$ is the identity operator. In the symbolic presentation above of the OPEs, the symbol $[B]$, represents the Virasoro algebra series of the operator $B$ and its descendants.
For instance, the symbolic expansion (\ref{eq1.6}) should be read, more explicitly, as :
\bea
\tilde{\psi}(z')\tilde{\psi}(z)=\frac{1}{(z'-z)^{\Delta_{\psi}+4}}\{\eta\psi^{+}(z)+
(z'-z)\eta\beta^{(1)}_{\tilde{\psi}\tilde{\psi},\psi^{+}}\partial\psi^{+}(z)\nn\\
+(z'-z)^{2}[\eta\beta^{(11)}_{\tilde{\psi}\tilde{\psi},\psi^{+}}\partial^{2}\psi^{+}(z)+\eta\beta^{(2)}_{\tilde{\psi}\tilde{\psi},\psi^{+}}
L_{-2}\psi^{+}(z)+\tilde{\lambda}
\tilde{\psi}(z)]+...\} \label{eq1.10}
\eea
There are 8 operator algebra constants entering the expansions (\ref{eq1.1})-(\ref{eq1.9}). They could be defined by the corresponding three-point functions as follows :
\bea
<\psi\psi\psi>=\lambda, \quad
<\psi\psi\tilde{\psi}>=\zeta, \quad
<\psi\tilde{\psi}\tilde{\psi}>=\eta, \quad
<\tilde{\psi}\tilde{\psi}\tilde{\psi}>=\tilde{\lambda}, \nn\\
<\psi\psi^{+}B>=\gamma, \quad
<\psi\tilde{\psi}^{+}B>=\mu, \quad
<\tilde{\psi}\tilde{\psi}^{+}B>=\tilde{\gamma}, \quad
<BBB>=b \label{eq1.18}
\eea
Here $<\psi\psi\psi> = <\psi(\infty)\psi(1)\psi(0)>$, etc. . 

As compared to [1] we have changed the notation of the constant 
$<BBB>=b$, this is instead of $\xi$ in [1]. This minor change 
in the notation will be helpful in the following, in [6], 
when we shall consider more general chiral
algebras, with two more bosonic fields, U and W, cf. (\ref{eq1.27}),
resulting in more operator algebra constants. 
At that point it will be more natural (simpler for the notations) 
to reserve only latin letters for the purely bosonic constants, like 
$<BBB>=b$, $<BUU>=a$, etc. .

The chiral algebra (\ref{eq1.1})-(\ref{eq1.9}) has to be associative, otherwise the multipoint correlation functions could not be defined.
The associativity imposes certain constraints, in the form of algebraic equations, on the constants (\ref{eq1.18}). The whole subject is if a non-trivial solution exists. It happens that in this case the solution does exist and with the central charge of the Virasoro algebra remaining a free parameter. We have found, in [1], the following values of the constants (\ref{eq1.18}):
\beq
\lambda=\frac{14}{27}\sqrt{3}\sqrt{\frac{c+32}{c}} \label{eq1.19}
\eeq
\beq
\zeta=\frac{8}{27}\sqrt{30}\sqrt{\frac{(c+56)(11c+14)}{c(784+57c)}} \label{eq1.20}
\eeq
\beq
\eta=\frac{7}{27}\frac{(349c+2688)\sqrt{3}}{784+57c}\sqrt{\frac{c+32}{c}}
\label{eq1.21}
\eeq
\beq
\tilde{\lambda}=\frac{98}{135}\sqrt{30}\frac{(c+32)(4877c^{2}+51466c+13104)}
{\sqrt{c(c+56)(11c+14)(784+57c)}(784+57c)} \label{eq1.22}
\eeq
\beq
\gamma=\frac{4}{27}\sqrt{15}\sqrt{\frac{(c+56)(11c+14)}{c(22+5c)}} \label{eq1.23}
\eeq
\beq
\mu=\frac{28}{9}\sqrt{6}\sqrt{\frac{(c+32)(22+5c)}{c(784+57c)}} \label{eq1.24}
\eeq
\beq
\tilde{\gamma}=\frac{7}{135}\sqrt{15}\frac{(20595c^{3}+823534c^2+5532912c+2121728)}
{\sqrt{c(22+5c)(c+56)(11c+14)}(784+57c)} \label{eq1.25}
\eeq
\beq
b=\frac{2}{15}\sqrt{15}\frac{(85c^{2}+2566c+976)}{\sqrt{c(22+5c)(c+56)(11c+14)}} \label{eq1.26}
\eeq
The methods and the calculations leading to the solution (\ref{eq1.19})-(\ref{eq1.26}) will be presented in the next section.

It could be mentioned that if the set of the chiral fields were limited to just $\psi(z)$ and $\psi^{+}(z)$, i.e. if we suppressed the fields $\tilde{\psi}$, $\tilde{\psi}^{+}$, $B$, then there would be no solution, for $\Delta_{\psi}=8/3$, except for isolated values of the central charge, $c=-14/11$ and $c=-56$. The minimal set of chiral fields for which the solution exist, with $c$ free, is the one stated above, that of $\psi$, $\psi^{+}$, $\tilde{\psi}$, $\tilde{\psi}^{+}$, $B$.

In the next publication we shall present, in addition, two non-minimal solutions, with $c$ free, for the  a bigger set of chiral fields :
\beq
\psi, \psi^{+}, \tilde{\psi}, \tilde{\psi^{+}}, U, B, W \label{eq1.27}
\eeq
In this set there are two additional bosonic fields, $U$ and $W$, with conformal dimensions  $\Delta_{U}=3$ and $\Delta_{W}=5$, the rest of 
the fields having the same dimensions: $\Delta_{\psi}=8/3$, 
$\Delta_{\tilde{\psi}}=\Delta_{\psi}+2$,
$\Delta_{B}=4$. 

It could be added that the solution 1, (\ref{eq1.19})-(\ref{eq1.26}), for the set of fields $\psi$, $\psi^{+}$, $\tilde{\psi}$, $\tilde{\psi}^{+}$,$B$, and the solutions 2 and 3, which will be given in the next publication, obtained for the set of fields $\psi$, $\psi^{+}$, $\tilde{\psi}$, $\tilde{\psi}^{+}$, $U$, $B$, $W$, these solutions are distinct : if we add the field $U$ to the $Z_{3}$ neutral channel, then the appearance of the fields $B$ and $W$ becomes necessary ; but if we start with just the field $B$ in the neutral channel, then the chiral algebra closes without $U$ and $W$, closes differently.

\vskip0.7cm

Before closing this introductory section we shall state the rules which we 
have followed in defining the chiral algebra and in its associativity calculations.

We start with the basic fields, $\psi(z)$, $\psi^{+}(z)$, which are $Z_{3}$ charged. For these fields there are two products to be expanded as OPE : $\psi(z')\psi(z)$ and $\psi(z')\psi^{+}(z)$, equations (\ref{eq1.1}) and (\ref{eq1.2}).
In each expansion should be made explicit all the terms which are singular (divergent, as $z'\rightarrow z$). This assumes that the operators should be given explicitly up the term $\sim (z'-z)^{2}$ (up to level 2) in the OPE of $\psi(z')\psi(z)$, and up to the term $\sim(z'-z)^{5}$ (up to level 5) in the OPE of $\psi(z')\psi^{+}(z)$, equations (\ref{eq1.1}), (\ref{eq1.2}).

This rule is required by the condition that the commutation rules of the modes of the operators $\psi(z)$, $\psi^{+}(z)$, which will follow from equations (\ref{eq1.1}), (\ref{eq1.2}) (see next Section), should define all matrix elements of the mode operators, of the highest weight type representations.

This implies that we have to decide on the operator content in the charged sector, in the expansion (\ref{eq1.1}), up to level 2, and up to level 5 in the neutral sector, expansion (\ref{eq1.2}). Filling these levels with just Virasoro descendants of the leading operator $(\psi^{+}(z)$, in (\ref{eq1.1}), identity $I$, in (\ref{eq1.2})), i.e. with no new Virasoro primary operators on those levels, leads to the chiral algebra proposal which does not pass the associativity constraints.

Looking for minimal solution (minimal in the content of chiral fields) we tried to add just one chiral boson $B(z)$, in the neutral sector, on level 4. This
proposal still gets blocked, at some point, by the associativity constraints. Finally the solution was found by adding one more operator, 
$\tilde{\psi}^{+}(z)$, on level 2 in the charged sector.

We have added this historical remark to make it more clear what sort of problems get involved when searching for a particular chiral algebra, for a given extended symmetry, in the associativity constraints approach.

New (Virasoro) primary operators which have been added in the neutral and charged sectors, in (\ref{eq1.1}) and in (\ref{eq1.2}), constitute the chiral algebra fields (local operators), together with the initial fields $\psi(z)$, $\psi^{+}(z)$. On has to consider next all their products, equations (\ref{eq1.3})-(\ref{eq1.9}).

Now, in the corresponding expansions, involving higher conformal dimension fields, $\tilde{\psi}$, $B$, one needs not to make explicit all the singular terms. One has to make explicit the terms up to second level in the charged sector and up to the fifth level in the neutral sector, as it has been done in the development of the products of basic fields, $\psi$, $\psi^{+}$, which are at the origin of the algebra. We have developed accordingly the products in (\ref{eq1.3})-(\ref{eq1.5}), (\ref{eq1.10}), and similarly it should have been done for the products in (\ref{eq1.7})-(\ref{eq1.9}).

Due to the fact that we have suppressed a certain number of singular terms, in (\ref{eq1.3})-(\ref{eq1.9}), the corresponding commutation rules of the modes will miss a certain number of matrix elements, of the mode operators of higher (dimension) fields. But, working the operator algebra commutation rules backwards, the modes of higher fields could be factorized into products 
of the modes of basic operators, $\psi$, $\psi^{+}$, and finally the missing matrix elements could be calculated, through the commutation rules of modes of 
the basic fields. Example of such "going through factorization" calculation will be given in Section 5, when calculating the correlation function $<B(\infty)\tilde{\psi}(1)\tilde{\psi}^{+}(z)B(0)>$.

In developing  products (\ref{eq1.3})-(\ref{eq1.9})  the additional constants get involved, 
$\eta$, $\tilde{\lambda}$, $\mu$, $\tilde{\gamma}$, $b$, in addition 
to $\zeta$ and $\gamma$ already appearing in (\ref{eq1.1}), (\ref{eq1.2}). They have all to be defined by the associativity calculations (next Section).

In the next Section, by a particular technique of analyzing different triple products of operators $\psi$, $\psi^{+}$, $\tilde{\psi}$, $\tilde{\psi}^{+}$, $B$, we shall define all the constants, as functions of the central charge $c$ 
which remains a free parameter, -- the equations (\ref{eq1.19})-(\ref{eq1.26}).

Still the question remains of the final proof for the associativity of the algebra. We take it that the proof is complete if all the four-point functions of the fields $\psi$, $\psi^{+}$, $\tilde{\psi}$, $\tilde{\psi}^{+}$, $B$ are consistently defined.

They are not so numerous in this theory. We have the following list of 14 functions :
\bea
<\psi\psi\psi^{+}\psi^{+}>,
<\tilde{\psi}\psi\psi^{+}\psi^{+}>,
<\tilde{\psi}\tilde{\psi}\psi^{+}\psi^{+}>,
<\psi\tilde{\psi}\tilde{\psi}^{+}\psi^{+}>,
<\tilde{\psi}\tilde{\psi}\tilde{\psi}^{+}\psi^{+}>\nn\\
<\tilde{\psi}\tilde{\psi}\tilde{\psi}^{+}\tilde{\psi}^{+}>,
<\psi\psi\psi B>,
<\tilde{\psi}\psi\psi B>,
<\tilde{\psi}\tilde{\psi}\psi B>,
<\tilde{\psi}\tilde{\psi}\tilde{\psi}B>\nn\\
<\psi\psi^{+}BB>,
<\tilde{\psi}\psi^{+}BB>,
<\tilde{\psi}\tilde{\psi}^{+}BB>,
<BBBB> \label{eq1.41}
\eea
Here $<\psi\psi\psi^{+}\psi^{+}>=<\psi(\infty)\psi(1)\psi^{+}(z)\psi^{+}(0)>$, etc. . Evidently, the particular positioning of four fields is not important, and we assume the $Z_{3}$ conjugation invariance of correlation functions :
\beq
<\psi\psi\psi^{+}\psi^{+}>=(<\psi\psi\psi^{+}\psi^{+}>)^{+}=<\psi^{+}\psi^{+}\psi\psi> \label{eq1.42}
\eeq
etc., -- the algebra (\ref{eq1.1})-(\ref{eq1.9}) being $Z_{3}$ conjugation invariant.

Although we evoke the definition of all the 4-point functions, as a final proof, which seems to be correct, the actual method, in fact,  is simpler. In general terms, for the moment, it could be described as follows.

In analyzing the triple products, as we shall do it in the next Section, there is a finite number of relations, for lowest matrix elements, in which, on one hand, all the matrix elements could be defined directly, using the expansions in (\ref{eq1.1})-(\ref{eq1.9}), on the other hand these matrix elements enter into the lowest commutation rules relations, containing only the directly definable matrix elements.
Consistency requires that these lowest commutation relations, 
a finite number of them, should be verified. This provides algebraic equations on the constants $\lambda$, $\zeta$, $\eta$, $\tilde{\lambda}$, $\gamma$, $\mu$, $\tilde{\gamma}$, $b$.

Considering different triple products and extracting, in each case, 
the lowest commutation relations, with all matrix elements definable 
directly, one finally determines all the constants (or gets blocked at some point, which was not the case with the algebra (\ref{eq1.1})-(\ref{eq1.9})). The procedure is well illustrated by the calculations in the next Section.

If all these lowest commutation rules relations, with all the matrix elements
definable directly, are satisfied, then this ensures that all 
four-point functions will be consistent automatically.

This assumes that the 4-point functions are finally calculated by using the higher commutation relations, involving matrix elements which are not definable directly. The letter matrix elements enter into
the series expansion of 4-point functions; they are defined exclusively by the commutation rules.

To summarize, all the consistencies are ensured 
(or not) when the lowest commutation relations are analyzed, 
those involving only the matrix elements definable directly.

In the next Section the methods and techniques are exposed and the values (\ref{eq1.19})-(\ref{eq1.26}) of the operator algebra constants are obtained.

In Section 3 we do the complete classification of the triple products
(into three distinct classes) 
and we define a subset of them, of those which provide a complete set of equations
to be satisfied to ensure the associativity of the chiral algebra.

Section 4 is devoted to various verifications, illustrations, by calculating 
4-point functions related to the second class of triple products.

In Section 5 the function   $<B(\infty)\tilde{\psi}(1)\tilde{\psi}^{+}(z)B(0)>$
is calculated, to illustrate the method of dealing with higher (conformal
dimension) fields and the triple products of the third class.

Section 6 contains some conclusions.

Various Appendices contain values of $\beta$ coefficients and
some more technical aspects of the calculations.  

\vskip0.7cm

We finish the Introduction with some suggestions for the readers of this paper, which is too long.

As has already been stated in the Abstract, in this paper, while dealing 
with the proof of the associativity of a particular 
non-standard parafermionic algebra, 
maybe the main result of this article is the new method of the associativity analysis of the chiral algebras of parafermionic type. This second aspect of the article might be of interest for a more general audience.

The presentation, of both aspects, was intended to be as complete as possible, with all the subtleties of the method to be unraveled. There are many calculations, in the Sections 4 and 5, which are intended to be 
the demonstrations and verifications of the general classification 
of Section 3, the Section which summarizes the method of the associativity analysis, 
on the example of our particular non-standard parafermionic algebra.

This explains the length of the paper.

But the opposite aspect of the very detailed presentation is, 
and the structure of the paper is such, that one is not supposed to read all the paper through, for the first, usual reading. We may suggest three possible 
levels of looking through the paper.

For the general reading, many of the detailed calculations can be omitted.
To understand the method and the structure of the associativity proof, in more or less general terms, it is sufficient to read the Introduction and the Sections 2 and 3. In  Section 2 the method is developed, progressively, 
"in the action", while the Section  3 summarizes and gives the general classification and the structure of the method and on the proof. 
The Sections 4 and 5 and all the Appendices could safely be dropped, 
for the first, general reading. 

For more specialized reading, when going into the subtleties of the statements made in Section 3, the Section 4 might be very useful.
It deals principally with the triple products of the second type 
and the related correlation functions.
The Appendices B and C might also be helpful at this point.

Finally the Section 5, and the Appendix D which goes with it, 
are intended for a very specialized reading. The Section 5 is very technical, containing most complicated aspects of the calculations. It is intended to show that the classification of the Section 3 gets confirmed 
for the third type of triple products, and the 4-point functions which are associated with them. It is also intended to demonstrate that, in fact, 
the correlation functions listed here, in the Introduction, eq.(1.21), and
in the Section 3, could all be calculated. 

The Section 5 might appear to be very useful for someone who will really work on the subject, or on closely related problems. Otherwise, it could safely be dropped without any damage for understanding the rest of the paper.

\numberwithin{equation}{subsection}

\section{Defining the constants of the operator\\ algebra (\ref{eq1.1})-(\ref{eq1.9})}

Let us consider the correlation function
\beq
<\psi^{+}(\infty)\psi(1)\psi(z)\psi^{+}(0)> \label{eq2.1}
\eeq

We shall try to show most of our methods of associativity calculations on the example of this function. We shall not define all the constants with it, but the methods will be exposed. In the following these same methods will be applied to other functions, to complete the definition of the operator algebra constants and to make further checks.

As has been stated in the last part of the previous Section, analyzing triple products is, in fact, a more efficient way of getting the associativity constraint equations, on the constants. We shall come to that, progressively. But let us take it, for the moment, that our problem is the 4-point function (\ref{eq2.1}).

By looking at the limits $z\rightarrow 0$ and $z\rightarrow 1$, and by using the operator algebra expansions (\ref{eq1.1})-(\ref{eq1.2}), we conclude that the analytic form of the function (\ref{eq2.1}) should be as follows :
\beq
<\psi^{+}(\infty)\psi(1)\psi(z)\psi^{+}(0)>=\frac{P_{n}(z)}{(1-z)^{\Delta_{\psi}}(z)^{2\Delta_{\psi}}} \label{eq2.2}
\eeq
where $P_{n}(z)$ is a polynomial, of degree $n$ to be determined,
\beq
P_{n}(z)=a_{0}+a_{1}z+a_{2}z^{2}+...+a_{n}z^{n} \label{eq2.3}
\eeq
To define $n$, let us look at the limit $z\rightarrow\infty$. To analyze this limit it is more convenient to rewrite the equation (\ref{eq2.2}) as follows :
\beq
<\psi^{+}(\infty)\psi(z)\psi(1)\psi^{+}(0)>=\frac{P_{n}(z)}{(z-1)^{\Delta_{\psi}}(z)^{2\Delta_{\psi}}} \label{eq2.4}
\eeq

One could think that in (\ref{eq2.2}) $z$ was positioned between $0$ and $1$, and that, going to (\ref{eq2.4}) we have continued $z$, around $1$, to the interval $(1,\infty)$. In doing so, on the l.h.s. of (\ref{eq2.2}), we continue the field $\psi(z)$ around $\psi(1)$. This produces a phase factor which is finally compensated by the phase factor produced on the r.h.s. of (\ref{eq2.2}) while continuing the factor $1/(1-z)^{\Delta_{\psi}}$. Analytical continuations of parafermionic fields, on one side, and analytical expressions on the other, with the effect of compensation, are discussed in greater detail in the paper [2], in the context of the second series of parafermions.

Now we can send directly $z$ to $\infty$, on both sides of (\ref{eq2.4}).

On the r.h.s. one finds, when $z\rightarrow\infty$,
\beq
\frac{P_{n}(z)}{(z-1)^{\Delta_{\psi}}(z)^{2\Delta_{\psi}}}\simeq\frac{a_{n}z^{n}}{(z)^{3\Delta_{\psi}}}=\frac{a_{n}}{(z)^{3\Delta_{\psi}-n}} \label{eq2.5}
\eeq

On the l.h.s. of (\ref{eq2.4}), by using the OPE (\ref{eq1.2}), one finds :
\beq
<\psi^{+}(\infty)\psi(z)\psi(1)\psi^{+}(0)>\simeq<\psi^{+}(\infty)\psi(z)\cdot 1>=1 \label{eq2.6}
\eeq
We have used the fact that $|z|>>1$, so that $\psi(1)\psi^{+}(0)$ are relatively close, one to another, and we can replace their product by 1, according to (\ref{eq1.2}). Next $<\psi^{+}(\infty)\psi(z)>=1$, according to the standard limiting procedure which defines the correlation functions with one of the fields at $\infty$.

Comparing (\ref{eq2.5}) and (\ref{eq2.6}) we have to admit that
\beq
3\Delta_{\psi}-n=0 \quad \rightarrow \quad n=8 \label{eq2.7}
\eeq
and
\beq
a_{8}=1 \label{eq2.8}
\eeq
so that our correlation function is of the following general form
\beq
<\psi^{+}(\infty)\psi(1)\psi(z)\psi^{+}(0)>=\frac{P_{8}(z)}{(1-z)^{\Delta_{\psi}}(z)^{2\Delta_{\psi}}} \label{eq2.9}
\eeq
Comparing the operator algebra developments in $z$ and in $1/z$, for the limits $z\rightarrow 0$ and $z\rightarrow\infty$ in (\ref{eq2.2}) and (\ref{eq2.4}), one finds easily that the coefficients of $P_{8}(z)$ have to be symmetric :
\beq
a_{0}=a_{8}=1 \label{eq2.10}
\eeq
\beq
a_{1}=a_{7},\quad a_{2}=a_{6},\quad a_{3}=a_{5} \label{eq2.11}
\eeq

Finally, by taking the limit $z\rightarrow 1$, one finds :
\bea
<\psi^{+}(\infty)\psi(1)\psi(z)\psi^{+}(0)>\simeq<\psi^{+}(\infty)\cdot\frac{\lambda}{(1-z)^{\Delta_{\psi}}}\psi^{+}(1)\cdot \psi^{+}(0)>\nn\\
=\frac{\lambda}{(1-z)^{\Delta_{\psi}}}<\psi^{+}(\infty)\psi^{+}(1)\psi^{+}(0)>=\frac{\lambda^{2}}{(1-z)^{\Delta_{\psi}}} \label{eq2.12}
\eea
-- for the l.h.s. of (\ref{eq2.2}) ( we have used (\ref{eq1.1})), and 
\beq
\frac{P_{8}(z)}{(1-z)^{\Delta_{\psi}}(z)^{2\Delta_{\psi}}}\simeq\frac{\sum^{8}_{k=0}a_{k}}{(1-z)^{\Delta_{\psi}}} \label{eq2.13}
\eeq
for the r.h.s., so that
\beq
\sum_{k=0}^{8}a_{k}=\lambda^{2} \label{eq2.14}
\eeq
The relations (\ref{eq2.10}), (\ref{eq2.11}), (\ref{eq2.14}) have to be verified by the coefficients of the polynomial $P_{8}(z)$, in the analytic form (\ref{eq2.9}) of our function.

Having established the analytic form of the function $<\psi^{+}\psi\psi\psi^{+}>$, we proceed now to its evaluation, by the way of the operator
algebra. Unknown still are the coefficients $a_{1}, ..., a_{7}$ 
of the polynomial $P_{8}(z)$.

\vskip1cm


\subsection{Calculation of the function $<\psi^{+}(\infty)\psi(1)\psi(z)\psi^{+}(0)>$ by its development in powers of $z$.}

We take a choice to do the calculation by first developing the function
$<\psi^{+}(\infty)\psi(1)\psi(z)\psi^{+}(0)>$ in powers of $z$, as if we consider the limit $z\rightarrow 0$, though this time we shall define the complete form of the function. In doing so, we have to develop first the product $\psi(z)\psi^{+}(0)$. Next we shall have to multiply this development, each term, by $\psi(1)$ and redevelop. Finally we shall have to project the obtained series onto $\psi^{+}(\infty)$. In this way we shall obtain a series for our function in powers of $z$, with coefficients expressed in terms of matrix elements of the parafermionic algebra : matrix elements of mode operators of our parafermionic fields. This expansion will have to be compared with the expansion in $z$ of the analytic form, in the r.h.s. of (\ref{eq2.9}). In this way we shall determine the coefficients of the polynomial $P_{8}(z)$, which have to verify all the required properties. This is our small program, for the beginning.

First we have to develop $\psi(z)\psi^{+}(0)$, by introducing the mode operators of $\psi(z)$. We have to evoke the structure of the modules in the third $Z_{3}$ parafermionic theory, our present one, with $\Delta_{\psi}=8/3$. We shall denote this theory, sometimes, as $Z_{3}^{(3)}$; the first two $Z_{3}$ theories are those having $\Delta_{\psi}=2/3$ 
and $\Delta_{\psi}=4/3$, [3,4].

The module of the identity operator is presented in the Fig.1. From the structure of this module we could read the structure of the charged operators module, with operators $\Phi^{(+1)}$, $\Phi^{(-1)}$ at the top, Fig.2.

\begin{figure}
\begin{center}
\epsfxsize=300pt\epsfysize=380pt{\epsffile{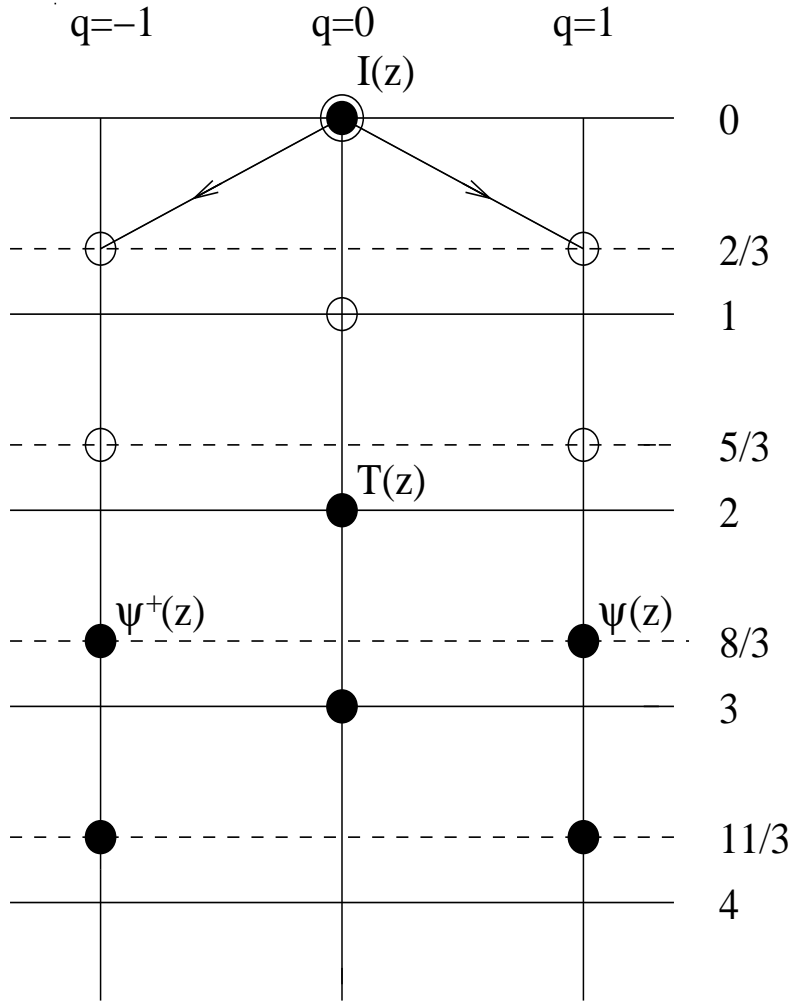}}
\caption{The structure of the identity operator module.
$q=0, \pm 1$ are the $Z_{3}$ charges and we have assumed that the
$Z_{3}$ charge of $\psi(z)$ is $q=1$ and that of $\psi^{+}(z)$ is
$q=-1$. In the case of the identity operator module, in the charged sectors
$q=\pm 1$, all the levels above $\psi(z)$ are empty (empty circles).
For other neutral fields $\Phi^{0}(z)$ ($q=0$) at the top, they should
be filled, in general, with parafermionic descendants. 
We resume that positioning of the operators $\psi(z)$, $\psi^{+}(z)$
dictates the structure of the identity operator module, and of the
neutral fields module more generally.
\label{fig1}
}
\end{center}
\end{figure}

\begin{figure}
\begin{center}
\epsfxsize=300pt\epsfysize=380pt{\epsffile{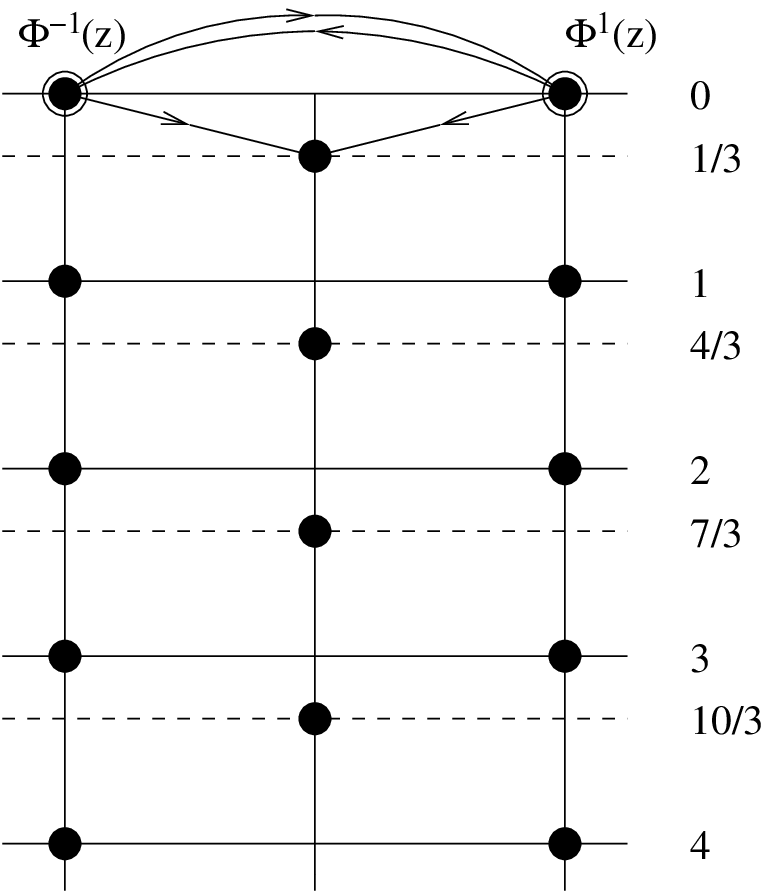}}
\caption{Charged operators module, with, charged fields $\Phi^{1}(z)$,
$\Phi^{-1}(z)$ at the top.  Its structure is obtained from the
identity module, Fig.1, by suppressing the level $0$, in Fig.1, and
starting with the level 2/3, at the top, which becomes level $0$ in
the present Figure.
\label{fig2}
}
\end{center}
\end{figure}

Important information for the present is that the "gap" in the neutral fields
module is 2/3, which corresponds to the first occupied (in principal) level
below the level $0$, Fig.1. In the charged fields module the gap is 1/3, 
Fig. 2. This implies the following developments of $\psi(z)$, $\psi^{+}(z)$ in mode operators, when they are applied to neutral and charged fields:
\beq
\psi(z)\Phi^{0}(0)=\sum_{n}\frac{1}{(z)^{\Delta_{\psi}-\frac{2}{3}+n}}\psi_{-\frac{2}{3}+n}\Phi^{0}(0) \label{eq2.15}
\eeq
\beq
\psi^{+}(z)\Phi^{0}(0)=\sum_{n}\frac{1}{(z)^{\Delta_{\psi}-\frac{2}{3}+n}}\psi^{+}_{-\frac{2}{3}+n}\Phi^{0}(0) \label{eq2.16}
\eeq
\beq
\psi(z)\Phi^{1}(0)=\sum_{n}\frac{1}{(z)^{\Delta_{\psi}+n}}\psi_{n}\Phi^{1}(0) \label{eq2.17}
\eeq
\beq
\psi(z)\Phi^{-1}(0)=\sum_{n}\frac{1}{(z)^{\Delta_{\psi}-\frac{1}{3}+n}}\psi_{-\frac{1}{3}+n}\Phi^{-1}(0) \label{eq2.18}
\eeq
\beq
\psi^{+}(z)\Phi^{1}(0)=\sum_{n}\frac{1}{(z)^{\Delta_{\psi}-\frac{1}{3}+n}}\psi^{+}_{-\frac{1}{3}+n}\Phi^{1}(0) \label{eq2.19}
\eeq
\beq
\psi^{+}(z)\Phi^{-1}(0)=\sum_{n}\frac{1}{(z)^{\Delta_{\psi}+n}}\psi^{+}_{n}\Phi^{-1}(0) \label{eq2.20}
\eeq
In the developments (\ref{eq2.17}), (\ref{eq2.20}) there are no gaps since the uppermost actions of mode operators are horizontal, Fig.2. Evidently, we are assuming the $Z_{3}$ addition of charges, so that $1+1=-1$ and $-1-1=+1$.

Now, for the product $\psi(z)\psi^{+}(0)$, we take it as if $\psi(z)$ is applied to the charged field $\psi^{+}(0)$, of the charge $q=-1$, and we have to use the development (\ref{eq2.18}). We obtain :
\beq
\psi(z)\psi^{+}(0)=\sum_{n}\frac{1}{(z)^{\Delta_{\psi}-\frac{1}{3}+n}}\psi_{-\frac{1}{3}+n}\psi^{+}(0) \label{eq2.21} 
\eeq
We have to compare this development (the highest weight representation type) with direct expansion of this product in (\ref{eq1.2}). First of all, from (\ref{eq1.2}) we observe that the most singular term in (\ref{eq2.21}) (singular, if $z\rightarrow 0$) should be :
\beq
\sim\frac{1}{(z)^{2\Delta_{\psi}}}=\frac{1}{(z)^{16/3}} \label{eq2.22}
\eeq
Comparing with the power $\Delta_{\psi}-\frac{1}{3}+n$ in (\ref{eq2.21}) we conclude that the series in (\ref{eq2.21}) starts with $n=3$. Comparing further, term by term, the expansions in (\ref{eq1.2}) and (\ref{eq2.21}), we have to identify the following actions of the $\psi$ mode operators on $\psi^{+}(0)$ :
\beq
\psi_{-\frac{1}{3}+n}\psi^{+}(0)=0,\quad n>3 \label{eq2.23}
\eeq
\beq
\psi_{-\frac{1}{3}+3}\psi^{+}(0)=\psi_{\frac{8}{3}}\psi^{+}(0)=1 \label{eq2.24}
\eeq
\beq
\psi_{\frac{5}{3}}\psi^{+}(0)=0 \label{eq2.25}
\eeq
\beq
\psi_{\frac{2}{3}}\psi^{+}(0)=\frac{2\Delta_{\psi}}{c}T(0) \label{eq2.26}
\eeq
\beq
\psi_{-\frac{1}{3}}\psi^{+}(0)=\frac{\Delta_{\psi}}{c}\partial T(0) \label{eq2.27}
\eeq
\beq
\psi_{-\frac{4}{3}}\psi^{+}(0)=\beta_{\psi\psi^{+},I}^{(112)}\cdot\partial^{2}T(0)+\beta^{(22)}_{\psi\psi^{+},I}\cdot\wedge(0)+\gamma\cdot B(0) \label{eq2.28}
\eeq
\beq
\psi_{-\frac{7}{3}}\psi^{+}(0)=\beta_{\psi\psi^{+},I}^{(1112)}\cdot\partial^{3}T(0)+\beta^{(122)}_{\psi\psi^{+},I}\cdot\partial\wedge(0)+\gamma\cdot \beta^{(1)}_{\psi\psi^{+},B}\cdot\partial B(0) \label{eq2.29}
\eeq
These actions of mode operators are know explicitly, they are dictated by the OPE in (\ref{eq1.2}). Actions of still lower modes, $\psi_{-10/3}\psi^{+}(0)$, etc., are not known explicitly, they are just descendant operators of the parafermionic algebra. Their matrix elements, needed in particular in the calculation of the correlation function (\ref{eq2.1}), they will have to be determined by the commutation rules of the mode operators, which we shall have to define.

Resuming it again, there is a certain number of actions of mode operators which are know explicitly, equations (\ref{eq2.23})-(\ref{eq2.29}). The rest are just the highest weight representation states, descendants, which should be dealt with by the commutation rules algebra of the mode operators.  

We observe, in addition, that $\psi^{+}(0)$ is not annihilated by the positifs modes $\psi_{8/5}$, $\psi_{2/3}$, eqs.(\ref{eq2.24}), (\ref{eq2.26}), which should be the case if we applied these modes to a primary field, certain field $\Phi^{-1}(0)$, primary with respect to the parafermionic algebra. This is because $\psi^{+}(0)$ is not a primary field, of the parafermion algebra. In fact, $\psi^{+}(0)$ should be viewed as a descendant of the identity $I$,
\beq
\psi^{+}(0)=\psi^{+}_{-\frac{8}{3}}I(0) \label{eq2.30}
\eeq	
in the module in Fig.1.

After these remarks we shall present the expansion in (\ref{eq2.21}) in a somewhat more explicit form, as
\bea
\psi(z)\psi^{+}(0)=\frac{1}{(z)^{2\Delta_{\psi}}}\times\{\psi_{\frac{8}{3}}\psi^{+}(0)+z\psi_{\frac{5}{3}}\psi^{+}(0)+z^{2}\psi_{\frac{2}{3}}\psi^{+}(0)\nn\\
+z^{3}\psi_{-\frac{1}{3}}\psi^{+}(0)+z^{4}\psi_{-\frac{4}{3}}\psi^{+}(0)+z^{5}\psi_{-\frac{7}{3}}\psi^{+}(0)+z^{6}\psi_{-\frac{10}{3}}\psi^{+}(0)+...\} \label{eq2.31}
\eea
Next we apply, to this development, the field $\psi(1)$, present in the correlation function (\ref{eq2.1}). We obtain :
\beq
\psi(1)\psi(z)\psi^{+}(0)=\frac{1}{(z)^{2\Delta_{\psi}}}\cdot\{\psi(1)\psi_{\frac{8}{3}}\psi^{+}(0)+z\psi(1)\psi_{\frac{5}{3}}\psi^{+}(0)+z^{2}\psi(1)\psi_{\frac{2}{3}}\psi^{+}(0)+...\} \label{eq2.32}
\eeq
Next, we have to develop $\psi(1)$, in mode operators, as applied to the $Z_{3}$ neutral fields $\psi_{8/3}\psi^{+}(0)$, $\psi_{5/3}\psi^{+}(0)$, $\psi_{2/3}\psi^{+}(0)$, etc. . We have to use the expansion in modes, around $0$, (\ref{eq2.15}).

For the first term in (\ref{eq2.32}) we obtain:
\beq
\psi(1)\psi_{\frac{8}{3}}\psi^{+}(0)=\sum_{n}\frac{1}{(1-0)^{\Delta_{\psi}-\frac{2}{3}+n}}\psi_{-\frac{2}{3}+n}\psi_{\frac{8}{3}}\psi^{+}(0) \label{eq2.33}
\eeq
The descendant field $\psi_{-2/3+n}\psi_{8/3}\psi^{+}(0)$ belongs to the charged sector, $q=1$, and stays on the level
\beq
(\frac{2}{3}-n)-\frac{8}{3}+\Delta_{\psi} \label{eq2.34}
\eeq
in the module of the identity Fig.1. (In general, dealing with products of the chiral algebra fields, we always stay in the module of the identity operator). In the module in Fig.1, the uppermost field, in the charged sector $q=1$, is $\psi(z)$, level $\Delta_{\psi}=8/3$. This implies that in the series (\ref{eq2.33}) the first non-trivial action, of $\psi_{-2/3+n}$ on $\psi_{8/3}\psi^{+}(0)$, is that for $n=-2$. So we could write, 
from (\ref{eq2.33}), that
\beq
\psi(1)\psi_{\frac{8}{3}}\psi^{+}(0)=\psi_{-\frac{8}{3}}\psi_{\frac{8}{3}}\psi^{+}(0)+\psi_{-\frac{11}{3}}\psi_{\frac{8}{3}}\psi^{+}(0)+\psi_{-\frac{14}{3}}\psi_{\frac{8}{3}}\psi^{+}(0)+... \label{eq2.35}
\eeq
In this series
\beq
\psi_{-\frac{8}{3}}\psi_{\frac{8}{3}}\psi^{+}(0)=\psi_{-\frac{8}{3}}I(0)=\psi(0) \label{eq2.36}
\eeq 
\beq
\psi_{-\frac{11}{3}}\psi_{\frac{8}{3}}\psi^{+}(0)=\psi_{-\frac{11}{3}}I(0)=\partial\psi(0)\label{eq2.37}
\eeq 
\beq
\psi_{-\frac{14}{3}}\psi_{\frac{8}{3}}\psi^{+}(0)=\psi_{-\frac{14}{3}}I(0)=\frac{1}{2}\partial^{2}\psi(0) \label{eq2.38}
\eeq 
etc. . In fact, comparing
\beq
\psi(z)=\psi(0)+z\partial\psi(0)+\frac{1}{2}z^{2}\partial^{2}\psi(0)+... \label{eq2.39}
\eeq
and
\beq
\psi(z)=\psi(z)I(0)=\sum_{n}\frac{1}{(z)^{\Delta_{\psi}-\frac{2}{3}+n}}\psi_{-\frac{2}{3}+n}I(0)=\sum_{n}\frac{1}{(z)^{2+n}}\psi_{-\frac{2}{3}+n}I(0) \label{eq2.40}
\eeq
we have to conclude that
\beq
\psi_{-\frac{8}{3}}I(0)=\psi(0) \label{eq2.41}
\eeq
\beq
\psi_{-\frac{11}{3}}I(0)=\partial\psi(0) \label{eq2.42}
\eeq
\beq
\psi_{-\frac{14}{3}}I(0)=\frac{1}{2}\partial^{2}\psi(0) \label{eq2.43}
\eeq
etc., which have been used in (\ref{eq2.36})-(\ref{eq2.38}).

Now, when the series in (\ref{eq2.35}) is finally projected onto $\psi^{+}(\infty)$, in the correlation function (\ref{eq2.1}), will remain, in (\ref{eq2.35}) just the first term, $\psi_{-8/3}\psi_{8/3}\psi^{+}(0)\sim\psi(0)$, the others will disappear. Assuming this projection, we could write
\beq
\psi(1)\psi_{\frac{8}{3}}\psi^{+}(0)=\psi_{-\frac{8}{3}}\psi_{\frac{8}{3}}\psi^{+}(0) \label{eq2.44}
\eeq
-- under projection.

Similarly, for the second term in (\ref{eq2.32}), we shall find that
\beq
\psi(1)\psi_{\frac{5}{3}}\psi^{+}(0)=\psi_{-\frac{5}{3}}\psi_{\frac{5}{3}}\psi^{+}(0) \label{eq2.45}
\eeq
-- under projection. And so on. So that the expansion for the triple product in (\ref{eq2.32}) takes the form :
\beq
\psi(1)\psi(z)\psi^{+}(0)=\frac{1}{(z)^{2\Delta_{\psi}}}\cdot\{\psi_{-\frac{8}{3}}\psi_{\frac{8}{3}}\psi^{+}(0)+z\psi_{-\frac{5}{3}}\psi_{\frac{5}{3}}\psi^{+}(0)+z^{2}\psi_{-\frac{2}{3}}\psi_{\frac{2}{3}}\psi^{+}(0)+...\} \label{eq2.46}
\eeq
-- under projection onto $\psi^{+}(\infty)$ in (\ref{eq2.1}). If we introduce the following notation for the matrix elements:
\beq
\psi_{-\frac{8}{3}}\psi_{\frac{8}{3}}\psi^{+}(0)=m_{0}\psi(0) \label{eq2.47}
\eeq
\beq
\psi_{-\frac{5}{3}}\psi_{\frac{5}{3}}\psi^{+}(0)=m_{1}\psi(0) \label{eq2.48}
\eeq
\beq
\psi_{-\frac{2}{3}}\psi_{\frac{2}{3}}\psi^{+}(0)=m_{2}\psi(0) \label{eq2.49}
\eeq
\beq
\psi_{\frac{1}{3}}\psi_{-\frac{1}{3}}\psi^{+}(0)=m_{3}\psi(0) \label{eq2.50}
\eeq
etc., then the $z$ expansion for the correlation function (\ref{eq2.1}) takes the form :
\beq
<\psi^{+}(\infty)\psi(1)\psi(z)\psi^{+}(0)>=\frac{1}{(z)^{2\Delta_{\psi}}}\cdot\{ m_{0}+m_{1}z+m_{2}z^{2}+m_{3}z^{3}+...\} \label{eq2.51}
\eeq
-- the coefficients in this series are to be calculated as matrix elements in (\ref{eq2.47})-(\ref{eq2.50}).

 When these matrix elements are defined, so that the coefficients $m_{0}$, $m_{1}$, $m_{2}$... are known, then the expansion in (\ref{eq2.51}) could be compared with the $z$  expansion of the analytic form (\ref{eq2.9}), for our function (\ref{eq2.1}), and the coefficients of the polynomial $P_{8}(z)$ could be determined.
 
 Resuming, we have to be able to define the matrix elements in (\ref{eq2.47})-(\ref{eq2.50}) (certainly not limited by the matrix element in (\ref{eq2.50})). For this purpose we need to derive the commutation rules for the modes of $\psi$ and $\psi$, when applied to $\psi^{+}(0)$. We shall denote theses commutation rules as $\{\psi,\psi\}\psi^{+}(0)$.
 
 The derivation will be based on the OPE (\ref{eq1.1}). We shall proceed in a standard way, as it has been done, for parafermionic algebras in [3,4,2,5]. We write three integrals :
 \beq
 I_{1}=\frac{1}{(2\pi i)^{2}}\oint_{C'_{0}}dz'(z')^{\Delta_{\psi}-\frac{1}{3}+n}\oint_{C_{0}}dz(z)^{_{\psi}-\frac{1}{3}+m}\times(z'-z)^{\Delta_{\psi}-3}\times\psi(z')\psi(z)\psi^{+}(0) \label{eq2.52}
 \eeq
\beq
I_{2}=\frac{1}{(2\pi i)^{2}}\oint_{C_{0}}dz(z)^{\Delta_{\psi}-\frac{1}{3}+m}\oint_{C'_{0}}dz'(z')^{\Delta_{\psi}-\frac{1}{3}+n}\times(z-z')^{\Delta_{\psi}-3}\times\psi(z)\psi(z')\psi^{+}(0) \label{eq2.53}
\eeq
\bea
I_{3}=\frac{1}{(2\pi i)^{2}}\oint_{C_{0}}dz(z)^{\Delta_{\psi}-\frac{1}{3}+m-1}\oint_{C_{z}}dz'(z')^{\Delta_{\psi}-\frac{1}{3}+n-1}\nn\\
\times\frac{1}{(z'-z)^{3}}\{ \lambda\psi^{+}(z)+(z'-z)\lambda\beta^{(1)}_{\psi\psi,\psi^{+}}\cdot\partial\psi^{+}(z)\nn\\
+(z'-z)^{2}[\lambda\beta^{(11)}_{\psi\psi,\psi^{+}}\cdot\partial^{2}\psi^{+}(z)+\lambda\beta^{(2)}_{\psi\psi,\psi^{+}}\cdot L_{-2}\psi^{+}(z)+\zeta\tilde{\psi}^{+}(z)]\}\psi^{+}(0) \label{eq2.54}
\eea
In the third integral, the integrand expression just replaces $(z'-z)^{\Delta_{\psi}-3}\cdot\psi(z')\psi(z)$ when $\psi(z')\psi(z)$ is developed as in (\ref{eq1.1}) and the higher terms in (\ref{eq1.1}), starting with the term $\sim(z'-z)^{3}$, are dropped because they do not contribute to the integral. The power $\Delta_{\psi}-3$ of $(z'-z)^{\Delta_{\psi}-3}$ in $I_{1}$, $I_{2}$, $I_{3}$ is chosen so that all the explicit terms in (\ref{eq1.1}) contribute and the non-explicit terms, dots in (\ref{eq1.1}), could be dropped when integrating over $z'$ around $z$, in the integral $I_{3}$.

The integration contour of $I_{1}$, $I_{2}$, $I_{3}$ are shown in Fig.3.

By the standard analytic continuation procedure (involving a deformation of integration contours and also the continuation of the fields $\psi(z')$, $\psi(z)$, one around another, and compensation of the phase factors produced) one establishes the following relation between the integrals $I_{1}$, $I_{2}$, $I_{3}$ :
\beq
I_{1}+I_{2}=I_{3} \label{eq2.55}
\eeq
We observe that we have "$+$" in the l.h.s. of the above relation, instead of more usual "$-$", because of the odd extra power (which is $- 3$) of the factor $(z'-z)^{\Delta_{\psi}-3}$ in the integrals.

\begin{figure}
\begin{center}
\epsfxsize=250pt\epsfysize=600pt{\epsffile{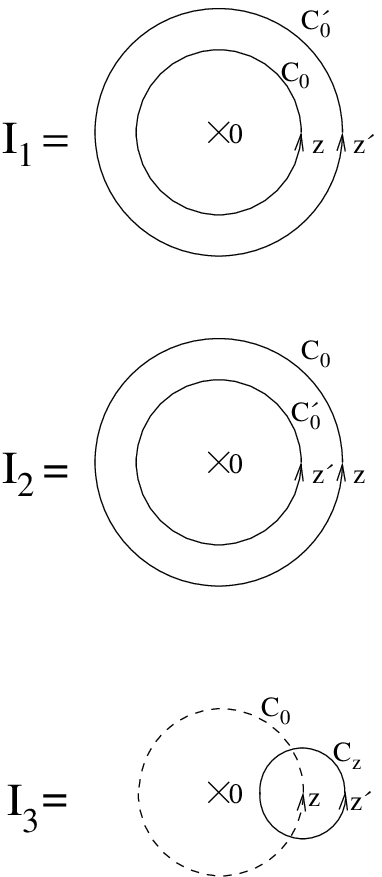}}
\caption{The integration contours of the integrals $I_{1}$, $I_{2}$, $I_{3}$ in equations 
(\ref{eq2.52})--(\ref{eq2.54}).
\label{fig3}
}
\end{center}
\end{figure}

We develop
\beq
(z'-z)^{\Delta_{\psi}-3}=(z')^{-\frac{1}{3}}(1-\frac{z}{z'})^{-\frac{1}{3}}=(z')^{-\frac{1}{3}}\sum_{l=0}^{\infty}D^{l}_{-\frac{1}{3}}(\frac{z}{z'})^{l} \label{eq2.56}
\eeq
-- in $I_{1}$,
\beq
(z-z')^{-\frac{1}{3}}=(z)^{-\frac{1}{3}}\sum_{l=0}^{\infty}D_{-\frac{1}{3}}^{l}(\frac{z'}{z})^{l} \label{eq2.57}
\eeq
-- in $I_{2}$, and we calculate $I_{3}$ by the theorem of residues. In the above, $D^{l}_{-\frac{1}{3}}$ are the binomial coefficients of the expansion :
\beq
(1-u)^{\alpha}=\sum_{l=0}^{\infty}D_{\alpha}^{l}\cdot u^{l} \label{eq2.58}
\eeq

Expressing the integrals $I_{1}$, $I_{2}$ in terms of the modes of the fieds $\psi(z')$, $\psi_(z)$ one finds, from (\ref{eq2.55}),
\beq
\sum_{l=0}^{\infty}D_{-\frac{1}{3}}^{l}(\psi_{-\frac{2}{3}+n-l}\psi_{-\frac{1}{3}+m+l}+\psi_{-\frac{2}{3}+m-l}\psi_{-\frac{1}{3}+n+l})\psi^{+}(0)=R(n,m) \label{eq2.59}
\eeq
\bea
R(n,m)=\{\frac{1}{2}\cdot(\Delta_{\Psi}-\frac{1}{3}+n-1)(\Delta_{\psi}-\frac{1}{3}+n-2)\lambda\psi^{+}_{n+m-1}\nn\\
-(\Delta_{\psi}-\frac{1}{3}+n-1)\cdot(\Delta_{\psi}+n+m-1)\cdot\lambda\beta^{(1)}_{\psi\psi,\psi^{+}}\cdot\psi^{+}_{n+m-1}\nn\\
+(\Delta_{\psi}+n+m)(\Delta_{\psi}+n+m-1)\lambda\beta^{(11)}_{\psi\psi,\psi^{+}}\cdot\psi^{+}_{n+m-1}\nn\\
+\lambda\beta^{(2)}_{\psi\psi,\psi^{+}}\cdot(L_{-2}\psi^{+})_{n+m-1}+\zeta\tilde{\psi}^{+}_{n+m-1}\}\psi^{+}(0) \label{eq2.60}
\eea
$R(n,m)$ is the result of integration in the integral $I_{3}$, eq.(\ref{eq2.54}). ($R$ for the r.h.s.). Equation (\ref{eq2.59}) represents the commutation rules $\{\psi,\psi\}\psi^{+}(0)$.

It could be added that the modes of the parafermionic fields in $R(n,m)$, eq.(\ref{eq2.60}), are defined as in eq.(\ref{eq2.20}) and the modes $\psi_{-\frac{2}{3}+n-l}$, $\psi_{-\frac{2}{3}+m-l}$ in the l.h.s. of (\ref{eq2.59}) are defined as in eq.(\ref{eq2.15}), as they act on neutral fields $\psi_{-\frac{1}{3}+m+l}\psi^{+}(0)$ and $\psi_{-\frac{1}{3}+n+l}\psi^{+}(0)$.

It could be added also that, in getting the modes in (\ref{eq2.59}), (\ref{eq2.60}), we have to use the integral definitions of modes, which follow from (\ref{eq2.15})-(\ref{eq2.20}). For example, from (\ref{eq2.15}) follows that
\beq
\psi_{-\frac{2}{3}+n}\Phi^{0}(0)=\frac{1}{2\pi i}\oint_{C_{0}}dz(z)^{\Delta_{\psi}-\frac{2}{3}+n-1}\psi(z)\Phi^{0}(0) \label{eq2.61}
\eeq
and similarly for the other equations in (\ref{eq2.15})-(\ref{eq2.20}).

Now we turn back to the matrix elements in (\ref{eq2.47})-(\ref{eq2.50}). To make them appear in the commutation rules (\ref{eq2.59}), (\ref{eq2.60}), in the l.h.s., we have to impose the constraint :
\beq
n+m-1=0\quad\rightarrow\quad m=1-n \label{eq2.62}
\eeq
With $m=1-n$, the commutation rules take the form :
\beq
\sum_{l=0}^{\infty}D_{-\frac{1}{3}}^{l}(\psi_{-\frac{2}{3}+n-l}\psi_{\frac{2}{3}-n+l}+\psi_{\frac{1}{3}-n-l}\psi_{-\frac{1}{3}+n+l})\psi^{+}(0)=R(n)\label{eq2.63}
\eeq
\bea
R(n)=R(n,1-n)=\{\frac{1}{2}\cdot(\Delta_{\psi}-\frac{1}{3}+n-1)(\Delta_{\psi}-\frac{1}{3}+n-2)\lambda\psi^{+}_{0}\nn\\
-(\Delta_{\psi}-\frac{1}{3}+n-1)\cdot(\Delta_{\psi})\cdot\lambda\beta^{(1)}_{\psi\psi,\psi^{+}}\cdot\psi^{+}_{0}\nn\\
+(\Delta_{\psi}+1)(\Delta_{\psi})\lambda\beta^{(11)}_{\psi\psi,\psi^{+}}\cdot\psi^{+}_{0}\nn\\
+\lambda\beta^{(2)}_{\psi\psi,\psi^{+}}\cdot(L_{-2}\psi^{+})_{0}+\zeta\tilde{\psi}^{+}_{0}\}\psi^{+}(0) \label{eq2.64}
\eea

Let us list now the equations, for matrix elements, which follow from (\ref{eq2.63}), starting from the lowest one (in the values of indexes of the modes) and going upwards.

The lowest equation is obtained for \underline{$n=1$} (or $n=0$, which is symmetric) :
\bea
(\psi_{\frac{1}{3}}\psi_{-\frac{1}{3}}+D^{1}_{-\frac{1}{3}}\psi_{-\frac{2}{3}}\psi_{\frac{2}{3}}+D^{2}_{-\frac{1}{3}}\psi_{-\frac{5}{3}}\psi_{\frac{5}{3}}+D^{3}_{-\frac{1}{3}}\psi_{-\frac{8}{3}}\psi_{\frac{8}{3}}\nn\\
+\psi_{-\frac{2}{3}}\psi_{\frac{2}{3}}+D^{1}_{-\frac{1}{3}}\psi_{-\frac{5}{3}}\psi_{\frac{5}{3}}+D^{2}_{-\frac{1}{3}}\psi_{-\frac{8}{3}}\psi_{\frac{8}{3}})\psi^{+}(0)=R(1) \label{eq2.65}
\eea
We observe that the sums over $l$ in (\ref{eq2.63}), written explicitly in (\ref{eq2.65}), are cutted at the matrix element $\psi_{-\frac{8}{3}}\psi_{\frac{8}{3}}\psi^{+}(0)$. This is because $\psi_{\frac{11}{3}}\psi^{+}(0)$, $\psi_{\frac{14}{3}}\psi^{+}(0)$, and so on, are all zero, eq.(\ref{eq2.23}).

Next to the lowest equation is obtained for \underline{$n=2$} (or $n=-1)$ :
\bea
(\psi_{\frac{4}{3}}\psi_{-\frac{4}{3}}+D^{1}_{-\frac{1}{3}}\psi_{\frac{1}{3}}\psi_{-\frac{1}{3}}+D^{2}_{-\frac{1}{3}}\psi_{-\frac{2}{3}}\psi_{\frac{2}{3}}+D^{3}_{-\frac{1}{3}}\psi_{-\frac{5}{3}}\psi_{\frac{5}{3}}\nn\\
+D^{4}_{-\frac{1}{3}}\psi_{-\frac{8}{3}}\psi_{\frac{8}{3}}+\psi_{-\frac{5}{3}}\psi_{\frac{5}{3}}+D^{1}_{-\frac{1}{3}}\psi_{-\frac{8}{3}}\psi_{\frac{8}{3}})\psi^{+}(0)=R(2) \label{eq2.66}
\eea
\underline{$n=3$}
\bea
(\psi_{\frac{7}{3}}\psi_{-\frac{7}{3}}+D^{1}_{-\frac{1}{3}}\psi_{\frac{4}{3}}\psi_{-\frac{4}{3}}+D^{2}_{-\frac{1}{3}}\psi_{\frac{1}{3}}\psi_{-\frac{1}{3}}+D^{3}_{-\frac{1}{3}}\psi_{-\frac{2}{3}}\psi_{\frac{2}{3}}\nn\\
+D^{4}_{-\frac{1}{3}}\psi_{-\frac{5}{3}}\psi_{\frac{5}{3}}+D^{5}_{-\frac{1}{3}}\psi_{-\frac{8}{3}}\psi_{\frac{8}{3}}+\psi_{-\frac{8}{3}}\psi_{\frac{8}{3}})\psi^{+}(0)=R(3) \label{eq2.67}
\eea
\underline{$n=4$}
\bea
(\psi_{\frac{10}{3}}\psi_{-\frac{10}{3}}+D^{1}_{-\frac{1}{3}}\psi_{\frac{7}{3}}\psi_{-\frac{7}{3}}+D^{2}_{-\frac{1}{3}}\psi_{\frac{4}{3}}\psi_{-\frac{4}{3}}+D^{3}_{-\frac{1}{3}}\psi_{\frac{1}{3}}\psi_{-\frac{1}{3}}\nn\\
+D^{4}_{-\frac{1}{3}}\psi_{-\frac{2}{3}}\psi_{\frac{2}{3}}+D^{5}_{-\frac{1}{3}}\psi_{-\frac{5}{3}}\psi_{\frac{5}{3}}+D^{6}_{-\frac{1}{3}}\psi_{-\frac{8}{3}}\psi_{\frac{8}{3}})\psi^{+}(0)=R(4) \label{eq2.68}
\eea
and so on.

We observe now that the matrix elements in the first 3 equations (those for $n=1,2,3$) could all be calculated directly, because $\psi_{\frac{8}{3}}\psi^{+}(0)$, $\psi_{\frac{5}{3}}\psi^{+}(0)$,..., $\psi_{-\frac{7}{3}}\psi^{+}(0)$ are all known explicitly, eqs(\ref{eq2.23})-(\ref{eq2.29}). Could also be calculated directly the actions of mode operators in $R(n)$, eq.(\ref{eq2.64}), $\psi^{+}_{0}\psi^{+}(0)$, $(L_{-2}\psi^{+})_{0}\psi^{+}(0)$, $\tilde{\psi}^{+}_{0}\psi^{+}(0)$.

This means that, when these low matrix elements
\bea
\psi_{-\frac{8}{3}}\psi_{\frac{8}{3}}\psi^{+}(0),  
\quad   \psi_{-\frac{5}{3}}\psi_{\frac{5}{3}}\psi^{+}(0), 
\quad  \psi_{-\frac{2}{3}}\psi_{\frac{2}{3}}\psi^{+}(0),\nn\\  
\psi_{\frac{1}{3}}\psi_{-\frac{1}{3}}\psi^{+}(0),  
\quad  \psi_{\frac{4}{3}}\psi_{-\frac{4}{3}}\psi^{+}(0),  
\quad   \psi_{\frac{7}{3}}\psi_{-\frac{7}{3}}\psi^{+}(0) \label{eq2.69}
\eea
plus
\beq
\psi^{+}_{0}\psi^{+}(0), \quad (L_{-2}\psi^{+})_{0}\psi^{+}(0),
\quad \tilde{\psi}^{+}_{0}\psi^{+}(0) \label{eq2.70}
\eeq
are all calculated and substituted, then the equations (\ref{eq2.65}), (\ref{eq2.66}), (\ref{eq2.67}) above become equations on operator algebra constants, $\lambda,\zeta,\eta,\tilde{\lambda},\gamma,\mu,\tilde{\gamma},b$ (on some of them, to be more precise). While the eq.(\ref{eq2.68}) and equations which follow, for higher values of $n$, define the unknown matrix elements, $\psi_{\frac{10}{3}}\psi_{-\frac{10}{3}}\psi^{+}(0)$, etc. . They are all needed in order to calculate the $z$ expansion of our correlation function, eq.(\ref{eq2.51}).

The calculation of the matrix elements (\ref{eq2.69}) and (\ref{eq2.70}) is done in the Appendix B. One finds the following values :
\beq
\psi_{-\frac{8}{3}}\psi_{\frac{8}{3}}\psi^{+}(0)=\psi(0) \label{eq2.71}
\eeq
\beq
\psi_{-\frac{5}{3}}\psi_{\frac{5}{3}}\psi^{+}(0)=0 \label{eq2.72}
\eeq
\beq
\psi_{-\frac{2}{3}}\psi_{\frac{2}{3}}\psi^{+}(0)=\frac{2\Delta_{\psi}^{2}}{c}\cdot\psi(0) \label{eq2.73}
\eeq
\beq
\psi_{\frac{1}{3}}\psi_{-\frac{1}{3}}\psi^{+}(0)=\frac{2\Delta_{\psi}^{2}}{c}\psi(0) \label{eq2.74}
\eeq
\beq
\psi_{\frac{4}{3}}\psi_{-\frac{4}{3}}\psi^{+}(0)=(\beta^{(112)}_{\psi\psi^{+},I}\cdot (\Delta_{\psi}+\frac{1}{3})(\Delta_{\psi}-\frac{2}{3}) \Delta_{\psi}+\beta^{(22)}_{\psi\psi^{+},I}\cdot (2\Delta_{\psi}-\frac{2}{3})\Delta_{\psi}+\gamma^{2})\psi(0) \label{eq2.75}
\eeq
\bea
\psi_{\frac{7}{3}}\psi_{-\frac{7}{3}}\psi^{+}(0)=(\beta^{(1112)}_{\psi\psi^{+},I}\cdot (\Delta_{\psi}+\frac{4}{3})(\Delta_{\psi}+\frac{1}{3})(\Delta_{\psi}-\frac{2}{3})\Delta_{\psi}\nn\\+\beta^{(122)}_{\psi\psi^{+},I}\cdot (\Delta_{\psi}+\frac{4}{3})(2\Delta_{\psi}-\frac{2}{3})\Delta_{\psi}+\gamma^{2}\beta^{(1)}_{\psi\psi^{+},B}\cdot (\Delta_{\psi}+\frac{4}{3}))\psi(0) \label{eq2.76}
\eea
-- for the matrix elements in (\ref{eq2.69}), and
\beq
\psi^{+}_{0}\psi^{+}(0)=\lambda\psi(0) \label{eq2.77}
\eeq
\beq
(L_{-2}\psi^{+})_{0}\psi^{+}(0)=2\Delta_{\psi}\lambda\psi(0)\label{eq2.78}
\eeq
\beq
\tilde{\psi}^{+}_{0}\psi^{+}(0)=\zeta\psi(0)\label{eq2.79}
\eeq
-- for the matrix elements in (\ref{eq2.70}).
We substitute (\ref{eq2.71})-(\ref{eq2.79}) into the equations (\ref{eq2.65}), (\ref{eq2.66}), (\ref{eq2.67}), with $\beta$ coefficients to be found in the Appendix A and the values of the coefficients $D^{l}_{-\frac{1}{3}}$ easily defined from their definition in (\ref{eq2.58}).

Putting all this on Mathematica, one obtains, from (\ref{eq2.65}), (\ref{eq2.66}), (\ref{eq2.67}), the following equations :
\beq
\frac{32 (84 + c)}{81 c} + \frac{8 (-116 + c) \lambda^2}{784 + 57 c} 
= \zeta^2\label{eq2.79a}
\eeq
\bea
\frac{4 (63168 + c (6926 + 145 c))}{243 c (22 + 5 c)} + \gamma^2 \nn\\
= \zeta^2 + \frac{(1712 + 49 c) \lambda^2}{784 + 57 c}\label{eq2.79b}
\eea
\bea
\frac{4 (333760 + c (23886 + 1025 c))}{729 c (22 + 5 c)} 
+ \frac{7 \gamma^2}{3}\nn\\
   = \zeta^2 + \frac{(3280 + 163 c) \lambda^2}{784 + 57 c}\label{eq2.79c}
\eea

Solving these equations one obtains
 the values (\ref{eq1.19}), (\ref{eq1.20}), (\ref{eq1.23}) of the constants $\lambda$, $\zeta$, $\gamma$.

To be more precise, into the equations (\ref{eq2.79a}), (\ref{eq2.79b}), (\ref{eq2.79c}) enter only squares of $\lambda, \zeta, \gamma$. So that by these equations the three constants are defined up to signs. In (\ref{eq1.19}), (\ref{eq1.20}), (\ref{eq1.23}) we have chosen these constants to be positives.

Finally we shall finish the calculation of the function $<\psi^{+}(\infty)\psi(1)\psi(z)\psi^{+}(0)>$. We remined that this function is given by the $z$ expansion in (\ref{eq2.51}), with the coefficients $m_{0}$, $m_{1}$, $m_{2}$, etc., defined
by the matrix elements in (\ref{eq2.47})-(\ref{eq2.50}), etc. . Using the results (\ref{eq2.71})-(\ref{eq2.76}), of the direct calculations in the Appendix B, we could state that we know the coefficients $m_{0}$,..., $m_{5}$, of the expansion in (\ref{eq2.51}) :
\beq
m_{0}=1\label{eq2.80}
\eeq
\beq
m_{1}=0\label{eq2.81}
\eeq
\beq
m_{2}=\frac{2\Delta_{\psi}^{2}}{c}\label{eq2.82}
\eeq
\beq
m_{3}=\frac{2\Delta_{\psi}^{2}}{c}\label{eq2.83}
\eeq
\beq
m_{4}=\beta^{(112)}_{\psi\psi^{+},I}\cdot (\Delta_{\psi}+\frac{1}{3})(\Delta_{\psi}-\frac{2}{3}) \Delta_{\psi}+\beta^{(22)}_{\psi\psi^{+},I}\cdot (2\Delta_{\psi}-\frac{2}{3})\Delta_{\psi}+\gamma^{2}\label{eq2.84}
\eeq
\bea
m_{5}=\beta^{(1112)}_{\psi\psi^{+},I}\cdot (\Delta_{\psi}+\frac{4}{3})(\Delta_{\psi}+\frac{1}{3})(\Delta_{\psi}-\frac{2}{3})\Delta_{\psi}\nn\\+\beta^{(122)}_{\psi\psi^{+},I}\cdot (\Delta_{\psi}+\frac{4}{3})(2\Delta_{\psi}-\frac{2}{3})\Delta_{\psi}+\gamma^{2}\beta^{(1)}_{\psi\psi^{+},B}\cdot (\Delta_{\psi}+\frac{4}{3})\label{eq2.85}
\eea
The higher coefficients, $m_{6}$, $m_{7}$, etc. , are to be calculated the commutation rules (\ref{eq2.63}). More specifically, the matrix element $\psi_{\frac{10}{3}}\psi_{-\frac{10}{3}}\psi^{+}(0)$, and so the coefficient $m_{6}$, is defined by the equation (\ref{eq2.68}), which follow from (\ref{eq2.63}) for $n=4$. With $n=5$ in (\ref{eq2.63}) one defines the matrix element $\psi_{\frac{13}{3}}\psi_{-\frac{13}{3}}\psi^{+}(0)$ and the coefficient $m_{7}$, and so on. One could write the following recurrence relation, which follows from (\ref{eq2.63}) for $n\geq 4$ :
\beq
m_{n+2}=R(n)-\sum_{k=1}^{n+2}D^{k}_{-\frac{1}{3}}m_{n+2-k}\label{eq2.86}
\eeq
It defines all the coefficients $\{m_{n}\}$, starting with $m_{6}$.

By performing these calculations with Mathematica one finds the following
values of the coefficients :
\beq
m_{6}=\frac{16(58584+1837c)}{6561c}\label{eq2.87}
\eeq
\beq
m_{7}=\frac{32(130072+3949c)}{19683c}\label{eq2.88}
\eeq
\beq
m_{8}=\frac{22(784256+23919c)}{59049c}\label{eq2.89}
\eeq
\beq
m_{9}=\frac{32(19135512+586789c)}{1594323c}\label{eq2.90}
\eeq
\beq
m_{10}=\frac{2992(259192+7981c)}{1594323c}\label{eq2.91}
\eeq
As compared to the expressions (\ref{eq2.80})-(\ref{eq2.85}) for $m_{0},...,m_{5}$, in getting the expressions (\ref{eq2.87})-(\ref{eq2.91}) we have already substituted the values of $\lambda$, $\zeta$, $\gamma$, defined above, and the values of all the $\beta$ coefficients.

In this way, the $z$ development of the function $<\psi^{+}(\infty)\psi(1)\psi(z)\psi^{+}(0)>$ in (\ref{eq2.51}) is defined, up to $z^{10}$. It has been defined by the operator algebra of parafermionic fields.

On the other hand, we develop the analytic form of this function in eq.(\ref{eq2.9}). We have to develop the factor $P_{8}(z)/(1-z)^{\Delta_{\psi}}$, in powers of $z$. The coefficients of this last expansion will appear as linear combinations of coefficients of the polynomial $P_{8}(z)$. They are slightly long to provide them here. By they are easily obtained with Mathematica.

In fact, we have developed the product $P_{10}(z)/(1-z)^{\Delta_{\psi}}$, extending the polynomial by terms $a_{9}z^{9}+a_{10}z^{10}$, to provide an additional check. If calculations by the operator algebra are consistent with the analytic form (\ref{eq2.9}) we should find, with calculations, $a_{9}=a_{10}=0$.

By comparing the coefficients of the two expansions, in (\ref{eq2.51}) and that of the analytic form (\ref{eq2.9}) (with $P_{8}(z)$ replaced by $P_{10}(z)$) we get a system of equations which determine the coefficients of the polynomial $P_{10}(z)$. Solving them we have found the following values of the coefficients :
\beq
a_{0}=1, \,\,
a_{1}=-\frac{8}{3}, \,\,
a_{2}=\frac{4(32+5c)}{9c}, \,\,
a_{3}=-\frac{40(48+c)}{81c}, \,\,
a_{4}=\frac{2(5440+83c)}{243c} \label{eq2.102}
\eeq
and next we find that $a_{5}=a_{3}$, $a_{6}=a_{2}$, $a_{7}=a_{1}$,
$a_{8}=a_{0}$, $a_{9}=0$, $a_{10}=0$. 

Resuming, we have found that the symmetry conditions (\ref{eq2.10}), (\ref{eq2.11}) are satisfied. One checks easily that the condition (\ref{eq2.14}), coming from the limit $z\rightarrow 1$, is also satisfied.

We tend to conclude that once the consistency equations (\ref{eq2.65}), (\ref{eq2.66}), (\ref{eq2.67}) are satisfied and the values of the constants $\lambda$, $\zeta$, $\gamma$, defined by these equations, are used, then the correlation function, calculated further with the use of  the commutation rules, will satisfy all the consistencies automatically.

In the next subsection we shall check this statement additionally by calculating the function $<\psi^{+}(\infty)\psi(1)\psi(z)\psi^{+}(0)>$ by its  development in powers of $(1-z)$. As if we use, fully this time, the limit $z\rightarrow 1$.

\subsection{ Calculation of the function $<\psi^{+}(\infty)\psi(1)\psi(z)\psi^{+}(0)>$ by its development in powers of $1-z$}.

The starting point is our function and its analytic form:
\beq
<\psi^{+}(\infty)\psi(1)\psi(z)\psi^{+}(0)>=\frac{Q_{8}(1-z)}{(1-z)^{\Delta_{\psi}}(z)^{2\Delta_{\psi}}}\label{eq2.2.1}
\eeq
With respect to eq.(\ref{eq2.9}), we have replaced $P_{8}(z)$, eq.(\ref{eq2.3}), by
\beq
Q_{8}(1-z)=b_{0}+b_{1}\cdot(1-z)+b_{2}\cdot(1-z)^{2}+...+b_{8}\cdot(1-z)^{8}\label{eq2.2.2}
\eeq
which is more convenient when analyzing the expansion in powers of $(1-z)$. Otherwise, at the end of the calculations, when the coefficients $b_{0}$, $b_{1}$,.., $b_{8}$ are all defined, we should find that
\beq
Q_{8}(1-z)=P_{8}(z)\label{eq2.2.3}
\eeq
if our calculations, in Section 2.1 and here, are consistent.

To prepare the development of (\ref{eq2.2.1}), by the operator algebra, in the limit $z\rightarrow 1$, we shall move the fields in the l.h.s. and the variables $0$, $z$, $1$ in the r.h.s., to the following positions :
\beq
<\psi^{+}(\infty)\psi^{+}(0)\psi(z)\psi(1)>=\frac{Q_{8}(1-z)}{(-1)^{2\Delta_{\psi}}(z-1)^{\Delta_{\psi}}(-z)^{2\Delta_{\psi}}}\label{eq2.2.4}
\eeq
The factor $1/(-1)^{2\Delta_{\psi}}$ in the equation above is due to the analytic continuation of $\psi^{+}(0)$ around $\psi(1)$ when the fields in the l.h.s. of (\ref{eq2.2.1}) are moved to their positions in (\ref{eq2.2.4}). Similarly, $1/(z)^{2\Delta_{\psi}}\rightarrow 1/(-z)^{2\Delta_{\psi}}$, when $\psi^{+}(0)$ is moved around $\psi(z)$, under the phase compensation convention, cf. the corresponding analytic continuation when moving from (\ref{eq2.2}) to (\ref{eq2.4}) in the previous subsection.

We could rewrite (\ref{eq2.2.4}) as follows :
\beq
<\psi^{+}(\infty)\psi^{+}(0)\psi(z)\psi(1)>=\frac{Q_{8}(1-z)}{(-1)^{\Delta_{\psi}}(z-1)^{\Delta_{\psi}}(z)^{2\Delta_{\psi}}}\label{eq2.2.5}
\eeq
using $(-1)^{4\Delta_{\psi}}=(-1)^{\Delta_{\psi}}$, for $\Delta_{\psi}=8/3$.

Now we imagine that $z$ is close to 1 and we develop first the product $\psi(z)\psi(1)$ ($\psi(z)$ is developed in modes which are applied to $\psi(1)$, as in (\ref{eq2.17}) but with $\Phi^{1}(0)\rightarrow\psi(1)$) and then we apply $\psi^{+}(0)$ and develop again, in powers of $(-1)=(0-1)$ this time. Finally this double development is projected onto $\psi^{+}(\infty)$.

We proceed similarly as it has been done in the previous subsection. In this way we find the following series :
\bea 
\psi^{+}(0)\psi(z)\psi(1)=\frac{1}{(-1)^{\Delta_{\psi}}}\cdot\frac{1}{(z-1)^{\Delta_{\psi}}}\cdot\{\psi^{+}_{0}\psi_{0}\psi(1)\nn\\
-(z-1)\cdot\psi_{1}^{+}\psi_{-1}\psi(1)+(z-1)^{2}\psi^{+}_{2}\psi_{-2}\psi(1)\nn\\
-(z-1)^{3}\psi^{+}_{3}\psi_{-3}\psi(1)+...\}\label{eq2.2.6}
\eea
-- under the projection onto $\psi^{+}(\infty)$, i.e. modulo terms which will disappear under the projection onto $\psi^{+}(\infty)$, and
\bea
<\psi^{+}(\infty)\psi^{+}(0)\psi(z)\psi(1)>=\nn\\
\frac{1}{(-1)^{\Delta_{\psi}}}\cdot\frac{1}{(z-1)^{\Delta_{\psi}}}\cdot\{m_{0}-m_{1}\cdot(z-1)+m_{2}\cdot(z-1)^{2}-m_{3}\cdot(z-1)^{3}+...\}\label{eq2.2.7}
\eea
In the series above the coefficients $\{m_{k}\}$ are defined by the corresponding matrix elements :
\beq
\psi^{+}_{0}\psi_{0}\psi(1)=m_{0}\psi(1)\label{eq2.2.8}
\eeq
\beq
\psi^{+}_{1}\psi_{-1}\psi(1)=m_{1}\psi(1)\label{eq2.2.9}
\eeq 
\beq
\psi^{+}_{2}\psi_{-2}\psi(1)=m_{2}\psi(1)\label{eq2.2.10}
\eeq
and so on.

It could be observed that the factor $1/(-1)^{\Delta_{\psi}}$ in the r.h.s. of (\ref{eq2.2.5}) is compensated by a similar factor produced in the l.h.s. of (\ref{eq2.2.5}) when the product $\psi^{+}(0)\psi(z)\psi(1)$ is decomposed. This could be seen, for instance, with the principal term coming from the product of three fields :
\bea
\psi^{+}(0)\psi(z)\psi(1)\simeq\psi^{+}(0)\frac{\lambda}{(z-1)^{\Delta_{\psi}}}\cdot\psi^{+}(1)\nn\\
=\frac{\lambda}{(z-1)^{\Delta_{\psi}}}\cdot\psi^{+}(0)\psi^{+}(1)\simeq\frac{\lambda}{(z-1)^{\Delta_{\psi}}}\cdot\frac{\lambda}{(-1)^{\Delta_{\psi}}}\cdot\psi(1)\nn\\
=\frac{1}{(-1)^{\Delta_{\psi}}}\cdot\frac{\lambda^{2}}{(z-1)^{\Delta_{\psi}}}\cdot\psi(1)\label{eq2.2.11}
\eea
The principal term, of this direct expansion of the product, corresponds to the matrix element $\psi_{0}^{+}\psi_{0}\psi(1)$ in (2.2.8), so that $m_{0}=\lambda^{2}$.

We conclude that, to calculate the $(z-1)$ expansion of the correlation function $<\psi^{+}(\infty)\psi^{+}(0)\psi(z)\psi(1)>$, eq.(\ref{eq2.2.7}), we have to be able to define the matrix elements 
(\ref{eq2.2.8})-(\ref{eq2.2.10}) and so on. For this we need the commutation rules $\{\psi^{+},\psi\}\psi(1)$.

We define the integrals
\bea
I_{1}=\frac{1}{(2\pi i)^{2}}\oint_{C'_{1}}dz'(z')^{\Delta_{\psi}-\frac{1}{3}+n-1}\oint_{C_{1}}dz(z)^{\Delta_{\psi}+m-1}\nn\\
\times(z'-z)^{2\Delta_{\psi}-6}\times\psi^{+}(z')\psi(z)\psi(1)\label{eq2.2.12}
\eea
\bea
I_{2}=\frac{1}{(2\pi i)^{2}}\oint_{C_{1}}dz(z)^{\Delta_{\psi}+m-1}\oint_{C'_{1}}dz'(z')^{\Delta_{\psi}-\frac{1}{3}+n-1}\nn\\
\times(z-z')^{2\Delta_{\psi}-6}\times\psi(z)\psi^{+}(z')\psi(1)\label{eq2.2.13}
\eea
\bea
I_{3}=\frac{1}{(2\pi i)^{2}}\oint_{C_{1}}dz(z)^{\Delta_{\psi}+m-1}\oint_{C_{z}}dz'(z')^{\Delta_{\psi}-\frac{1}{3}+n-1}\nn\\
\times(z'-z)^{2\Delta_{\psi}-6}\times\psi^{+}(z')\psi(z)\psi(1)\nn\\
=\frac{1}{(2\pi i)^{2}}\oint_{c_{1}}dz(z)^{\Delta_{\psi}+m-1}\oint_{C_{z}}dz'(z')^{\Delta_{\psi}-\frac{1}{3}+n-1}\nn\\
\times\frac{1}{(z'-z)^{6}}\{1+(z'-z)^{2}\frac{2\Delta_{\psi}}{c}T(z)+(z'-z)^{3}\frac{\Delta_{\psi}}{c}\partial T(z)\nn\\
+(z'-z)^{4}[\beta^{(112)}_{\psi\psi^{+},I}\partial^{2}T(z)+\beta^{(22)}_{\psi\psi^{+},I}\wedge(z)+\gamma\cdot B(z)]\nn\\
+(z'-z)^{5}\cdot[\beta^{(1112)}_{\psi\psi^{+},I}\cdot\partial^{3}T(z)+\beta^{(122)}_{\psi\psi^{+},I}\partial\wedge(z)+\gamma\beta^{(1)}_{\psi\psi^{+},B}\cdot\partial B(z)]\}\psi(1)\label{eq2.2.14}
\eea
In $I_{3}$ we have used the expansion (\ref{eq1.2}) for the product $\psi^{+}(z')\psi(z)$. We observe that the fields in the explicit part of this decomposition, up to terms $\sim(z'-z)^{5}$, are assumed to be $Z_{3}$ conjugation invariant. In particular, $B^{+}(z)=B(z)$. So that $\psi^{+}(z')\psi(z)=\psi(z')\psi^{+}(z)$, modulo terms higher than $(z'-z)^{5}$.

The integration contours for $I_{1}$, $I_{2}$, $I_{3}$ are the same as 
those in Fig.3, with a replacement, everywhere, 0 by 1, including 
in the indeces of the contours : $C_{0}$, $C'_{0}$ to replaced 
by $C_{1}$, $C'_{1}$.

This time these integrals are related, by the analytic continuation of integration contours, as follows :
\beq
I_{1}-I_{2}=I_{3}\label{eq2.2.15}
\eeq
We have the minus sign in the l.h.s., instead of plus sign in (\ref{eq2.55}), in the previous subsection, because the extra power in $(z'-z)^{2\Delta_{\psi}-6}$, which is 6, is even. This extra power has been chosen so that all the explicit terms in (\ref{eq1.2}) contribute to the integral $I_{3}$.

Next we develop
\beq
(z'-z)^{2\Delta_{\psi}-6}=(z'-z)^{-\frac{2}{3}}=(z')^{-\frac{2}{3}}\sum_{l=0}^{\infty}D^{l}_{-\frac{2}{3}}(\frac{z}{z'})^{l}\label{2.2.16}
\eeq
-- in the integral $I_{1}$, and
\beq
(z-z')^{2\Delta_{\psi}-6}=(z)^{-\frac{2}{3}}\sum_{l=0}^{\infty}D^{l}_{-\frac{2}{3}}(\frac{z'}{z})^{l}\label{eq2.2.17}
\eeq
-- in the integral $I_{2}$, and we express these integrals in terms of modes of the fields $\psi^{+}(z')$ and $\psi(z)$.

The integral $I_{3}$ is calculated by the theorem of residues, for the integration over $z'$, and then expressed in modes, for the integration over $z$.

In this way we find that the $\{\psi^{+},\psi\}\psi(1)$ commutation rules are given by the expressions :
\beq
\sum_{l=0}^{\infty}D^{l}_{-\frac{2}{3}}(\psi^{+}_{n-l-1}\psi_{m+l}-\psi_{-\frac{2}{3}+m-l}\psi^{+}_{-\frac{1}{3}+n+l})\psi(1)=R(n,m)\label{eq2.2.18}
\eeq

$R(n,m)$ in this equation is different from $R(n,m)$ in eq.(\ref{eq2.60}), in the case of the commutation rules $\{\psi,\psi\}\psi^{+}(0)$, eq.(\ref{eq2.59}). Here, in  eq.(\ref{eq2.2.18}), and in the following, we shall use the same notation for the r.h.s. of different commutation rules. In general, the expressions of the corresponding $R(n,m)$ are used only immediately after the corresponding commutation rules. 

This time, for (\ref{eq2.2.18}), $R(n,m)$ is of the form :
\bea
R(n,m)=\{\frac{1}{5!}(\Delta_{\psi}-\frac{1}{3}+n-1)(\Delta_{\psi}-\frac{1}{3}+n-2)(\Delta_{\psi}-\frac{1}{3}+n-3)\nn\\(\Delta_{\psi}-\frac{1}{3}+n-4)(\Delta_{\psi}-\frac{1}{3}+n-5)\cdot\delta_{n+m-1,0}\nn\\
+\frac{1}{3!}(\Delta_{\psi}-\frac{1}{3}+n-1)(\Delta_{\psi}-\frac{1}{3}+n-2)(\Delta_{\psi}-\frac{1}{3}+n-3)\cdot\frac{2\Delta_{\psi}}{c}L_{n+m-1}\nn\\
-\frac{1}{2!}(\Delta_{\psi}-\frac{1}{3}+n-1)(\Delta_{\psi}-\frac{1}{3}+n-2)(n+m+1)\cdot\frac{\Delta_{\psi}}{c}\cdot L_{n+m-1}\nn\\
+(\Delta_{\psi}-\frac{1}{3}+n-1)[(n+m+2)(n+m+1)\beta^{(112)}_{\psi\psi^{+},I}\cdot L_{n+m-1}\nn\\
+\beta^{(22)}_{\psi\psi^{+},I}\cdot\wedge_{n+m-1}+\gamma\cdot B_{n+m-1}]\nn\\
-[(n+m+3)(n+m+2)(n+m+1)\beta^{(1112)}_{\psi\psi^{+},I}\cdot L_{n+m-1}\nn\\
+(n+m+3)\cdot \beta^{(122)}_{\psi\psi^{+},I}\cdot\wedge_{n+m-1}
+(n+m+3)\gamma\beta^{(1)}_{\psi\psi^{+},B}\cdot B_{n+m-1}]\}\psi(1)\label{eq2.2.19}
\eea

We have to calculate the matrix elements (\ref{eq2.2.8})-(\ref{eq2.2.10}), which appear in the series (\ref{eq2.2.6}), (\ref{eq2.2.7}). To make them appear in the commutation rules (\ref{eq2.2.18}), (\ref{eq2.2.19}) we have to impose the constraint :
\beq
n+m-1=0,\quad m=1-n\label{eq2.2.20}
\eeq
With this constraint the commutation rules take the form :
\beq
\sum_{l=0}^{\infty}D_{-\frac{2}{3}}^{l}(\psi^{+}_{n-l-1}\psi_{1-n+l}-\psi_{\frac{1}{3}-n-l}\psi_{-\frac{1}{3}+n+l})\psi(1)=R(n)\label{eq2.2.21}
\eeq
\bea
R(n)=R(n,1-n)=
\{\frac{1}{5!}(\Delta_{\psi}-\frac{1}{3}+n-1)(\Delta_{\psi}-\frac{1}{3}+n-2)\nn\\(\Delta_{\psi}-\frac{1}{3}+n-3)(\Delta_{\psi}-\frac{1}{3}+n-4)(\Delta_{\psi}-\frac{1}{3}+n-5)\nn\\
+\frac{1}{3!}(\Delta_{\psi}-\frac{1}{3}+n-1)(\Delta_{\psi}-\frac{1}{3}+n-2)(\Delta_{\psi}-\frac{1}{3}+n-3)\cdot\frac{2\Delta_{\psi}}{c}L_{0}\nn\\
-\frac{1}{2!}(\Delta_{\psi}-\frac{1}{3}+n-1)(\Delta_{\psi}-\frac{1}{3}+n-2)2\cdot\frac{\Delta_{\psi}}{c}\cdot L_{0}\nn\\
+(\Delta_{\psi}-\frac{1}{3}+n-1)[3\cdot 2\cdot\beta^{(112)}_{\psi\psi^{+},I}\cdot L_{0}
+\beta^{(22)}_{\psi\psi^{+},I}\cdot\wedge_{0}+\gamma\cdot B_{0}]\nn\\
-[4\cdot 3\cdot 2\cdot\beta^{(1112)}_{\psi\psi^{+},I}\cdot L_{0}+4\cdot \beta^{(122)}_{\psi\psi^{+},I}\cdot\wedge_{0}
+4\cdot\gamma\beta^{(1)}_{\psi\psi^{+},B}\cdot B_{0}]\}\psi(1)\label{eq2.2.22}
\eea

Now we are going to list the equations which follow from (\ref{eq2.2.21}). But first we shall list the actions of the modes, of $\psi_{1-n+l}$ on $\psi(1)$ and of $\psi^{+}_{-\frac{1}{3}+n+l}$ on $\psi(1)$, in (\ref{eq2.2.21}), whose actions which are known explicitly.

By comparing the expansions (\ref{eq1.1}) (with $z'\rightarrow z$, $z\rightarrow 1$) and (\ref{eq2.17}) (with $\Phi^{1}(0)\rightarrow\psi(1)$, and the series expansion in $z$, in the r.h.s., replaced by the series in $z-1$), 
we find :
\beq
\psi_{n}\psi(1)=0,\quad n>0\label{eq2.2.23}
\eeq
\beq
\psi_{0}\psi(1)=\lambda\psi^{+}(1)\label{eq2.2.24}
\eeq
\beq
\psi_{-1}\psi(1)=\lambda\beta^{(1)}_{\psi\psi,\psi^{+}}
\partial\psi^{+}(1)\label{eq2.2.25}
\eeq
\beq
\psi_{-2}\psi(1)=\lambda\beta^{(11)}_{\psi\psi,\psi^{+}}\cdot\partial^{2}\psi^{+}(1)+\lambda\beta_{\psi\psi,\psi^{+}}^{(2)}\cdot L_{-2}\psi^{+}(1)+\zeta\tilde{\psi}^{+}(1)\label{eq2.2.26}
\eeq
Similarly, by comparing the expansions (\ref{eq1.2}) (its form $Z_{3}$ conjugate, with $\psi(z')\rightarrow\psi^{+}(z)$ $\psi^{+}(z)\rightarrow\psi(1)$) and (\ref{eq2.19}) (with $\Phi^{1}(0)\rightarrow\psi(1)$), we find :
\beq
\psi^{+}_{-\frac{1}{3}+n}\psi(1)=0, \quad n>3\label{eq2.2.27}
\eeq
\beq
\psi^{+}_{-\frac{1}{3}+3}\psi(1)=\psi^{+}_{\frac{8}{3}}\psi(1)=1\label{eq2.2.28}
\eeq
\beq
\psi^{+}_{\frac{5}{3}}\psi(1)=0\label{eq2.2.29}
\eeq
\beq
\psi^{+}_{\frac{2}{3}}\psi(1)=\frac{2\Delta_{\psi}}{c}\cdot T(1)\label{eq2.2.30}
\eeq
\beq
\psi^{+}_{-\frac{1}{3}}\psi(1)=\frac{\Delta_{\psi}}{c}\partial T(1)\label{eq2.2.31}
\eeq
\beq
\psi^{+}_{-\frac{4}{3}}\psi(1)=\beta^{(112)}_{\psi\psi^{+},I}\partial^{2}T(1)+\beta^{(22)}_{\psi\psi^{+},I}\wedge(1)+\gamma\cdot B(1)\label{eq2.2.32}
\eeq
\beq
\psi^{+}_{-\frac{7}{3}}\psi(1)=\beta^{(1112)}_{\psi,\psi^{+}I}\cdot\partial^{3}T(1)+\beta^{(122)}\cdot \partial\wedge(1)+\gamma\beta^{(1)}_{\psi\psi^{+},B}\cdot\partial B(1)\label{eq2.2.33}
\eeq

Knowing the actions of the modes, in (\ref{eq2.2.23})-(\ref{eq2.2.26}) and in (\ref{eq2.2.27})-(\ref{eq2.2.33}), we can list now the equations which follow from the commutation rules (\ref{eq2.2.21}). They are as follows :

\noindent\underline{$n=-2.$}
\bea
(-\psi_{+\frac{7}{3}}\psi^{+}_{-\frac{7}{3}}-D^{1}_{-\frac{2}{3}}\cdot\psi_{\frac{4}{3}}\psi^{+}_{-\frac{4}{3}}-D^{2}_{-\frac{2}{3}}\cdot\psi_{\frac{1}{3}}\psi^{+}_{-\frac{1}{3}}-D^{3}_{-\frac{2}{3}}\cdot\psi_{-\frac{2}{3}}\psi^{+}_{\frac{2}{3}}
\nn\\
-D^{4}_{-\frac{2}{3}}\cdot\psi_{-\frac{5}{3}}\psi^{+}_{\frac{5}{3}}-D^{5}_{-\frac{2}{3}}\cdot\psi_{-\frac{8}{3}}\psi^{+}_{\frac{8}{3}})\psi(1)=R(-2)\label{eq2.2.34}
\eea
\underline{$n=-1.$}
\bea
(-\psi_{\frac{4}{3}}\psi^{+}_{-\frac{4}{3}} 
- D^{1}_{-\frac{2}{3}}\cdot\psi_{\frac{1}{3}}\psi^{+}_{-\frac{1}{3}}
- D^{2}_{-\frac{2}{3}}\cdot\psi_{-\frac{2}{3}}\psi_{\frac{2}{3}}^{+}
-D^{3}_{-\frac{2}{3}}\cdot\psi_{-\frac{5}{3}}\psi^{+}_{\frac{5}{3}}\nn\\
-D^{4}_{-\frac{2}{3}}\cdot\psi_{-\frac{8}{3}}\psi^{+}_{\frac{8}{3}})\psi(1)
=R(-1)\label{eq2.2.35}
\eea
\underline{$n=0.$}
\beq
(-\psi_{\frac{1}{3}}\psi^{+}_{-\frac{1}{3}}-D^{1}_{-\frac{2}{3}}\cdot\psi_{-\frac{2}{3}}\psi^{+}_{\frac{2}{3}}-D^{2}_{-\frac{2}{3}}\cdot\psi_{-\frac{5}{3}}\psi^{+}_{\frac{5}{3}}-D^{3}_{-\frac{2}{3}}\cdot\psi_{-\frac{8}{3}}\psi^{+}_{\frac{8}{3}})\psi(1)=R(0)\label{eq2.2.36}
\eeq
\underline{$n=1.$}
\beq
(\psi^{+}_{0}\psi_{0}  - \psi_{-\frac{2}{3}}\psi^{+}_{\frac{2}{3}}
-D^{1}_{-\frac{2}{3}}\cdot\psi_{-\frac{5}{3}}\psi^{+}_{\frac{5}{3}}
-D^{2}_{-\frac{2}{3}}\cdot\psi_{-\frac{8}{3}}\psi^{+}_{\frac{8}{3}})\psi(1)
=R(1)\label{eq2.2.37}
\eeq
\underline{$n=2.$}
\beq
(\psi^{+}_{1}\psi_{-1}+D^{1}_{-\frac{2}{3}}\psi^{+}_{0}\psi_{0}-\psi_{-\frac{5}{3}}\psi^{+}_{\frac{5}{3}}-D^{1}_{-\frac{2}{3}}\cdot\psi_{-\frac{8}{3}}\psi^{+}_{\frac{8}{3}})\psi(1)=R(2)\label{eq2.2.38}
\eeq
\underline{$n=3.$}
\beq
\psi^{+}_{2}\psi_{-2}+D^{1}_{-\frac{2}{3}}\cdot\psi^{+}_{1}\psi_{-1}+D^{2}_{-\frac{2}{3}}\psi^{+}_{0}\psi_{0}-\psi_{-\frac{8}{3}}\psi^{+}_{\frac{8}{3}})\psi(1)=R(3)\label{eq2.2.39}
\eeq
\underline{$n=4.$}
\beq
\psi^{+}_{3}\psi_{-3}+D^{1}_{-\frac{2}{3}}\cdot\psi^{+}_{2}\psi_{-2}+D^{2}_{-\frac{2}{3}}\psi^{+}_{1}\psi_{-1}+D^{3}_{-\frac{2}{3}}\psi^{+}_{0}\psi_{0})\psi(1)=R(4)\label{eq2.2.40}
\eeq

The equations (\ref{eq2.2.34})-(\ref{eq2.2.39}) contain the low matrix elements which could all be determined directly. One finds, with the direct calculations, the following values for these matrix elements :
\beq
\psi_{0}^{+}\psi_{0}\psi(1)=\lambda^{2}\psi(1)\label{eq2.2.41}
\eeq
\beq
\psi_{1}^{+}\psi_{-1}\psi(1)=\lambda^{2}\cdot\beta^{(1)}_{\psi\psi,\psi^{+}}\cdot\Delta_{\psi}\psi(1)\label{eq2.2.42}
\eeq
\beq
\psi^{+}_{2}\psi_{-2}\psi(1)=[\lambda^{2}\cdot\beta^{(11)}_{\psi\psi,\psi^{+}}\cdot\Delta_{\psi}(\Delta_{\psi}+1)+\lambda^{2}\cdot\beta^{(2)}_{\psi\psi,\psi^{+}}\cdot2\Delta_{\psi}+\zeta^{2}]\psi(1)\label{eq2.2.43}
\eeq
next,
\beq
\psi_{-\frac{8}{3}}\psi^{+}_{\frac{8}{3}}\psi(1)=\psi(1)\label{eq2.2.44}
\eeq
\beq
\psi_{-\frac{5}{3}}\psi^{+}_{\frac{5}{3}}\psi(1)=0\label{eq2.2.45}
\eeq
\beq
\psi_{-\frac{2}{3}}\psi^{+}_{\frac{2}{3}}\psi(1)=\frac{2\Delta^{2}_{\psi}}{c}\psi(1)
\label{eq2.2.46}
\eeq
\beq
\psi_{\frac{1}{3}}\psi^{+}_{-\frac{1}{3}}\psi(1)=\frac{\Delta^{2}_{\psi}}{c}(\Delta_{\psi}-\frac{2}{3})\psi(1)\label{eq2.2.47}
\eeq
\bea
\psi_{\frac{4}{3}}\psi^{+}_{-\frac{4}{3}}\psi(1)=[\beta^{(112)}_{\psi\psi^{+},I}\cdot(\Delta_{\psi}+\frac{1}{3})(\Delta_{\psi}-\frac{2}{3})\Delta_{\psi}\nn\\
+\beta^{(22)}_{\psi\psi^{+},I}\cdot(2\Delta_{\psi}-\frac{2}{3})\Delta_{\psi}+\gamma^{2}]\psi(1)\label{eq2.2.48}
\eea
\bea
\psi_{\frac{7}{4}}\psi^{+}_{-\frac{7}{4}}\psi(1)=[\beta^{(1112)}_{\psi\psi^{+},I}\cdot(\Delta_{\psi}+\frac{4}{3})(\Delta_{\psi}+\frac{1}{3})(\Delta_{\psi}-\frac{2}{3})\Delta_{\psi}\nn\\
+\beta^{(122)}_{\psi\psi^{+},I}\cdot(\Delta_{\psi}+\frac{3}{4})(2\Delta_{\psi}-\frac{2}{3})\Delta_{\psi}+\gamma^{2}\cdot\beta^{(1)}_{\psi\psi^{+},B}\cdot(\Delta_{\psi}+\frac{3}{4})]\psi(1)\label{eq2.2.49}
\eea
Should be added the actions of modes which enter $R(n)$, eq.(\ref{eq2.2.22}) :
\beq
L_{0}\psi(1)=\Delta_{\psi}\Psi(1)\label{eq2.2.50}
\eeq
\beq
\wedge_{0}\psi(1)=\Delta_{\psi}(\Delta_{\psi}+2)\psi(1)\label{eq2.2.51}
\eeq
\beq
B_{0}\psi(1)=\gamma\cdot\psi(1)\label{eq2.2.52}
\eeq
(\ref{eq2.2.41})-(\ref{eq2.2.43}) and (\ref{eq2.2.44})-(\ref{eq2.2.49}) have been calculated by using the actions of modes in (\ref{eq2.2.24})-(\ref{eq2.2.26}) and (\ref{eq2.2.28})-(\ref{eq2.2.33}) plus some commutations, the way it has been done in the Appendix B for the matrix elements (\ref{eq2.71})-(\ref{eq2.76}). In fact, (\ref{eq2.2.44})-(\ref{eq2.2.49}) and (\ref{eq2.71})-(\ref{eq2.76}) are identical, except for the replacement
(0) by (1).

Putting the expressions (\ref{eq2.2.41})-(\ref{eq2.2.52}) into the equations (\ref{eq2.2.34})-(\ref{eq2.2.39}) and analyzing them with Mathematica one finds that out of six equations only three are independent. Solving them, one obtains the same values of the constants $\lambda$, $b$, $\gamma$ as those obtained in the previous subsection. No new constraints are being produced.

Now we shall finish the calculation of the correlation function $<\psi^{+}(\infty)\psi(1)\psi(z)\psi^{+}(0)>$ in eq.(\ref{eq2.2.1}), or in its analytically continued configuration, eq.(\ref{eq2.2.5}).

Its development by the operator algebra is given in (\ref{eq2.2.7}), with the coefficients $m_{0}$, $m_{1}$, $m_{2}$ defined by the matrix elements in (\ref{eq2.2.8})-(\ref{eq2.2.10}). The first three coefficients, $m_{0}$, $m_{1}$, $m_{2}$, are known from the direct calculations, with the results in (\ref{eq2.2.41})-(\ref{eq2.2.43}). The following ones are to be determined by the commutation rules (\ref{eq2.2.21}). In particular, the matrix element $\psi^{+}_{3}\psi_{-3}\psi(1)$, and the corresponding coefficient $m_{3}$, is determined by the equation (\ref{eq2.2.40}).

In general, from (\ref{eq2.2.21}), follows the recurrence relation :
\beq
m_{n}=R(n+1)-\sum^{n}_{k=1}D_{-\frac{2}{3}}^{k}\cdot m_{n-k}\label{eq2.2.53}
\eeq
which determines the coefficients $m_{3}$, $m_{4}$, $m_{5}$, etc., when $m_{0}$, $m_{1}$, $m_{2}$ are given (predetermined), eqs. (\ref{eq2.2.8})-(\ref{eq2.2.10}) and (\ref{eq2.2.41})-(\ref{eq2.2.43}).

In this way one finds the following values :
\beq
m_{0}=\frac{196(32+c)}{243c}\label{eq2.2.54}
\eeq
\beq
m_{1}=\frac{784(32+c)}{729c}\label{eq2.2.55}
\eeq
\beq
m_{2}=\frac{8(20048+451c)}{2187c}\label{eq2.2.56}
\eeq
\beq
m_{3}=\frac{112(21448+407c)}{19683c}\label{eq2.2.57}
\eeq
\beq
m_{4}=\frac{4(2828512+136829c)}{59049c}\label{eq2.2.58}
\eeq
\beq
m_{5}=\frac{16(3159512+272041c)}{177147c}\label{eq2.2.59}
\eeq
\beq
m_{6}=\frac{8(80847728+10947211c)}{1594323c}\label{eq2.2.60}
\eeq
\beq
m_{7}=\frac{208(12749344+2495669c)}{4782969c}\label{eq2.2.61}
\eeq
\beq
m_{8}=\frac{19(553789184+147687265c)}{14348907c}\label{eq2.2.62}
\eeq
In getting these values, we have already substituted the values of the constants $\lambda$, $\zeta$, $\gamma$, eqs. (\ref{eq1.19}), (\ref{eq1.20}), (\ref{eq1.23}).

On the other hand the analytic form for this correlation function, eq.(\ref{eq2.2.5}), could also be developed in $z-1$. The coefficients of this last expansion are expressed as linear combinations of the coefficients $\{b_{k}\}$ of the polynomial $Q_{8}(1-z)$, eq.(\ref{eq2.2.2}). Putting equal 
the coefficients of the two expansions one obtains a system of linear equations on the coefficients $\{b_{k}\}$.

In this way one finds the following values, with the help of Mathematica :
\bea
b_{0}=\frac{196(32+c)}{243c}, \quad
b_{1}=-\frac{784(32+c)}{243c}, \quad
b_{2}=\frac{8(1904+53c)}{81c}, \nn\\
b_{3}=-\frac{16(3080+67c)}{243c}, \quad
b_{4}=\frac{4(8480+499c)}{243c}, \quad
b_{5}=-\frac{208(24+5c)}{81c}, \nn\\
b_{6}=\frac{8(16+13c)}{9c}, \quad
b_{7}=-\frac{16}{3}, \quad
b_{8}=1\label{eq2.2.71}
\eea

It is easy to verify that the equation (\ref{eq2.2.3}) on $Q_{8}(1-z)$ and $P_{8}(z)$ is verified, so that the two calculations, in the previous subsection (expansion in $z$, the s-channel expansion in the old terminology of dual amplitudes) and in the present one (expansion in $1-z$, the t-channel expansion) are consistent, one with another. In the present context it signifies that particular associativity relations, of the parafermionic operators algebra, are being verified.

It could also be concluded, from the above calculations, that without the fields $\tilde{\psi}(z)$ and $B(z)$ the algebra of the initial fields, $\psi(z)$, $\psi^{+}(z)$, would not be consistent. In fact, with the above calculations we have found the non-zero values of the constants $\zeta$ and $\gamma$, eqs.(\ref{eq1.20}), (\ref{eq1.23}). We remind that these constants enter into the expansions (\ref{eq1.1}) and (\ref{eq1.2}) for the products $\psi(z')\psi(z)$ and $\psi(z')\psi^{+}(z)$.

It could be observed  that to define the constants $\lambda$, $\zeta$, $\gamma$ it was sufficient to analyze the expansion of the product
\beq
\psi(1)\psi(z)\psi^{+}(0)\label{eq2.2.72}
\eeq
We have calculated the first coefficients of this expansion (matrix elements), by doing it directly, and we have substituted them into the lowest commutation relations. This gave equations on the constants of the operator algebra. Solving them, we have found the values of the constants. And next, doing further calculations with the obtained values of the constants, we have found that the correlation  function $<\psi^{+}(\infty)\psi(1)\psi(z)\psi^{+}(0)>$ was automatically consistent, in its expansions with respect to all limits : $z\rightarrow 0$, $z\rightarrow 1$, $z\rightarrow\infty$.

So far we have determined only three constants, $\lambda$, $\zeta$, $\gamma$, out of eight, eqs.(\ref{eq1.18}). We have determined the constants which enter into the expansion of the product (\ref{eq2.2.72}). In the following subsections we shall determine the remaining five constants. The analyses will be concentrated on expansions of different triple products, and not on actual calculations of the corresponding four-point functions, which are secondary objects for these analyses. Still, in all cases, the corresponding correlation functions have been checked to be consistent, once the equations, coming from products and lowest commutation relations,     
have been satisfied. The way this has  been done, in the previous and in the present subsections, for the product (\ref{eq2.2.72}) and the function $<\psi^{+}(\infty)\psi(1)\psi(z)\psi^{+}(0)>$.

\subsection{Analysis of the product $\psi(1)\psi(z)\tilde{\psi}^{+}(0).$}

As we shall see, into the expansion of this product enter the constants $\eta, \tilde{\lambda}, \mu$, which we shall determine. These constants are defined in (\ref{eq1.18}).

If we think of developing the product
\beq
\psi(1)\psi(z)\tilde{\psi}^{+}(0)\label{eq2.3.1}
\eeq
in powers of $z$, then the corresponding commutation relation is
\beq
\{\psi,\psi\}\tilde{\psi}^{+}(0)\label{2.3.2}
\eeq
It is of the type $\{\psi,\psi\}\Phi^{-1}(0)$ and it has already been defined in the Section 2.1, with $\psi^{+}(0)$ in place of $\tilde{\psi}^{+}(0)$. So we shall copy it here, from eqs. (\ref{eq2.63}), (\ref{eq2.64}) :
\beq
\sum^{\infty}_{l=0}D^{l}_{-\frac{1}{3}}(\psi_{-\frac{2}{3}+n-l}\psi_{\frac{2}{3}-n+l}+\psi_{\frac{1}{3}-n-l}\psi_{-\frac{1}{3}+n+l})\tilde{\psi}^{+}(0)=R(n)\label{eq2.3.3}
\eeq
\bea
R(n)=\{\frac{1}{2}(\Delta_{\psi}-\frac{1}{3}+n-1)(\Delta_{\psi}-\frac{1}{3}+n-2)\lambda\psi^{+}_{0}\nn\\
-(\Delta_{\psi}-\frac{1}{3}+n-1)(\Delta_{\psi})\cdot\lambda\beta^{(1)}_{\psi\psi,\psi^{+}}\cdot\psi^{+}_{0}\nn\\
+(\Delta_{\psi}+1)(\Delta_{\psi})\cdot\lambda\cdot\beta^{(11)}_{\psi\psi,\psi^{+}}\cdot\psi^{+}_{0}\nn\\
+\lambda\beta^{(2)}_{\psi\psi,\psi^{+}}\cdot(L_{-2}\psi^{+})_{0}+\zeta\tilde{\psi}^{+}_{0}\}\tilde{\psi}^{+}(0)\label{eq2.3.4}
\eea

To identify the matrix elements which could be calculated directly, we consider the actions of modes of $\psi(z)$ on $\tilde{\psi}^{+}(0)$. Comparing the direct expansion of the product $\psi(z)\tilde{\psi}^{+}(0)$, eq.(\ref{eq1.4}), and the formal expansion in modes (of the type (\ref{eq2.18})) :
\beq
\psi(z)\tilde{\psi}^{+}(0)=\sum_{n}\frac{1}{(z)^{\Delta_{\psi}-\frac{1}{3}+n}}\cdot\psi_{-\frac{1}{3}+n}\tilde{\psi}^{+}(0)\label{eq2.3.5}
\eeq
we find, first, that the dominant term corresponds to $n=1$ :
\bea
2\Delta_{\psi}-2=\Delta_{\psi}-\frac{1}{3}+n\nn\\
n=\Delta_{\psi}+\frac{1}{3}-2=1\label{eq2.3.6}
\eea
that the expansion in (\ref{eq2.3.5}) is of the form :
\bea
\psi(z)\tilde{\psi}^{+}(0)=\frac{1}{(z)^{2\Delta_{\psi}-2}}\{\psi_{\frac{2}{3}}\tilde{\psi}^{+}(0)+z\psi_{-\frac{1}{3}}\tilde{\psi}^{+}(0)\nn\\
+z^{2}\psi_{-\frac{4}{3}}\tilde{\psi}^{+}(0)+...\}\label{eq2.3.7}
\eea
and that the actions of the modes of $\psi$ on $\tilde{\psi}^{+}(0)$ are as follows :
\beq
\psi_{-\frac{1}{3}+n}\tilde{\psi}^{+}(0)=0,\quad n>1\label{eq2.3.8}
\eeq
\beq
\psi_{\frac{2}{3}}\tilde{\psi}^{+}(0)=\mu B(0)\label{eq2.3.9}
\eeq
\beq
\psi_{-\frac{1}{3}}\tilde{\psi}^{+}(0)=\mu\beta^{(1)}_{\psi\tilde{\psi}^{+},B}\cdot\partial B(0)\label{eq2.3.10}
\eeq
Starting with $\psi_{-\frac{4}{3}}\tilde{\psi}^{+}(0)$, in (\ref{eq2.3.7}), these operators are just the descendant states, to be dealt with the commutation rules, when their matrix elements are to be defined.

With the commutation rules $\{\psi,\psi\}\psi^{+}(0),$ in their particular form in (\ref{eq2.3.3}), we are going to consider the matrix elements of the type :
\beq
\psi_{\mu}\psi_{-\mu}\tilde{\psi}^{+}(0)\label{eq2.3.11}
\eeq
with $\mu=-\frac{2}{3}$, $\frac{1}{3}$, $\frac{4}{3}$, etc. .
By counting the dimensions ($\psi_{-\mu}$ is of the dimension $\mu$, $\tilde{\psi}^{+}(0)$ is of the dimension $\tilde{\Delta}_{\psi}=\Delta_{\psi}+2$, both with respect to the dilatation, global scaling) we know in advance that the triple product matrix elements in (\ref{eq2.3.11}) will produce operators in the $Z_{3}$ sector $q=1$ of (global) dimension $\Delta_{\psi}+2$ :
\beq
\psi_{\mu}\psi_{-\mu}\tilde{\psi}^{+}(0)\rightarrow\partial^{2}\psi(0),\,\,\,L_{-2}\psi(0),\,\,\,\tilde{\psi}(0)\label{eq2.3.12}
\eeq

We know, in addition, by looking at the actions (\ref{eq2.3.8})-(\ref{eq2.3.10}), that the matrix elements
\beq
\psi_{-\frac{2}{3}}\psi_{\frac{2}{3}}\tilde{\psi}^{+}(0),\quad \psi_{\frac{1}{3}}\psi_{-\frac{1}{3}}\tilde{\psi}^{+}(0)\label{eq2.3.13}
\eeq
could be calculated directly.

Now we have to look at the low commutation relations in (\ref{eq2.3.3}). They are the following ones :

\underline{n=1}
\beq
(\psi_{\frac{1}{3}}\psi_{-\frac{1}{3}}+D^{1}_{-\frac{1}{3}}\psi_{-\frac{2}{3}}\psi_{\frac{2}{3}}+\psi_{-\frac{2}{3}}\psi_{\frac{2}{3}})\tilde{\psi}^{+}(0)=R(1)\label{eq2.3.14}
\eeq

\underline{n=2}
\beq
(\psi_{\frac{4}{3}}\psi_{-\frac{4}{3}}+D^{1}_{-\frac{1}{3}}\psi_{\frac{1}{3}}\psi_{-\frac{1}{3}}+D^{2}_{-\frac{1}{3}}\psi_{-\frac{2}{3}}\psi_{\frac{2}{3}})\tilde{\psi}^{+}(0)=R(2)\label{eq2.3.15}
\eeq
and so on. We find that the first commutation relation, eq.(\ref{eq2.3.14}), in which all the terms could be calculated directly, provides an associativity relation to be  verified. It gives an equation on the constants of the parafermionic algebra. 

The next commutation relation, eq.(\ref{eq2.3.15}), and the commutation relations which follow, define the matrix elements $\psi_{\frac{4}{3}}\psi_{-\frac{4}{3}}\tilde{\psi}^{+}(0)$, and so on, which could not be calculated directly.

We conclude that (\ref{eq2.3.14}) is the equation on the constants which we were looking for, and that we need the expressions for the matrix elements  in eq.(\ref{eq2.3.13}). We need also the expressions for the actions of modes $\psi^{+}_{0},\,\,\, (L_{-2}\psi^{+})_{0},\,\,\, \tilde{\psi}^{+}_{0}$ on $\tilde{\psi}^{+}(0)$,
\beq
\psi^{+}_{0}\tilde{\psi}^{+}(0),\,\,\,(L_{-2}\psi^{+})_{0}\tilde{\psi}^{+}(0),\,\,\,\tilde{\psi}^{+}_{0}\tilde{\psi}^{+}(0)\label{eq2.3.16}
\eeq
which enter into $R(1)$ in the r.h.s. of (\ref{eq2.3.14}), cf.eq.(\ref{eq2.3.4}) for $R(n)$.

We do these calculations in the Appendix C where we find the following results :
\bea
\psi_{-\frac{2}{3}}\psi_{\frac{2}{3}}\tilde{\psi}^{+}(0)=\mu(\gamma\beta^{(11)}_{\psi B,\psi}\cdot\partial^{2}\psi(0)\nn\\
+\gamma\beta^{(2)}_{\psi B,\psi}\cdot L_{-2}\psi(0)+\mu\tilde{\psi}(0))\label{eq2.3.17}
\eea
\bea
\psi_{\frac{1}{3}}\psi_{-\frac{1}{3}}\tilde{\psi}^{+}(0)=\mu\beta^{(1)}_{\psi\tilde{\psi}^{+},B}\cdot[(\Delta_{\psi}-\frac{2}{3})\cdot(\gamma\beta^{(11)}_{\psi B,\psi}\cdot\partial^{2}\psi(0)\nn\\
+\gamma\beta^{(2)}_{\psi B,\psi}\cdot L_{-2}\psi(0)+\mu\tilde{\psi}(0))+\gamma\beta^{(1)}_{\psi B,\psi}\cdot\partial^{2}\psi(0)]\label{eq2.3.18}
\eea
\beq
\psi^{+}_{0}\tilde{\psi}^{+}(0)=\zeta\beta^{(11)}_{\psi\tilde{\psi},\psi^{+}}\cdot\partial^{2}\psi(0)+\zeta\beta^{(2)}_{\psi\tilde{\psi},\psi^{+}}L_{-2}\psi(0)+\eta\tilde{\psi}(0)\label{eq2.3.19}
\eeq
\bea
(L_{-2}\psi^{+})_{0}\tilde{\psi}^{+}(0)=\zeta\cdot L_{-2}\psi(0)+2(\Delta_{\psi}+1)\cdot(\zeta \beta^{(11)}_{\psi\tilde{\psi},\psi^{+}}\cdot\partial^{2}\psi(0)\nn\\
+\zeta \beta^{(2)}_{\psi\tilde{\psi},\psi^{+}}\cdot L_{-2}\psi(0)+\eta\tilde{\psi}(0))+\zeta\beta^{(1)}_{\psi\tilde{\psi},\psi^{+}}\cdot \partial^{2}\psi(0)\label{eq2.3.20}
\eea
\beq
\tilde{\psi}^{+}_{0}\tilde{\psi}^{+}(0)=\eta\beta^{(11)}_{\tilde{\psi}\tilde{\psi},\psi^{+}}\cdot\partial^{2}\psi(0)+\eta\beta^{(2)}_{\tilde{\psi}\tilde{\psi},\psi^{+}}L_{-2}\psi(0)+\tilde{\lambda}\psi(0)\label{eq2.3.21}
\eeq
Substituting these expressions into (\ref{eq2.3.14}) one obtains on equation in which
appear the operators $\partial^{2}\psi(0), \,L_{-2}\psi(0)$ and $\tilde{\psi}(0)$. Collecting the corresponding terms and demanding that the coefficients at each of these operators vanish, one obtains three equations on the operator algebra constants :
\bea
66(-8+3c)(784+57c)\zeta\cdot\eta\nn\\
-12(158816+c(-9052+15c))\zeta\cdot\lambda\nn\\
-(784+57c)\cdot(-1112+201c)\gamma\cdot\mu=0\label{eq2.3.22}
\eea
\bea
26(784+57c)\zeta\cdot\eta+4(9492+191c)\zeta\cdot\lambda\nn\\
-31(784+57c)\gamma\cdot\mu=0\label{eq2.3.23}
\eea
\bea
6(784+57c)\zeta\cdot\tilde{\lambda}-48(-202+c)\eta\cdot\lambda\nn\\
-11(784+57c)\mu^{2}=0\label{eq2.3.24}
\eea
The constants $\lambda,\zeta,\gamma$ have already been defined in the Section 2.1 and confirmed in the Section 2.2. We substitute their values (eqs.(\ref{eq1.19}), (\ref{eq1.20}), (\ref{eq1.23})) into the three equations above. This gives us three equations for the three remaining constants : $\eta, \mu, \tilde{\lambda}$. Solving them we obtain the values for these constants which are provided in (\ref{eq1.21}), (\ref{eq1.24}), (\ref{eq1.22}).

It could be observed that in the present analysis of the triple product $\psi(1)\psi(z)\tilde{\psi}^{+}(0)$ we have analyzed, effectively, three channels of the decomposition of this product :
\beq
\psi(1)\psi(z)\tilde{\psi}^{+}(0)\rightarrow\partial^{2}\psi(0),\,\,\,L_{-2}\psi(0),\,\,\,\tilde{\psi}(0)\label{eq2.3.25}
\eeq
Virasoro algebra operators $\partial^{2}$, $L_{-2}$ do not interfere with the associativity constraints. (One way to justify this statement is to remind that the correlation functions of Virasoro descendants could be expressed by the correlation functions of the corresponding Virasoro primaries to which the differential operators are applied). So, effectively, we have analyzed the associativity constraints related to the 4-point correlation functions
\beq
<\psi^{+}(\infty)\psi(1)\psi(z)\tilde{\psi}^{+}(0)>\label{eq2.3.26}
\eeq
and
\beq
<\tilde{\psi}^{+}(\infty)\psi(1)\psi(z)\tilde{\psi}^{+}(0)>\label{eq2.3.27}
\eeq
These constraints are expressed by the equations (\ref{eq2.3.22})-(\ref{eq2.3.24}). Once these equations had been satisfied, by fixing the values of the operator algebra constants, and the values found had been used in the further calculations of the functions (\ref{eq2.3.26}), (\ref{eq2.3.27}), these functions will then verify all the consistencies automatically. This has been checked for both functions, (\ref{eq2.3.26}) and (\ref{eq2.3.27}), by analyzing their expansion in $z$ and $1-z$, the way this has been done for the function $<\psi^{+}(\infty)\psi(1)\psi(z)\psi^{+}(0)>$ in the Sections 2.1 and 2.2.

We could summarize that out of eight constants in (\ref{eq1.18}) we have already defined six of them : $\lambda$, $\zeta$, $\gamma$, $\eta$, $\mu$, $\tilde{\lambda}$. It remains to define the constants $\tilde{\gamma}$ and $b$.

\subsection{Analysis of the product $\psi(1)\psi(z)\tilde{\psi}(0)$.}

We shall be looking at the $z$ expansion of this product. The corresponding commutation relation is $\{\psi,\psi\}\tilde{\psi}(0)$. It is of the following form :
\beq
\sum_{l=0}^{\infty}D^{l}_{-\frac{1}{3}}(\psi_{-\frac{1}{3}+n-l}\psi_{m+l}+\psi_{-\frac{1}{3}+m-l}\psi_{n+l})\tilde{\psi}(0)=R(n,m)\label{eq2.4.1}
\eeq
\bea
R(n,m)=\{\frac{1}{2}(\Delta_{\psi}+n-1)(\Delta_{\psi}+n-2)\lambda\cdot\psi^{+}_{-\frac{1}{3}+n+m}\nn\\
-(\Delta_{\psi}+n-1)(\Delta_{\psi}-\frac{1}{3}+n+m)\lambda\beta^{(1)}_{\psi\psi,\psi^{+}}\cdot\psi^{+}_{-\frac{1}{3}+n+m}\nn\\
+(\Delta_{\psi}+\frac{2}{3}+n+m)(\Delta_{\psi}-\frac{1}{3}+n+m)\lambda\beta^{(11)}_{\psi\psi,\psi^{+}}\cdot\psi^{+}_{-\frac{1}{3}+n+m}\nn\\
+\lambda\beta^{(2)}_{\psi\psi,\psi^{+}}\cdot(L_{-2}\psi^{+})_{-\frac{1}{3}+n+m}+\zeta\tilde{\psi}^{+}_{-\frac{1}{3}+n+m}\}\tilde{\psi}(0)\label{eq2.4.2}
\eea
It is calculated in a similar way, as the commutation relations $\{\psi,\psi\}\psi^{+}(0)$ and $\{\psi^{+},\psi\}\psi(0)$ in the previous subsections.

The product $\psi\psi\tilde{\psi}$ is $Z_{3}$ neutral, so it will be developed in the $Z_{3}$ neutral fields. We are going to analyze the level 4 channel, 
containing the dimension 4 fields of the $Z_{3}$ neutral sector, 
coming from the product $\psi(1)\psi(z)\tilde{\psi}(0)$. 
In these channel and sector we find the fields 
$\partial^{2}T$, $\wedge$, $B$:
\beq
\psi(1)\psi(z)\tilde{\psi}(0)\rightarrow\partial^{2}T(0),\,\,\,\wedge(0),\,\,\,B(0)\label{eq2.4.3}
\eeq
The reason to look at level 4 channel is to have the field $B$ present, which will make appear the constant $\tilde{\gamma}$ in the equations, as we shall see below. $\tilde{\gamma}$ is one of the two constants which are not  determined yet.

On lower levels, in the expansion of the product $\psi\psi\tilde{\psi}$, will appear just the Virasoro descendants of the identity operator. They will not produce new constraints. On levels higher than 4 will appear, in addition, the Virasoro descendants of the field $B$. But this will not lead to extra constraints as compared to the level 4 channel (\ref{eq2.4.3}). So, in fact, the channel (\ref{eq2.4.3}) is the only relevant one, in the decomposition of the product $\psi\psi\tilde{\psi}$.

To make appear the level 4 decomposition fields in the commutation relation (\ref{eq2.4.1}), (\ref{eq2.4.2}) we have to require that
\beq
\frac{1}{3}-n-m+\tilde{\Delta}_{\psi}=4\label{eq2.4.4}
\eeq
The l.h.s. of (\ref{eq2.4.4}) is the global dimension of the fields in (\ref{eq2.4.1}), (\ref{eq2.4.2}). So we have to require that
\beq
\frac{1}{3}-n-m+\Delta_{\psi}=2,\quad n+m=1,\quad m=1-n\label{eq2.4.5}
\eeq
With this restriction on the values of $m$, the commutation relation (\ref{eq2.4.1}), (\ref{eq2.4.2}) takes the form :
\beq
\sum^{\infty}_{l=0}D^{l}_{-\frac{1}{3}}(\psi_{-\frac{1}{3}+n-l}\psi_{1-n+l}+\psi_{\frac{2}{3}-n-l}\psi_{n+l})\tilde{\psi}(0)=R(n)\label{eq2.4.6}
\eeq
\bea
R(n)=R(n,1-n)=\{\frac{1}{2}(\Delta_{\psi}+n-1)(\Delta_{\psi}+n-2)\lambda\psi^{+}_{\frac{2}{3}}\nn\\
-(\Delta_{\psi}+n-1)(\Delta_{\psi}+\frac{2}{3})\lambda\beta^{(1)}_{\psi\psi,\psi^{+}}\cdot\psi^{+}_{\frac{2}{3}}\nn\\
+(\Delta_{\psi}+\frac{5}{3})(\Delta_{\psi}+\frac{2}{3})\lambda\beta^{(11)}_{\psi\psi,\psi^{+}}\cdot\psi^{+}_{\frac{2}{3}}\nn\\
+\lambda\beta^{(2)}_{\psi\psi,\psi^{+}}\cdot(L_{-2}\psi^{+})_{\frac{2}{3}}+\zeta\tilde{\psi}^{+}_{\frac{2}{3}}\}\tilde{\psi}(0)\label{eq2.4.7}
\eea

To make use of the commutation relations (\ref{eq2.4.6}), (\ref{eq2.4.7}) we need to know the actions of the modes of $\psi$, $\psi^{+}$, $\tilde{\psi}^{+}$ on the field $\tilde{\psi}(0)$. They are as follows :
\beq
\psi_{2+n}\tilde{\psi}(0)=0,\quad n>0\label{eq2.4.8}
\eeq
\beq
\psi_{2}\tilde{\psi}(0)=\zeta\psi^{+}(0)\label{eq2.4.9}
\eeq
\beq
\psi_{1}\tilde{\psi}(0)=\zeta\beta^{(1)}_{\psi\tilde{\psi},\psi^{+}}\cdot\partial\psi^{+}(0)\label{eq2.4.10}
\eeq
\beq
\psi_{0}\tilde{\psi}(0)=\zeta\beta^{(11)}_{\psi\tilde{\psi},\psi^{+}}\cdot\partial^{2}\psi^{+}(0)+\zeta\beta^{(2)}_{\psi\tilde{\psi},\psi^{+}}\cdot L_{-2}\psi^{+}(0)+\eta\tilde{\psi}^{+}(0)\label{eq2.4.11}
\eeq
and next the actions of modes which do not have an explicit (alternative) form : $\psi_{-1}\tilde{\psi}(0)$, $\psi_{-2}\tilde{\psi}(0)$, ... ;
\beq
\psi^{+}_{\frac{2}{3}+n}\tilde{\psi}(0)=0,\quad n>0\label{eq2.4.12}
\eeq
\beq
\psi^{+}_{\frac{2}{3}}\tilde{\psi}(0)=\mu B(0)\label{eq2.4.13}
\eeq
\beq
\psi^{+}_{-\frac{1}{3}}\tilde{\psi}(0)=\mu\beta^{(1)}_{\psi^{+}\tilde{\psi},B}\cdot\partial B(0)\label{eq2.4.14}
\eeq
$\psi^{+}_{-\frac{4}{3}}\tilde{\psi}(0),\, \psi^{+}_{-\frac{7}{3}}\tilde{\psi}(0),$..., are non-explicit;
\beq
\tilde{\psi}^{+}_{\frac{2}{3}}\tilde{\psi}(0)=\tilde{\gamma}\cdot B(0)+\beta^{(112)}_{\tilde{\psi}\tilde{\psi}^{+},I}\cdot\partial^{2}T(0)+\beta^{(22)}_{\tilde{\psi}\tilde{\psi}^{+},I}\cdot\wedge(0)\label{eq2.4.16}
\eeq
\bea
\tilde{\psi}^{+}_{-\frac{1}{3}}\tilde{\psi}(0)=\tilde{\gamma}\beta^{(1)}_{\tilde{\psi}^{+}\tilde{\psi},B}\cdot\partial B(0)
+\beta^{(1112)}_{\tilde{\psi}^{+}\tilde{\psi},I}\cdot\partial^{3}T(0)+\beta^{(122)}_{\tilde{\psi}^{+}\tilde{\psi},I}\cdot\partial\wedge(0)\label{eq2.4.17}
\eea
$\tilde{\psi}^{+}_{-\frac{4}{3}}\tilde{\psi}(0)$,\, $\tilde{\psi}^{+}_{-\frac{7}{3}}\tilde{\psi}(0)$,..., are non-explicit.

Returning to the commutation relation (\ref{eq2.4.6}) we observe that there is just one, lowest, commutation relation which will produce an equation on the constants, the commutation relation in which all the matrix elements could be calculated explicitly. It corresponds to $n=1$ (or $n=0$) in (\ref{eq2.4.6}) :
\beq
(\psi_{\frac{2}{3}}\psi_{0}+D^{1}_{-\frac{1}{3}}\psi_{-\frac{1}{3}}\psi_{1}+D^{2}_{-\frac{1}{3}}\psi_{-\frac{4}{3}}\psi_{2}+\psi_{-\frac{1}{3}}\psi_{1}+D^{1}_{-\frac{1}{3}}\psi_{-\frac{4}{3}}\psi_{2})\tilde{\psi}(0)=R(1)\label{eq2.4.18}
\eeq

We calculate the matrix elements which appear in the above equation. We get :
\bea
\psi_{-\frac{4}{3}}\psi_{2}\tilde{\psi}(0)=\psi_{-\frac{4}{3}}\zeta\psi^{+}(0)=\zeta\psi_{-\frac{4}{3}}\psi^{+}(0)\nn\\
=\zeta\{\beta^{(112)}_{\psi\psi^{+},I}\cdot\partial^{2} T(0)+\beta^{(22)}_{\psi\psi^{+},I}\cdot\wedge(0)+\gamma B(0)\}\label{eq2.4.19}
\eea
Calculation of other matrix elements will require, in addition, some commutations, similarly like it is being done is the Appendices B and C for other matrix elements.

With some calculations of this type one finds the following expressions :
\beq
\psi_{-\frac{1}{3}}\psi_{1}\tilde{\psi}(0)=\zeta\beta^{(1)}_{\psi\tilde{\psi},\psi^{+}}\cdot\{(\Delta_{\psi}-\frac{4}{3})\psi_{-\frac{4}{3}}\psi^{+}(0)+\frac{\Delta_{\psi}}{c}\partial^{2}T(0)\}\label{eq2.4.20}
\eeq
where
\beq
\psi_{-\frac{4}{3}}\psi^{+}(0)=\beta^{(112)}_{\psi\psi^{+},I}\cdot\partial^{2}T(0)+\beta^{(22)}_{\psi\psi^{+},I}\cdot\wedge(0)+\gamma B(0)\label{eq2.4.21}
\eeq
Next,
\bea
\psi_{\frac{2}{3}}\psi_{0}\tilde{\psi}(0)=\zeta\beta^{(11)}_{\psi\tilde{\psi},\psi^{+}}\cdot\psi_{\frac{2}{3}}\partial^{2}\psi^{+}(0)+\zeta\beta^{(2)}_{\psi\tilde{\psi},\psi^{+}}\cdot\psi_{\frac{2}{3}}L_{-2}\psi^{+}(0)\nn\\
+\eta\cdot\psi_{\frac{2}{3}}\tilde{\psi}^{+}(0)\label{eq2.4.22}
\eea
where
\bea
\psi_{\frac{2}{3}}\partial^{2}\psi^{+}(0)=(\Delta_{\psi}-\frac{1}{3})(\Delta_{\psi}-\frac{4}{3})\psi_{-\frac{4}{3}}\psi^{+}(0)+2(\Delta_{\psi}-\frac{1}{3})\frac{\Delta_{\psi}}{c}\cdot\partial^{2}T(0)\nn\\
+\frac{2\Delta_{\psi}}{c}\partial^{2}T(0)\label{eq2.4.23}
\eea
$\psi_{-\frac{4}{3}}\psi^{+}(0)$ is given by (\ref{eq2.4.21}),
\beq
\psi_{\frac{2}{3}}L_{-2}\psi^{+}(0)=(2\Delta_{\psi}-\frac{4}{3})\psi_{-\frac{4}{3}}\psi^{+}(0)+\frac{2\Delta_{\psi}}{c}\wedge(0)\label{eq2.4.24}
\eeq
\beq
\psi_{\frac{2}{3}}\tilde{\psi}^{+}(0)=\mu B(0)\label{eq2.4.25}
\eeq
We need also the actions of modes which appear in $R(1)$, cf. eq. (\ref{eq2.4.7}). They are found to be given by :
\beq
\psi^{+}_{\frac{2}{3}}\tilde{\psi}(0)=\mu B(0)\label{eq2.4.26}
\eeq
\beq
\tilde{\psi}^{+}_{\frac{2}{3}}\tilde{\psi}(0)=\tilde{\gamma}\cdot B(0)+\beta^{(112)}_{\tilde{\psi}\tilde{\psi}^{+},I}\cdot\partial^{2}T(0)+\beta^{(22)}_{\tilde{\psi}\tilde{\psi}^{+},I}\cdot\wedge(0)\label{eq2.4.27}
\eeq
\beq
(L_{-2}\psi^{+})_{\frac{2}{3}}\tilde{\psi}(0)=(2\Delta_{\psi}+\frac{8}{3})\mu\cdot B(0)\label{eq2.4.28}
\eeq
Putting the above expressions, equations (\ref{eq2.4.19})-(\ref{eq2.4.28}), into the equation (\ref{eq2.4.18}) one obtains, with Mathematica, an expression in which appears only the operator $B(0)$. The coefficients of the two other operators, $\partial^{2}T(0)$ and $\wedge(0)$, vanish without producing constraints. This is natural, the two operators are Virasoro descendants of the identity operator, the corresponding correlation function is the three points one :
\beq
<I\psi(1)\psi(z)\tilde{\psi}(0)>=<\psi(1)\psi(z)\tilde{\psi}(0)>\label{eq2.4.29}
\eeq
It produces no associativity constraints.

Requiring that the coefficient of the operator $B(0)$ vanishes one obtains a single equation on the constants, the following one :
\bea
7\cdot(1024+27c)\zeta\gamma-4\cdot(784+57c)\zeta\tilde{\gamma}\nn\\
+2\cdot(2(784+57c)\eta+(-2944+15c)\lambda)\mu=0\label{eq2.4.30}
\eea
Solving this equation for $\tilde{\gamma}$ and substituting the values of the other constants, $\lambda$, $\zeta$, $\eta$, $\gamma$, $\mu$, which are already known (have been determined in the subsections \ref{eq2.1} and \ref{eq2.3}), we obtain the value of $\tilde{\gamma}$ given in (\ref{eq1.25}).

We observe that the channel of $B(0)$ sorting out of the product $\psi(1)\psi(z)\tilde{\psi}(0)$ (which means that we are projecting it onto $B(\infty)$), this channel corresponds to the 4-point function
\beq
<B(\infty)\psi(1)\psi(z)\tilde{\psi}(0)>\label{eq2.4.31}
\eeq
We did further calculations of this function by developing it in $z$ and also in $1-z$, like we did it for the function $<\psi^{+}(\infty)\psi(1)\psi(z)\psi^{+}(0)>$ in subsections \ref{eq2.1} and \ref{eq2.2}. We have checked that the function (\ref{eq2.4.31}) is automatically consistent, once we have used the values of the constants found.

The only constant which remains unknown, out of eight, is $b$. It is defined in eq.(\ref{eq1.18}). We shall find its value in the next subsection.

\subsection{Analysis of the product $\psi(1)\psi^{+}(z)B(0)$.}

We shall analyze the $z$ expansion of this product. The corresponding commutation relation, $\{\psi,\psi^{+}\}B(0)$, is of the form :
\beq
\sum^{\infty}_{l=0}D^{l}_{-\frac{2}{3}}(\psi_{-\frac{4}{3}+n-l}\psi^{+}_{-\frac{2}{3}+m+l}-\psi^{+}_{-\frac{4}{3}+m-l}\psi_{-\frac{2}{3}+n+l})B(0)=R(n,m)\label{eq2.5.1}
\eeq
\bea
R(n,m)=\{\frac{1}{5!}(\Delta_{\psi}-\frac{2}{3}+n-1)(\Delta_{\psi}-\frac{2}{3}+n-2)(\Delta_{\psi}-\frac{2}{3}+n-3)\nn\\
\times(\Delta_{\psi}-\frac{2}{3}+n-4)(\Delta_{\psi}-\frac{2}{3}+n-5)\cdot\delta_{n+m-2,0}\nn\\
+\frac{1}{3!}(\Delta_{\psi}-\frac{2}{3}+n-1)(\Delta_{\psi}-\frac{2}{3}+n-2)(\Delta_{\psi}-\frac{2}{3}+n-3)\frac{2\Delta_{\psi}}{c}L_{n+m-2}\nn\\
-\frac{1}{2}(\Delta_{\psi}-\frac{2}{3}+n-1)(\Delta_{\psi}-\frac{2}{3}+n-2)\cdot(n+m)\frac{\Delta_{\psi}}{c}L_{n+m-2}\nn\\
+(\Delta_{\psi}-\frac{2}{3}+n-1)[\beta^{(112)}_{\psi\psi^{+},I}\cdot(n+m+1)(n+m)\cdot L_{n+m-2}\nn\\+\beta^{(22)}_{\psi\psi^{+},I}\cdot\wedge_{n+m-2}
+\gamma\cdot B_{n+m-2}]\nn\\
-\beta^{(1112)}_{\psi\psi^{+},I}\cdot(n+m+2)(n+m+1)(n+m)\cdot L_{n+m-2}
\nn\\-\beta^{(122)}_{\psi\psi^{+},I}\cdot(n+m+2)\cdot\wedge_{n+m-2}
-\gamma\cdot\beta^{(1)}_{\psi\psi^{+},B}\cdot(n+m+2)\cdot B_{n+m-2}\}B(0)\label{eq2.5.2}
\eea

We shall be looking for the dimension 4 operators channel, neutral sector, in the decomposition of the product $\psi\psi^{+}B$. This is to make appear the constant $b=<BBB>$, eq.(\ref{eq1.18}). In this channel
\beq
\psi(1)\psi^{+}(z)B(0)\rightarrow\partial^{2}T(0),\,\,\,\wedge(0),\,\,\,B(0)\label{eq2.5.3}
\eeq
In the commutation relation above, the dimension 4 operators channel corresponds to the constraint
\beq
n+m-2=0,\quad m=2-n\label{eq2.5.4}
\eeq
In this case (\ref{eq2.5.1}) and (\ref{eq2.5.2}) take the form :
\beq
\sum^{\infty}_{l=0}D^{l}_{-\frac{2}{3}}(\psi_{-\frac{4}{3}+n-l}\psi^{+}_{\frac{4}{3}-n+l}-\psi^{+}_{\frac{2}{3}-n-l}\psi_{-\frac{2}{3}+n+l})B(0)=R(n)\label{eq2.5.5}
\eeq
\bea
R(n)=R(n,2-n)\nn\\=\{\frac{1}{5!}(\Delta_{\psi}-\frac{2}{3}+n-1)(\Delta_{\psi}-\frac{2}{3}+n-2)(\Delta_{\psi}-\frac{2}{3}+n-3)\nn\\
\times(\Delta_{\psi}-\frac{2}{3}+n-4)(\Delta_{\psi}-\frac{2}{3}+n-5)\nn\\
+\frac{1}{3!}(\Delta_{\psi}-\frac{2}{3}+n-1)(\Delta_{\psi}-\frac{2}{3}+n-2)(\Delta_{\psi}-\frac{2}{3}+n-3)\frac{2\Delta_{\psi}}{c}L_{0}\nn\\
-\frac{1}{2}(\Delta_{\psi}-\frac{2}{3}+n-1)(\Delta_{\psi}-\frac{2}{3}+n-2)\cdot 2\cdot\frac{\Delta_{\psi}}{c}L_{0}\nn\\
+(\Delta_{\psi}-\frac{2}{3}+n-1)(\beta^{(112)}_{\psi\psi^{+},I}\cdot 3\cdot 2\cdot L_{0}
+\beta^{(22)}_{\psi\psi^{+},I}\cdot\wedge_{0}
+\gamma\cdot B_{0})\nn\\
-\beta^{(1112)}_{\psi\psi^{+},I}\cdot4\cdot3\cdot2\cdot L_{0}
-\beta^{(122)}_{\psi\psi^{+},I}\cdot4\cdot\wedge_{0}
-\gamma\cdot\beta^{(1)}_{\psi\psi^{+},B}\cdot4\cdot B_{0}\}B(0)\label{eq2.5.6}
\eea

Next we observe that we have the following actions of the modes of $\psi$ (and  $\psi^{+}$) on $B(0)$ :
\beq
\psi_{\frac{4}{3}+n}B(0)=0,\quad n>0\label{eq2.5.7}
\eeq
\beq
\psi_{\frac{4}{3}}B(0)=\gamma\cdot\psi(0)\label{eq2.5.8}
\eeq
\beq
\psi_{\frac{1}{3}}B(0)=\gamma\cdot\beta^{(1)}_{\psi B,\psi}\cdot\partial\psi(0)
\label{eq2.5.9}
\eeq
\beq
\psi_{-\frac{2}{3}}B(0)=\gamma\beta^{(11)}_{\psi B,\psi}\cdot\partial^{2}\psi(0)+\beta^{(2)}_{\psi B,\psi}\cdot L_{-2}\psi(0)+\mu\tilde{\psi}(0)\label{eq2.5.10}
\eeq

Now we can write down the lowest commutation relations, from (\ref{eq2.5.5}), those which lead to equations on the operator algebra constants.

The lowest one corresponds to $n=1$ :
\beq
(\psi_{-\frac{1}{3}}\psi^{+}_{\frac{1}{3}}+D^{1}_{-\frac{2}{3}}\psi_{-\frac{4}{3}}\psi^{+}_{\frac{4}{3}}-\psi^{+}_{-\frac{1}{3}}\psi_{\frac{1}{3}}-D^{1}_{-\frac{2}{3}}\psi^{+}_{-\frac{4}{3}}\psi_{\frac{4}{3}})B(0)=R(1)\label{eq2.5.11}
\eeq
This equation is satisfied identically : l.h.s. and r.h.s. vanish, separately.

The next one, for $n=2$ in (\ref{eq2.5.5}), is as follows :
\beq
(\psi_{\frac{2}{3}}\psi^{+}_{-\frac{2}{3}}+D^{1}_{-\frac{2}{3}}\psi_{-\frac{1}{3}}\psi^{+}_{\frac{1}{3}}+D^{2}_{-\frac{2}{3}}\psi_{-\frac{4}{3}}\psi^{+}_{\frac{4}{3}}-\psi^{+}_{-\frac{4}{3}}\psi_{\frac{4}{3}})B(0)=R(2)\label{eq2.5.12}
\eeq
This commutation relation leads to the equation on the constants.

With some calculations, we find the following expressions for the matrix elements in (\ref{eq2.5.12}) :
\beq
L_{0}B(0)=4\cdot B(0)\label{eq2.5.13}
\eeq
\beq
\wedge_{0}B_{0}=(2+L_{0})L_{0}B(0)=24B(0)\label{eq2.5.14}
\eeq
\beq
B_{0}B(0)=\beta^{(112)}_{BB,I}\cdot\partial^{2}T(0)+\beta^{(22)}_{BB,I}\cdot\wedge(0)+b\cdot B(0)\label{eq2.5.15}
\eeq
\beq
\psi_{-\frac{4}{3}}\psi^{+}_{\frac{4}{3}}B(0)=\gamma(\beta^{(112)}_{\psi\psi^{+},I}\cdot\partial^{2}T(0)+\beta^{(22)}_{\psi \psi^{+},I}\cdot\wedge(0)+\gamma\cdot B(0))
\label{eq2.5.16}
\eeq
\bea
\psi_{-\frac{1}{3}}\psi^{+}_{\frac{1}{3}}B(0)=\gamma\beta^{(1)}_{\psi B,\psi}\cdot\{\frac{4}{3}(\beta^{(112)}_{\psi\psi^{+},I}\cdot\partial^{2}T(0)+\beta^{(22)}_{\psi\psi^{+},I}\cdot\wedge(0)+\gamma\cdot B(0))\nn\\
+\frac{\Delta_{\psi}}{c}\partial^{2}T(0)\}\label{eq2.5.17}
\eea
\bea
\psi_{\frac{2}{3}}\psi^{+}_{-\frac{2}{3}}B(0)=\gamma\beta^{(11)}_{\psi B,\psi}\cdot\{\frac{7}{3} \cdot \frac{4}{3} (\beta^{(112)}_{\psi\psi^{+},I}\cdot\partial^{2}T(0)+\beta^{(22)}_{\psi\psi^{+},I}\cdot\wedge(0)+\gamma\cdot B(0))\nn\\
+2 \cdot \frac{7}{3} \cdot \frac{\Delta_{\psi}}{c}\partial^{2}T(0)
+\frac{2\Delta_{\psi}}{c}\partial^{2}T(0)\}\nn\\
+\gamma\beta^{(2)}_{\psi B,\psi}\cdot\{4(\beta^{(112)}_{\psi\psi^{+},I}\cdot\partial^{2}T(0)+\beta^{(22)}_{\psi\psi^{+},I}\cdot\wedge(0)+\gamma B(0))+\frac{2\Delta_{\psi}}{c}\wedge(0)\}\nn\\+\mu^{2}B(0)\label{eq2.5.18}
\eea
We observe in addition that
\beq
\psi^{+}_{-\frac{4}{3}}\psi_{\frac{4}{3}}B(0)=\psi_{-\frac{4}{3}}\psi^{+}_{\frac{4}{3}}B(0)\label{eq2.5.19}
\eeq
\beq
\psi^{+}_{-\frac{1}{3}}\psi_{\frac{1}{3}}B(0)=\psi_{-\frac{1}{3}}\psi^{+}_{\frac{1}{3}}B(0)\label{eq2.5.20}
\eeq
\beq
\psi^{+}_{\frac{2}{3}}\psi_{-\frac{2}{3}}B(0)=\psi_{\frac{2}{3}}\psi^{+}_{-\frac{2}{3}}B(0)\label{eq2.5.21}
\eeq
Now, putting the expressions (\ref{eq2.5.13})-(\ref{eq2.5.21}) into the commutation relation (\ref{eq2.5.12}), and simplifying it with Mathematica, one obtains an equation in which appears only the operator $B(0)$. The coefficients of $\partial^{2}T(0)$ and of $\wedge(0)$ vanish by themselves, which is natural (cf. the corresponding remarks in the subsection \ref{eq2.4}). Requiring that the coefficient of $B(0)$ vanish, we get the following equation on the constants :
\bea
-64(-116+c)(784+57c)+27(-256+c)c(22+5c)\gamma^{2}\nn\\
+6c(22+5c)(784+57c)(b\cdot\gamma-\mu^{2})=0\label{eq2.5.22}
\eea
Solving this equation with respect to the constant $b$, the last one which was still unknown, substituting the values of $\mu$ and $\gamma$ and simplifying, one gets the value of $b$ given in eq.(\ref{eq1.26}).

We remark finally that our analysis of the product $\psi\psi^{+}B$, with the channel (\ref{eq2.5.3}) being analyzed, corresponds to the associativity analysis of the 4-point function
\beq
<B(\infty)\psi(1)\psi^{+}(z)B(0)>\label{eq2.5.23}
\eeq
Having defined the constant $b$, the way this has been done above, we have continued calculating the function (\ref{eq2.5.23}), both in the $z$ and in the $1-z$ expansions. Like with the other tests, we have found this function to be automatically consistent once the defined value of $b$, and those of $\gamma$, $\mu$, have been used.

\numberwithin{equation}{section}

\section{Classification of the triple products. Complete proof of the associativity.}

First we shall list all the triple products, together with their decomposition channels. We start with principal fields ("light fields"), which are $\psi$,$\psi^{+}$, and adding, progressively the "secondary fields", ("heavy fields"), which are $\tilde{\psi}$, $\tilde{\psi}^{+}$, $B$. We call $\psi$, $\psi^{+}$ "light" and $\tilde{\psi}$, $\tilde{\psi}^{+}$, $B$ "heavy", because the dimensions of $\psi$, $\psi^{+}$ are smaller than those of $\tilde{\psi}$, $\tilde{\psi}^{+}$, $B$. 

The general rule is that three or two light fields in the triple product, $((lll)$ or $(llh))$, produces the associativity constraints on the operator algebra constants, while the triple products with one or zero light fields, $((lhh)$ or $(hhh))$, do not lead to constraints (equations) on the constants. This is being checked by first defining the corresponding commutation rules : symbolically,
the commutation rule $\{l,l\}l$, to be used for the triple product $(lll)$, 
or $\{l,l\}h$, for $(llh)$. Then writing down the lowest commutation relations for the modes of the operators and checking if there are  equations, so to say commutation relations in which all the matrix elements could be calculated independently, directly. These checks show that, in the present theory, equations are produced by the products $(lll)$ and $(llh)$, and that no equations are produced by the products $(lhh)$ and $(hhh)$.

We observe that some products of the types $(lll)$ and $(llh)$ have already been studied in subsections 2.1 $((\psi\psi\psi^{+}))$, 2.2. $((\psi^{+}\psi\psi))$, 2.3 $((\psi\psi\tilde{\psi}^{+}))$, 2.4 $((\psi\psi\tilde{\psi}))$, 2.5 $((\psi\psi^{+}B))$.

Products of types $(lhh)$ and $(hhh)$ will be studied in detail, particular cases of them, in the sections which follow.

Now we give the list, starting with the lightest and progressing to the heaviest products. First the products which give equations :
\beq
\psi\psi\psi\rightarrow B;\quad <B\psi\psi\psi>\label{eq3.1}
\eeq
\beq
\psi\psi\psi^{+}\rightarrow\psi,\tilde{\psi};\quad<\psi^{+}\psi\psi\psi^{+}>,\,\,<\tilde{\psi}^{+}\psi\psi\psi^{+}>\label{eq3.2}
\eeq
\beq
\psi\psi\tilde{\psi}\rightarrow B;\quad<B\psi\psi\tilde{\psi}>\label{eq3.3}
\eeq
\beq
\psi\psi\tilde{\psi}^{+}\rightarrow\psi,\tilde{\psi};\quad<\psi^{+}\psi\psi\tilde{\psi}^{+}>,\,\,<\tilde{\psi}^{+}\psi\psi\tilde{\psi}^{+}>\label{eq3.4}
\eeq
\beq
\psi\psi^{+}\tilde{\psi}\rightarrow\psi,\tilde{\psi};\quad<\psi^{+}\psi\psi^{+}\tilde{\psi}>,\,\,<\tilde{\psi}^{+}\psi\psi^{+}\tilde{\psi}>\label{eq3.5}
\eeq
\beq
\psi\psi B\rightarrow\psi^{+},\tilde{\psi}^{+};\quad<\psi\psi\psi B>,\,\,<\tilde{\psi}\psi\psi B>\label{eq3.6}
\eeq
\beq
\psi\psi^{+} B\rightarrow B;\quad<B\psi\psi^{+}B>,\label{eq3.7}
\eeq
Next the products, of the type $(lhh)$, which do not give equations on the constants :
\beq
\psi\tilde{\psi}\tilde{\psi}\rightarrow B ;\quad<B\psi\tilde{\psi}\tilde{\psi}>\label{eq3.8}
\eeq
\beq
\psi\tilde{\psi}\tilde{\psi}^{+}\rightarrow\psi,\tilde{\psi};\quad<\psi^{+}\psi\tilde{\psi}\tilde{\psi}^{+}>,\,\,<\tilde{\psi}^{+}\psi\tilde{\psi}\tilde{\psi}^{+}>\label{eq3.9}
\eeq
\beq
\psi^{+}\tilde{\psi}\tilde{\psi}\rightarrow\psi,\tilde{\psi};\quad<\psi^{+}\psi^{+}\tilde{\psi}\tilde{\psi}>,\,\,<\tilde{\psi}^{+}\psi^{+}\tilde{\psi}\tilde{\psi}>\label{eq3.10}
\eeq
\beq
\psi\tilde{\psi}B\rightarrow\psi^{+},\tilde{\psi}^{+};\quad<\psi\psi\tilde{\psi}B>,\,\,<\tilde{\psi}\psi\tilde{\psi}B>\label{eq3.11}
\eeq
\beq
\psi^{+}\tilde{\psi}B\rightarrow B;\quad< B\psi^{+}\tilde{\psi}B>\label{eq3.12}
\eeq
\beq
\psi BB\rightarrow\psi,\tilde{\psi};\quad<\psi^{+}\psi BB>,\,\,<\tilde{\psi}^{+}\psi BB>\label{eq3.13}
\eeq
And finally the products of the type $(hhh)$ (which do not give equations either) :
\beq
\tilde{\psi}\tilde{\psi}\tilde{\psi}\rightarrow B;\quad<B\tilde{\psi}\tilde{\psi}\tilde{\psi}>\label{eq3.14}
\eeq
\beq
\tilde{\psi}\tilde{\psi}\tilde{\psi}^{+}\rightarrow\psi,\tilde{\psi};\quad<\psi^{+}\tilde{\psi}\tilde{\psi}\tilde{\psi}^{+}>,\,\,<\tilde{\psi}^{+}\tilde{\psi}\tilde{\psi}\tilde{\psi}^{+}>\label{eq3.15}
\eeq
\beq
\tilde{\psi}\tilde{\psi}B\rightarrow\psi^{+},\tilde{\psi}^{+};\quad<\psi\tilde{\psi}\tilde{\psi}B>,\,\,<\tilde{\psi}\tilde{\psi}\tilde{\psi}B>\label{eq3.16}
\eeq
\beq
\tilde{\psi}\tilde{\psi}^{+}B\rightarrow B; \quad<B\tilde{\psi}\tilde{\psi}^{+}B>\label{eq3.17}
\eeq
\beq
\tilde{\psi}BB\rightarrow\psi,\tilde{\psi};\quad<\psi^{+}\tilde{\psi}BB>,\,\,<\tilde{\psi}^{+}\tilde{\psi}BB>\label{eq3.18}
\eeq
\beq
BBB\rightarrow B;\quad<BBBB>\label{eq3.19}
\eeq

Several remarks are in order.

1. Evidently, we have not listed the products which are obtained from those which are listed by the $Z_{3}$ conjugation. Our chiral algebra, eq.(\ref{eq1.1})-(\ref{eq1.9}), is supposed to be $Z_{3}$ conjugation symmetric.

2. The associativity constraints on the constants, which we organize and define in a particular way by analyzing the triple products, in each case, for each particular triple product with its decomposition channel, the associativity constraints are induced, in fact, by the corresponding 4-point function : by the problem of defining a given 4-point function which is to be consistent with the chiral algebra.

E.g. when we study the product $\psi\psi\psi^{+}\rightarrow\psi$, the associativity constraints obtained are those of the 4-point function $<\psi^{+}\psi\psi\psi^{+}>$.

One consequence of this observation is that the products with different positioning of the fields in the triple product and with the same decomposition channel (like $\psi\psi^{+}\psi\rightarrow\psi$ 
and $\psi\psi\psi^{+}\rightarrow\psi$) will produce same constraints on the constants.

Another consequence is that different triple products which are associated with the same 4-point function will also produce same constraints. E.g. the products $\psi\psi\psi\rightarrow B$ and $\psi\psi B$ with the lower (lighter) channel, $\psi^{+}$, eq. (\ref{eq3.1}) and (\ref{eq3.6}), are equivalent. Equivalent also are the product $\psi\psi\psi^{+}$ with the higher (heavier) channel, $\tilde{\psi}$, eq.(\ref{eq3.2}), and the product $\psi\psi^{+}\tilde{\psi}$, eq.(\ref{eq3.5}), with the lower channel. This is because the corresponding 4-point functions, $<\tilde{\psi}^{+}\psi\psi\psi^{+}>$ and $<\psi^{+}\psi\psi^{+}\tilde{\psi}>$ are the same, up to the $Z_{3}$ conjugation and repositionning of the fields.

3. As a consequence of these equivalences, in the above list of products, eq.(\ref{eq3.1})-(\ref{eq3.19}), we have to consider, systematically, only the higher channels of the decomposition, the lower channels corresponding to 4-point functions already considered. This is with the exception of the product $\psi\psi\psi^{+}$ eq.(\ref{eq3.2}) where both channels are to be considered. This is because it is at the very beginning of the list.

4. There are some additional equivalences in the list. E.g. when we have analyzed the product $\psi\psi\tilde{\psi}^{+}$, in the Sec. 2.3, with the higher decomposition channel, $\tilde{\psi}$, we have obtained equations where have appeared the operator $\tilde{\psi}(0)$, 
but also the Virasoro descendants of $\psi(0)$, $\partial^{2}\psi(0)$ and $L_{-2}\psi(0)$. So that, effectively, we have analyzed at the same time the functions $<\psi^{+}\psi\psi\tilde{\psi}^{+}>$ and $<\tilde{\psi}^{+}\psi\psi\tilde{\psi}^{+}>$. Which means that we didn't have to analyze, in Sections 2.1, 2.2, the product $\psi\psi\psi^{+}$ with the higher channel, $\tilde{\psi}$. We didn't, in fact.

Some other equivalences could be found in the list, if one compares the corresponding 4-point functions.
\vskip0.5cm
After these remarks we could give the short list of triple products which provide the complete set of associativity constraints in this theory. They are the following ones :
\beq
\psi\psi\psi\rightarrow B;\quad<B\psi\psi\psi>\label{eq3.20}
\eeq
\beq
\psi\psi\psi^{+}\rightarrow\psi;\quad<\psi^{+}\psi\psi\psi^{+}>\label{eq3.21}
\eeq
\beq
\psi\psi\tilde{\psi}\rightarrow B;\quad<B\psi\psi\tilde{\psi}>\label{eq3.22}
\eeq
\beq
\psi\psi\tilde{\psi}^{+}\rightarrow\tilde{\psi},\,\,\mbox{plus}\,\,\partial^{2}\psi,L_{-2}\psi;\quad <\tilde{\psi}^{+}\psi\psi\tilde{\psi}^{+}>,\,\,
\mbox{plus}\,\,<\psi^{+}\psi\psi\tilde{\psi}^{+}>
\label{eq3.23}
\eeq
\beq
\psi\psi^{+}\tilde{\psi}\rightarrow \tilde{\psi};\quad<\tilde{\psi}^{+}\psi\psi^{+}\tilde{\psi}>\label{eq3.24}
\eeq
\beq
\psi\psi^{+}B\rightarrow B;\quad<B\psi\psi^{+}B>\label{eq3.25}
\eeq
Still one remark is in order : we have added the Virasoro descendants $\partial^{2}\psi$, $L_{-2}\psi$ in eq.(\ref{eq3.23}), while we have suppressed them in(\ref{eq3.24}). This is because, if we solve the constraints successively, from (\ref{eq3.20}) to (\ref{eq3.25}), in this case, when we consider (\ref{eq3.24}), the correlation function $<\psi^{+}\psi\psi^{+}\tilde{\psi}>$ (due to $\partial^{2}\psi$, $L_{-2}\psi$) have already been considered in (\ref{eq3.23}). The values of the constants have already been chosen so that the function $<\psi^{+}\psi\psi^{+}\tilde{\psi}>$ were consistent. As a consequence, when we consider (\ref{eq3.24}), the coefficients of the operators $\partial^{2}\psi(0)$, $L_{-2}\psi(0)$, in the equations, will vanish automatically. They do vanish automatically.

We claim that the equations coming from the triple products (\ref{eq3.20}) - (\ref{eq3.25}) provide the complete set of associativity constraints in the present theory.

We have analyzed, in Sections 2.1 - 2.5, the products (\ref{eq3.21}), (\ref{eq3.23}), (\ref{eq3.22}), (\ref{eq3.25}), which allowed us to fix the values of all the eight constants.

Then, the equations coming from the two remaining relevant products, (\ref{eq3.20}) and (\ref{eq3.24}), serve as tests of the solution found. We did these tests : with the values of the constants found, the equations coming from (\ref{eq3.20}), (\ref{eq3.24}) all get satisfied.

We take it as a complete proof of associativity of the present chiral algebra.

Having made this conclusion, we haven't explained yet a certain number of aspects of the theory. We shall do it in the rest of the paper.

The second set of triple products, in (\ref{eq3.8}) - (\ref{eq3.13}), all of the type $(lhh)$, they do not produce equations, if we consider only the higher channels for the reasons explained above. The lowest commutation relations start defining the non-explicit matrix elements, needed to calculate the corresponding 4-point correlation function.

Particular aspect of this second set is that, on one hand, they provide no constraints, on the other hand, all the matrix elements, needed to calculate the correlation function, get defined by the commutation relations. In this case one obtains the 4-point function which is automatically consistent with the operator algebra, for all constants unconstrained.

We have calculated all the 4-point functions which appear in (\ref{eq3.8}) - (\ref{eq3.13}). We observe that the number of functions in (\ref{eq3.8}) - (\ref{eq3.13}) is smaller than it appears. We observe again that we have to consider only the higher channels, in (\ref{eq3.8}) - (\ref{eq3.13}), and the corresponding higher correlation function. Then we find that the function in (\ref{eq3.10}) is equivalent to the function in (\ref{eq3.9}), (\ref{eq3.11}) is equivalent to (\ref{eq3.8}), (\ref{eq3.13}) is equivalent to (\ref{eq3.12}). So that we have to calculate and verify the consistency with the operator algebra, just in the cases (\ref{eq3.8}), (\ref{eq3.9}), (\ref{eq3.12}). This will be done in the next Section.

The triple products in the last set, eq.(\ref{eq3.14}) - (\ref{eq3.19}), they all have the following common features. On one hand, they do not produce constraints (equations). On the other, the commutation relations for the modes of operators, involved in these products, they do not define all the matrix elements needed to calculate the corresponding 4-point functions. The lowest (non explicit) matrix element, in each case, is not defined, while the rest of (non explicit) matrix elements get expressed in terms of the lowest one. This is natural and, in fact,  could have been expected : the dimensions of the operators are too high, we are missing a number of singular terms
when calculating the commutation relations for the modes of these operators,
according to the rules described in the section 1.

But all the operators in the set (\ref{eq3.14}) - (\ref{eq3.19}) are the secondary ones, with respect to the operator algebra products. They all could be "factorized" into the products of principal fields, with some subtractions. E.g., the modes of $B$ could be expressed through the products of modes of $\psi$, $\psi^{+}$, with some subtractions, using, in addition, the commutation relation $\{\psi,\psi^{+}\}$ applied to the field to which $B$ is applied.

Modes of $\tilde{\psi}^{+}$ could be factorized into products of modes of $\psi$, $\psi$.

Then, after factorization of one of the secondary fields (of a given product in (\ref{eq3.14}) - (\ref{eq3.19})) into the product the principal fields (all is done for the modes of operators, to be more precise), the missing lowest matrix elements can be defined.

To show this technique in details, in the Section 5 the product (\ref{eq3.17}) will be analyzed, the correlation function $<B\tilde{\psi}\tilde{\psi}^{+}B>$ will be calculated and verified to be consistent with the operator algebra of chiral fields, with the constants staying unconstrained.

We observe finally that in the list (\ref{eq3.14}) - (\ref{eq3.19}) there are equivalences, like in the previous set, eq.(\ref{eq3.8}) - (\ref{eq3.13}). There are just four different functions in (\ref{eq3.14}) - (\ref{eq3.19}). They are :
\beq
<B\tilde{\psi}\tilde{\psi}\tilde{\psi}>\label{eq3.26}
\eeq
\beq
<\tilde{\psi}^{+}\tilde{\psi}\tilde{\psi}\tilde{\psi}^{+}>\label{eq3.27}
\eeq
\beq
<B\tilde{\psi}\tilde{\psi}^{+}B>\label{eq3.28}
\eeq
\beq
<BBBB>\label{eq3.29}
\eeq
 We shall calculate, in Section 5, the third one of these functions, 
 to illustrate the techniques.

\numberwithin{equation}{subsection}

\section{Calculation of the correlation functions $<B\psi\tilde{\psi}\tilde{\psi}>$, $<\tilde{\psi}^{+}\psi\tilde{\psi}\tilde{\psi}^{+}>$, $<B\psi^{+}\tilde{\psi}B>$.}

We shall calculate first the last of the above functions. We shall calculate it in the configuration $<B\tilde{\psi}^{+}\psi B>$. This is for no particular reason, just because for this configuration we have obtained ("historically",
in the process of calculations) somewhat more developed expressions. 

We have calculated this function by developing it in both channels :
in the channel $s$ (development in $z$) and in the channel $t$ (development in $z-1$).


\subsection{Correlation function $<B(\infty)\tilde{\psi}^{+}(1)\psi(z)B(0)>$, calculated in the $z$ expansion.}

We are considering the triple product $\tilde{ \psi}^{+}(1)\psi(z)B(0)\rightarrow B(0)$ which is equivalent to the product in (\ref{eq3.12}). 

To perform the $z$ expansion, of the product, and of the function afterwards, we need the commutation relation $\{\tilde{\psi}^{+},\psi\}B(0)$. It is of the form :
\beq
\sum^{\infty}_{l=0}D^{l}_{\frac{4}{3}}(\tilde{\psi}^{+}_{\frac{2}{3}+n-l}\psi_{-\frac{2}{3}+m+l}-\psi_{\frac{2}{3}+m-l}\tilde{\psi}^{+}_{-\frac{2}{3}+n+l})B(0)=R(n,m)\label{eq4.1}
\eeq
\bea
R(n,m)=\{(\tilde{\Delta}-\frac{2}{3}+n-1)\mu B_{n+m}\nn\\
-(4+n+m)\mu\beta^{(1)}_{\tilde{\psi}^{+}\psi,B}\cdot B_{n+m}\}B(0)\label{eq4.2}
\eea
(above, $\tilde{\Delta} = \Delta_{\tilde{\psi}}$).  Projection of this commutation relation on the level 4, level of B, corresponds to $n+m=0$, $m=-n$. One obtains :
\beq
\sum^{\infty}_{l=0}D^{l}_{\frac{4}{3}}(\tilde{\psi}^{+}_{\frac{2}{3}+n-l}\psi_{-\frac{2}{3}-n+l}-\psi_{\frac{2}{3}-n-l}\tilde{\psi}^{+}_{-\frac{2}{3}+n+l})B(0)=R(n)\label{eq4.3}
\eeq
\bea
R(n)=R(n,-n) = \{(\tilde{\Delta}-\frac{2}{3}+n-1)\mu B_{0}\nn\\
-4\mu\beta^{(1)}_{\tilde{\psi}^{+}\psi,B}\cdot B_{0}\}B(0)\label{eq4.4}
\eea
To use this commutation relation one needs to know the actions of the first modes of $\psi$ and $\tilde{\psi}^{+}$ on $B(0)$.
They are as follows :
\beq
\psi_{\frac{4}{3}+n}B(0)=0,\quad n>0\label{eq4.5}
\eeq
\beq
\psi_{\frac{4}{3}}B(0)=\gamma\psi(0)\label{eq4.6}
\eeq
\beq
\psi_{\frac{1}{3}}B(0)=\gamma\beta^{(1)}_{\psi B,\psi}\cdot\partial\psi(0)\label{eq4.7}
\eeq
\beq
\psi_{-\frac{2}{3}}B(0)=\gamma\beta^{(11)}_{\psi B,\psi}\cdot\partial^{2}\psi(0)+\gamma\beta^{(2)}_{\psi B,\psi}\cdot L_{-2}\psi(0)+\mu\tilde{\psi}(0)\label{eq4.8}
\eeq
and
\beq
\tilde{\psi}^{+}_{\frac{4}{3}+n}B(0)=0,\quad n>0\label{eq4.9}
\eeq
\beq
\tilde{\psi}^{+}_{\frac{4}{3}}B(0)=\mu\psi^{+}(0)\label{eq4.10}
\eeq
\beq
\tilde{\psi}^{+}_{\frac{1}{3}}B(0)=\mu\beta^{(1)}_{\tilde{\psi}B,\psi}\cdot\partial\psi^{+}(0)\label{eq4.11}
\eeq
\beq
\tilde{\psi}^{+}_{-\frac{2}{3}}B(0)=\mu\beta^{(11)}_{\tilde{\psi} B,\psi}\cdot\partial^{2}\psi^{+}(0)+\mu\beta^{(2)}_{\tilde{\psi} B,\psi}\cdot L_{-2}\psi^{+}(0)+\tilde{\gamma}\tilde{\psi}^{+}(0)\label{eq4.12}
\eeq

The lowest commutation relation in (\ref{eq4.3}), (\ref{eq4.4}) corresponds to \underline{$n=0$} :
\bea
(\tilde{\psi}_{\frac{2}{3}}^{+}\psi_{-\frac{2}{3}}+D^{1}_{\frac{4}{3}}\tilde{\psi}^{+}_{-\frac{1}{3}}\psi_{\frac{1}{3}}+D^{2}_{\frac{4}{3}}\tilde{\psi}^{+}_{-\frac{4}{3}}\psi_{\frac{4}{3}}\nn\\
-\psi_{\frac{2}{3}}\tilde{\psi}^{+}_{-\frac{2}{3}}-D^{1}_{\frac{4}{3}}\psi_{-\frac{1}{3}}\tilde{\psi}^{+}_{\frac{1}{3}}-D^{2}_{\frac{4}{3}}\psi_{-\frac{4}{3}}\tilde{\psi}^{+}_{\frac{4}{3}})B(0)=R(0)\label{eq4.13}
\eea
All the matrix elements in this relation could be defined directly. So it appears, that it provides an equation on the constants. In fact it is easily verified, with the expressions of the matrix elements that we shall define below, that the l.h.s. and the r.h.s. of (\ref{eq4.13}) vanish separately, for unconstrained values of the constants. There is no equation on the constants produced by the product $\tilde{\psi}^{+}\psi B\rightarrow B$, as has been stated in the Section 3.

For \underline{$n=1$} one obtains, from (\ref{eq4.3}) : 
\bea
(\tilde{\psi}_{\frac{5}{3}}^{+}\psi_{-\frac{5}{3}}+D^{1}_{\frac{4}{3}}\tilde{\psi}^{+}_{\frac{2}{3}}\psi_{-\frac{2}{3}}+D^{2}_{\frac{4}{3}}\tilde{\psi}^{+}_{-\frac{1}{3}}\psi_{\frac{1}{3}}\nn\\
+D^{3}_{\frac{4}{3}}\tilde{\psi}^{+}_{-\frac{4}{3}}\psi_{\frac{4}{3}}-\psi_{-\frac{1}{3}}\tilde{\psi}^{+}_{\frac{1}{3}}-D^{1}_{\frac{4}{3}}\psi_{-\frac{4}{3}}\tilde{\psi}^{+}_{\frac{4}{3}})B(0)=R(1)\label{eq4.14}
\eea
and for \underline{$n=2$} one gets :
\bea
(\tilde{\psi}_{\frac{8}{3}}^{+}\psi_{-\frac{8}{3}}+D^{1}_{\frac{4}{3}}\tilde{\psi}^{+}_{\frac{5}{3}}\psi_{-\frac{5}{3}}+D^{2}_{\frac{4}{3}}\tilde{\psi}^{+}_{\frac{2}{3}}\psi_{-\frac{2}{3}}\nn\\
+D^{3}_{\frac{4}{3}}\tilde{\psi}^{+}_{-\frac{1}{3}}\psi_{\frac{1}{3}}+D^{4}_{\frac{4}{3}}\tilde{\psi}^{+}_{-\frac{4}{3}}\psi_{\frac{4}{3}}-\psi_{-\frac{4}{3}}\tilde{\psi}^{+}_{\frac{4}{3}})B(0)=R(2)\label{eq4.15}
\eea
Starting from $\tilde{\psi}^{+}_{\frac{11}{3}}\psi_{-\frac{11}{3}}B(0)$, the matrix elements could be calculated by the recurrence relation :
\beq
m(n)=R(n-2)-\sum^{n}_{k=1}D^{k}_{\frac{4}{3}}m(n-k)\label{eq4.16}
\eeq
 where we use the notation :
 \bea
 m(0)=\tilde{\psi}^{+}_{-\frac{4}{3}}\psi_{\frac{4}{3}}B(0),\quad m(1)=\tilde{\psi}^{+}_{-\frac{1}{3}}\psi_{\frac{1}{3}}B(0),\nn\\
 m(2)=\tilde{\psi}^{+}_{\frac{2}{3}}\psi_{-\frac{2}{3}}B(0),\quad m(3)=\tilde{\psi}^{+}_{\frac{5}{3}}\psi_{-\frac{5}{3}}B(0),\nn\\
 m(4)=\tilde{\psi}^{+}_{\frac{8}{3}}\psi_{-\frac{8}{3}}B(0),\quad m(5)=\tilde{\psi}^{+}_{\frac{11}{3}}\psi_{-\frac{11}{3}}B(0), \quad \mbox{etc.}\label{eq4.17}
 \eea
We shall also note, in the following, the opposite ("conjugate") matrix elements, present in (\ref{eq4.13}) - (\ref{eq4.15}), as
\bea
mc(0)=\psi_{-\frac{4}{3}}\tilde{\psi}^{+}_{\frac{4}{3}}B(0),\quad mc(1)=\psi_{-\frac{1}{3}}\tilde{\psi}^{+}_{\frac{1}{3}}B(0),\nn\\
mc(2)=\psi_{\frac{2}{3}}\tilde{\psi}^{+}_{-\frac{2}{3}}B(0)\label{eq4.18}
\eea
\vskip0.5cm
We can conclude that all the matrix elements get defined.
The lowest ones, $m(0)$, $m(1)$, $m(2)$ in (\ref{eq4.17}), they can be calculated directly. The higher matrix elements, starting from $m(3)$, eq.(\ref{eq4.17}), they all get defined by the commutation relations (\ref{eq4.14}), (\ref{eq4.15}), (\ref{eq4.16}).

With some calculations, one finds the following expressions for the lowest matrix elements :
\beq
m(0)=\tilde{\psi}^{+}_{-\frac{4}{3}}\psi_{\frac{4}{3}}B(0)=\gamma\mu B(0)\label{eq4.19}
\eeq
\beq
m(1)=\tilde{\psi}^{+}_{-\frac{1}{3}}\psi_{\frac{1}{3}}B(0)=\gamma\beta^{(1)}_{\psi B,\psi}\cdot(\tilde{\Delta}-\frac{4}{3})\mu B(0)\label{eq4.20}
\eeq
\bea
m(2)=\tilde{\psi}^{+}_{\frac{2}{3}}\psi_{-\frac{2}{3}}B(0)=\gamma\beta^{(11)}_{\psi B,\psi}\cdot(\tilde{\Delta}-\frac{1}{3})(\tilde{\Delta}-\frac{4}{3})\mu B(0)\nn\\
+\gamma\beta^{(2)}_{\psi B,\psi}\cdot(2\tilde{\Delta}-\frac{4}{3})\mu B(0)\nn\\
+\mu\cdot(\beta^{(112)}_{\tilde{\psi}\tilde{\psi}^{+},I}\cdot\partial^{2}T(0)+\beta^{(22)}_{\tilde{\psi}\tilde{\psi}^{+},I}\cdot\wedge(0)+\tilde{\gamma}\cdot B(0))\label{eq4.21}
\eea
and
\beq
mc(0)=\psi_{-\frac{4}{3}}\tilde{\psi}^{+}_{\frac{4}{3}}B(0)=\mu\psi_{-\frac{4}{3}}\psi^{+}(0)\label{eq4.22}
\eeq
where
\beq
\psi_{-\frac{4}{3}}\psi^{+}(0)=\beta^{(112)}_{\psi\psi^{+},I}\cdot\partial^{2}T(0)+\beta^{(22)}_{\psi\psi^{+},I}\cdot\wedge(0)+\gamma\cdot B(0)\label{eq4.23}
\eeq
Next
\beq
mc(1)=\psi_{-\frac{1}{3}}\tilde{\psi}^{+}_{\frac{1}{3}}B(0)=\mu\beta^{(1)}_{\tilde{\psi}B,\psi}[(\Delta-\frac{4}{3})\psi_{-\frac{4}{3}}\psi^{+}(0)+\frac{\Delta}{c}\partial^{2}T(0)]\label{eq4.24}
\eeq
\bea
mc(2)=\psi_{\frac{2}{3}}\tilde{\psi}^{+}_{-\frac{2}{3}}B(0)=\mu\beta^{(11)}_{\tilde{\psi}B,\psi}\cdot[(\Delta-\frac{1}{3})(\Delta-\frac{4}{3})\psi_{-\frac{4}{3}}\psi(0)\nn\\
+2(\Delta-\frac{1}{3})\frac{\Delta}{c}\partial^{2}T(0)+\frac{2\Delta}{c}\partial^{2}T(0)]\nn\\
+\mu\beta^{(2)}_{\tilde{\psi}B,\psi}\cdot[(2\Delta-\frac{4}{3})\psi_{-\frac{4}{3}}\psi^{+}(0)+\frac{2\Delta}{c}\wedge(0)]+\tilde{\gamma}\mu B(0)\label{eq4.25}
\eea
In the above we have used the notations : $\Delta=\Delta_{\psi}$, 
$\tilde{\Delta}=\Delta_{\tilde{\psi}}$.
We need, in addition, the expression for $B_{0}B(0)$, which enters into $R(n)$, eq.(\ref{eq4.4}). It is of the form :
\beq
B_{0}B(0)=\beta^{(112)}_{BB,I}\cdot\partial^{2}T(0)+\beta^{(22)}_{BB,I}\cdot\wedge(0)+b\cdot B(0)\label{eq4.26}
\eeq

Now we can define the correlation function $<B(\infty)\tilde{\psi}^{+}(1)\psi(z)B(0)>$.

This function has the following analytic form :
\beq
<B(\infty)\tilde{\psi}^{+}(1)\psi(z)B(0)>=\frac{P_{6}(z)}{(1-z)^{2\Delta-2}(z)^{4}}\label{eq4.27}
\eeq
\beq
P_{6}(z)=\sum^{6}_{k=0}a_{k}z^{k}\label{eq4.28}
\eeq
with
\beq
a_{0}=a_{6}=\gamma\mu\label{eq4.29}
\eeq
and
\beq
\sum^{6}_{k=0}a_{k}=\mu b\label{eq4.30}
\eeq
In fact, for $z\rightarrow 0$,
\bea
<B(\infty)\tilde{\psi}^{+}(1)\psi(z)B(0)>\simeq<B(\infty)\tilde{\psi}^{+}(1)\frac{\gamma}{(z)^{4}}\psi(0)>\nn\\
=\frac{\gamma}{(z)^{4}}<B(\infty)\tilde{\psi}^{+}(1)\psi(0)>\simeq\frac{\gamma}{(z)^{4}}\cdot\mu\label{eq4.31}
\eea
For $z\rightarrow 1$,
\bea
<B(\infty) \tilde{\psi}^{+}(1)\psi(z)B(0)>
\simeq<B(\infty)\frac{\mu}{(1-z)^{\tilde{\Delta}+\Delta-4}} B(1)\cdot B(0)>\nn\\
=\frac{\mu}{(1-z)^{2\Delta-2}}\cdot<B(\infty)B(1)B(0)>=\frac{\mu}{(1-z)^{2\Delta-2}}\cdot b\label{eq4.32}
\eea
For $z\rightarrow\infty$,
\beq
<B(\infty)\psi(z)\tilde{\psi}^{+}(1)B(0)>=\frac{P_{6}(z)}{(z-1)^{2\Delta-2}(z)^{4}}\label{eq4.33}
\eeq
\bea
<B(\infty)\psi(z)\tilde{\psi}^{+}(1)B(0)>
\simeq<B(\infty)\psi(z)\cdot\mu\psi^{+}(0)>\nn\\
=\mu<B(\infty)\psi(z)\psi^{+}(0)>=\mu\cdot\frac{\gamma}{(z)^{2\Delta-4}}\label{eq4.34}
\eea
\beq
\frac{P_{6}(z)}{(z-1)^{2\Delta-2}z^{4}}\simeq\frac{a_{6}z^{6}}{(z)^{2\Delta+2}}=\frac{a_{6}}{(z)^{2\Delta-4}}\label{eq4.35}
\eeq
From (\ref{eq4.34}), (\ref{eq4.35}) the order 6 of the polynomial $P_{6}(z)$ gets confirmed and one obtains $a_{6}=\mu\gamma$, eq.(\ref{eq4.29}).

All these analysis eq.(\ref{eq4.27}) - (\ref{eq4.35}), based on the operator algebra (\ref{eq1.1}) - (\ref{eq1.9}), have to be confirmed by the function $<B(\infty)\tilde{\psi}^{+}(1)\psi(z)B(0)>$ :  by its expansion in the matrix elements of modes of the operators and  by the use of the commutation relations in (\ref{eq4.3}), (\ref{eq4.4}) to define the values 
of these matrix elements.

Expanding in modes of the operators, one gets the following expansion of the correlation function :
\bea
<B(\infty)\tilde{\psi}^{+}(1)\psi(z)B(0)>\nn\\
=\frac{1}{(z)^{4}}\times\{m_{B}(0)+z\cdot m_{B}(1)+z^{2}\cdot m_{B}(2)+z^{3}\cdot m_{B}(3)+...\}\label{eq4.35a}
\eea
Here
\beq
m_{B}(0)=<B(\infty)\tilde{\psi}^{+}_{-\frac{4}{3}}\psi_{\frac{4}{3}}B(0)>\label{eq4.36}
\eeq
\beq
m_{B}(1)=<B(\infty)\tilde{\psi}^{+}_{-\frac{1}{3}}\psi_{\frac{1}{3}}B(0)>\label{eq4.37}
\eeq
\beq
m_{B}(2)=<B(\infty)\tilde{\psi}^{+}_{\frac{2}{3}}\psi_{-\frac{2}{3}}B(0)>\label{eq4.38}
\eeq
\beq
m_{B}(3)=<B(\infty)\tilde{\psi}^{+}_{\frac{5}{3}}\psi_{-\frac{5}{3}}B(0)>\label{eq4.39}
\eeq
etc. .

$m_{B}(n)$, $n=0,1,2,3,...$ are the projections of the matrix elements in (\ref{eq4.17}) - (\ref{eq4.25}) onto $B(\infty)$. This means that $m_{B}(n)$ is obtained from $m(n)$ in (\ref{eq4.17}) - (\ref{eq4.25}) by keeping only the terms $\propto B(0)$, and further keeping just their coefficients. Like e.g. :
\beq
m_{B}(0)=(\tilde{\psi}^{+}_{-\frac{4}{3}}\psi_{\frac{4}{3}}B(0))_{B}=(\gamma\mu B(0))_{B}=\gamma \mu\ \label{eq4.40}
\eeq
\beq
mc_{B}(0)=(\psi_{-\frac{4}{3}}\tilde{\psi}^{+}_{\frac{4}{3}}B(0))_{B}=\mu(\beta^{(112)}_{\psi\psi^{+},I}\cdot\partial^{2}T(0)+\beta^{(22)}_{\psi\psi^{+},I}\cdot\wedge(0)+\gamma\cdot B(0))_{B}=\mu\gamma\label{eq4.41}
\eeq

We repeat it again that all the coefficients in (\ref{eq4.35a}) get defined :

1) $m_{B}(0)$, $m_{B}(1)$, $m_{B}(2)$, by the direct calculation, eq.(\ref{eq4.19}) -(\ref{eq4.21});

2) $m_{B}(n)$, $n=3.4,...,$ by the commutation relations, eq.(\ref{eq4.14}) -(\ref{eq4.16}).

No constraints on the constants of the operator algebra are being obtained, in the process of calculation of this correlation function. We observe it again that it is wrong to think that, calculated in this way, by the expansion in $z$, so to say by the $z\rightarrow 0$ limit expansion, the obtained expression for the correlation function will have to be confronted next with the limits $z\rightarrow 1$ and $z\rightarrow\infty$, which have to be consistent with the operator algebra, and that at this point the constraints on the values of the constants will be produced.

This is wrong. The constraints appear earlier, as consistencies of the commutation relations, for the lowest matrix elements, as it has been explained earlier. Once these consistencies are satisfied (or, as in the present case, are not being produced by the lowest commutation relations), the correlation function calculated further with the commutation relations, exclusively, will automatically satisfy all the consistencies with the other limits, the associativity relations so to say. This is because the commutation relations, by their definition, encodes, by themselves, the associativity relations between different limits.

In the present case, of the function $<B\tilde{\psi}^{+}\psi B>$, the above statement implies that the function, calculated in the $z\rightarrow 0$ limit expansion, should automatically satisfy all the consistencies with the other limits, \underline{for unconstrained values} of the operator algebra
constants. We are going to verify this conclusion.

Going back to the $z$ expansion in (\ref{eq4.35a}), of our correlation function, we assume that all the coefficients $m_{B}(n)$, $n=0,1,2,3,...$ had been defined. We then expand, in $z$, also the analytic form (\ref{eq4.27}) of this function, and we compare the expansion with that in (\ref{eq4.35a}), coefficient by coefficient. In this this way we get a set of the linear equations on the coefficients $\{a_{k},\,\,k=0,1,..., 6\}$ of the polynomial $P_{6}(z)$. Solving them, we get the coefficients and so we get the complete expression for the correlation function.

As with the previous calculations, in Section 2, we did an additional consistency verification by taking the polynomial with more terms than required by the operator algebra analysis. We would take $P_{8}(z)$, in place of $P_{6}(z)$, to check if, with our calculation of the coefficients, we get $a_{7}=a_{8}=0$.

Realizing these calculations with Mathematica, with the $\beta$ coefficients to be found in the Appendix A, we have obtained the following values for the coefficients $\{a_{k}\}$ :
\bea
a_{0}=\gamma\mu, \quad
a_{1}=-\frac{5}{2}\gamma\mu, \quad
a_{2}=\frac{[9(1376+45c)\gamma+4(784+57c)\tilde{\gamma}]\mu}{4(784+57c)}, \nn\\
a_{3}=\frac{[2b(784+57c)
-3(2560+21c)\gamma-4(784+57c)\tilde{\gamma}]\mu}{2(784+57c)}\label{eq4.45}
\eea
Next we have found that $a_{4}=a_{2}$, $a_{5}=a_{1}$, $a_{6}=a_{0}$, 
$a_{7}=0$, $a_{8}=0$.

We observe that $a_{6}=\gamma\mu$, as it should, eq.(\ref{eq4.29}). One could also check, with the expressions for $\{a_{k}\}$ above, that
\beq
\sum^{6}_{k=0}a_{k}=b\mu\label{eq4.51}
\eeq
as it should be the case, eq.(\ref{eq4.30}).

\subsection{Correlation function $<B(\infty)\tilde{\psi}^{+}(1)\psi(z)B(0)>$, calculated in the $z-1$ expansion.}

In this Section we shall do a similar test for the function $<B\tilde{\psi}^{+}\psi B>$, as we have done it in Sec.2.2 for the function $<\psi^{+}\psi\psi\psi^{+}>$ : we shall calculate the above function alternatively, in the $z\rightarrow 1$ limit expansion.

We shall put the operators of the function $<B(\infty)\tilde{\psi}^{+}(1)\psi(z)B(0)>$ into the configuration $<B(\infty)B(0)\psi(z)\tilde{\psi}^{+}(1)>$, which is more convenient for the $z\rightarrow 1$ limit calculations. We get the following analytic form for our function, in a new configuration:
\beq
<B(\infty)B(0)\psi(z)\tilde{\psi}^{+}(1)>=\frac{P_{6}(z)}{(z-1)^{2\Delta-2}\cdot(z)^{4}}\label{eq4.2.1}
\eeq
(\ref{eq4.2.1}) is obtained from (\ref{eq4.27}) by the analytic continuation of the fields, in the l.h.s., and of $0$ and $z$, in the r.h.s. , similarly as we have it done in Sec. 2.2 for the function $<\psi^{+}(\infty)\psi(1)\psi(z)\psi^{+}(0)>$.

Still, to make it more convenient for the $z\rightarrow 1$ limit expansion, we put (\ref{eq4.2.1}) into the form :
\beq
<B(\infty)B(0)\psi(z)\tilde{\psi}^{+}(1)>=\frac{Q_{6}(1-z)}{(z-1)^{2\Delta-2}\cdot(z)^{4}}\label{eq4.2.2}
\eeq
where
\beq
Q_{6}(1-z)=\sum^{6}_{k=0}b_{k}\cdot(1-z)^{k}\label{eq4.2.3}
\eeq
The coefficients $\{b_{k}\}$ are to be defined by the present calculation.

Evidently, with $\{b_{k}\}$ to be defined by the independent calculation, we have to verify finally that 
\beq
Q_{6}(1-z)=P_{6}(z)\label{4.2.4}
\eeq
The coefficients $\{a_{k}\}$ of $P_{6}(z)$ have been obtained in the previous Section.

By the operator algebra, expressed in modes of the operators, one obtains the following expansion of our correlation function, taken in the configuration of the l.h.s. of (\ref{eq4.2.2})) :
\bea
<B(\infty
)B(0)\psi(z)\tilde{\psi}^{+}(1)>\nn\\
=\frac{1}{(z-1)^{2\Delta-2}}\cdot\{<B(\infty)B_{0}\psi_{\frac{2}{3}}\tilde{\psi}(1)>\nn\\
-(z-1)<B(\infty)B_{1}\psi_{-\frac{1}{3}}\tilde{\psi}^{+}(1)>\nn\\
+(z-1)^{2}<B(\infty)B_{2}\psi_{-\frac{4}{3}}\tilde{\psi}^{+}(1)>\nn\\
-(z-1)^{3}<B(\infty)B_{3}\psi_{-\frac{7}{3}}\tilde{\psi}^{+}(1)>+...\}\label{eq4.2.5}
\eea
This expansion is obtained in a way similar to that for the function $<\psi^{+}\psi^{+}\psi\psi>$ in Sec.2.2.

We have to define the matrix elements :
\beq
B_{0}\psi_{\frac{2}{3}}\tilde{\psi}^{+}(1),\,\,\,B_{1}\psi_{-\frac{1}{3}}\tilde{\psi}^{+}(1),\,\,\,B_{2}\psi_{-\frac{4}{3}}\tilde{\psi}^{+}(1),\,\,\, \mbox{etc.}\label{eq4.2.6}
\eeq
and project them next onto $B(\infty)$. For this we need the commutation relation $\{B,\psi\}\tilde{\psi}^{+}(1)$. It is found to be of the following form :
\beq
(B_{n+1}\psi_{-\frac{1}{3}+m}-B_{n}\psi_{\frac{2}{3}+m}+\psi_{\frac{2}{3}+m}B_{n}-\psi_{-\frac{1}{3}+m}B_{n+1})\tilde{\psi}^{+}(1)=R(n,m)\label{eq4.2.7}
\eeq
\bea
R(n,m)=\{\frac{1}{2}(4+n-1)(4+n-2)\gamma\psi_{\frac{2}{3}+n+m}\nn\\
-(4+n-1)(\Delta-\frac{1}{3}+n+m+1)\gamma\beta^{(1)}_{B\psi,\psi}\cdot\psi_{\frac{2}{3}+n+m}\nn\\
+[(\Delta-\frac{1}{3}+n+m+2)(\Delta-\frac{1}{3}+n+m+1)\gamma\beta^{(11)}_{B\psi,\psi}\cdot\psi_{\frac{2}{3}+n+m}\nn\\
+\gamma \beta ^{(2)}_{B\psi,\psi}\cdot(L_{-2}\psi)_{\frac{2}{3}+n+m}+\mu\tilde{\psi}_{\frac{2}{3}+n+m}]\}\tilde{\psi}^{+}(1)\label{eq4.2.8}
\eea
Projecting this commutation relation onto level 4 operators, level of $B(1)$, requires $n+m=0$, $m=-n$. With this constraint on the values of $n,m$, we obtain, from (\ref{eq4.2.7}), (\ref{eq4.2.8}), the following (restricted) commutation relation :
\beq
B_{n+1}\psi_{-\frac{1}{3}-n}-B_{n}\psi_{\frac{2}{3}-n}+\psi_{\frac{2}{3}-n}B_{n}-\psi_{-\frac{1}{3}-n}B_{n+1})\tilde{\psi}^{+}(1)=R(n)\label{eq4.2.9}
\eeq
\bea
R(n)=R(n,-n)=\{\frac{1}{2}(4+n-1)(4+n-2)\gamma\psi_{\frac{2}{3}}\nn\\
-(4+n-1)(\Delta+\frac{2}{3})\gamma\beta^{(1)}_{B\psi,\psi}\cdot\psi_{\frac{2}{3}}\nn\\
+[(\Delta+\frac{5}{3})(\Delta+\frac{2}{3})\gamma\beta^{(11)}_{B\psi,\psi}\cdot\psi_{\frac{2}{3}}\nn\\
+\gamma \beta ^{(2)}_{B\psi,\psi}\cdot(L_{-2}\psi)_{\frac{2}{3}}+\mu\tilde{\psi}_{\frac{2}{3}}]\}\tilde{\psi}^{+}(1)\label{eq4.2.10}
\eea
We need to know the actions of the modes of operators onto $\tilde{\psi}^{+}(1)$ :
\beq
\psi_{\frac{2}{3}+n}\tilde{\psi}^{+}(1)=0,\quad n>0\label{eq4.2.11}
\eeq
\beq
\psi_{\frac{2}{3}}\tilde{\psi}^{+}(1)=\mu B(1)\label{eq4.2.12}
\eeq
\beq
\psi_{-\frac{1}{3}}\tilde{\psi}^{+}(1)=\mu\beta^{(1)}_{\psi\tilde{\psi}^{+},B}\cdot\partial B(1)\label{eq4.2.13}
\eeq
and
\beq
B_{2+n}\tilde{\psi}^{+}(1)=0,\quad n>0\label{eq4.2.14}
\eeq
\beq
B_{2}\tilde{\psi}^{+}(1)=\mu\psi^{+}(1)\label{eq4.2.15}
\eeq
\beq
B_{1}\tilde{\psi}^{+}(1)=\mu\beta^{(1)}_{B\tilde{\psi},\psi}\cdot\partial\psi^{+}(1)\label{eq4.2.15a}
\eeq
\beq
B_{0}\tilde{\psi}^{+}(1)=\mu\beta^{(11)}_{B\tilde{\psi},\psi}\cdot\partial^{2}\psi^{+}(1)+\mu\beta^{(2)}_{B\tilde{\psi},\psi}\cdot L_{-2}\psi^{+}(1)+\tilde{\gamma}\tilde{\psi}^{+}(1)\label{eq4.2.16}
\eeq

From (\ref{eq4.2.9}) we obtain the lowest commutation relation, for 
\underline{$n=0$} :
\beq
(B_{1}\psi_{-\frac{1}{3}}-B_{0}\psi_{\frac{2}{3}}+\psi_{\frac{2}{3}}B_{0}-\psi_{-\frac{1}{3}}B_{1})\tilde{\psi}^{+}(1)=R(0)\label{eq4.2.17}
\eeq
With the expressions for the explicit matrix elements given below, one could check that (\ref{eq4.2.17}) is satisfied in a trivial way, one gets $0=0$, for unconstraint values of the constants.

The next commutation relation, which, in fact, is the lowest one,
is obtained for \underline{$n=1$}:
\beq
(B_{2}\psi_{-\frac{4}{3}}-B_{1}\psi_{-\frac{1}{3}}+\psi_{-\frac{1}{3}}B_{1}-\psi_{-\frac{4}{3}}B_{2})\tilde{\psi}^{+}(1)=R(1)\label{eq4.2.18}
\eeq
For \underline{$n=2$} :
\beq
(B_{3}\psi_{-\frac{7}{3}}-B_{2}\psi_{-\frac{4}{3}}+\psi_{-\frac{4}{3}}B_{2})\tilde{\psi}^{+}(1)=R(2)\label{eq4.2.19}
\eeq
For \underline{$n=3$} :
\beq
(B_{4}\psi_{-\frac{10}{3}}-B_{3}\psi_{-\frac{7}{3}})\tilde{\psi}^{+}(1)=R(3)\label{eq4.2.20}
\eeq
and so on.

We observe that no equations (constraints) on the operator algebra constants are being produced, by the lowest commutation relations :
(\ref{eq4.2.17}), is trivial, and the first relevant one, eq.(\ref{eq4.2.18}), already defines an unknown matrix element (which could not be calculated directly)
$B_{2}\psi_{-\frac{4}{3}}\tilde{\psi}^{+}(1)$. Next, (\ref{eq4.2.19}) defines $B_{3}\psi_{-\frac{7}{3}}\tilde{\psi}^{+}(1)$, and so an.

Similarly to the previous section, no constraints are being produced, but all the matrix elements get defined.

We have to complete the relations (\ref{eq4.2.17}), (\ref{eq4.2.18}), (\ref{eq4.2.19}) by the expressions for the lowest matrix elements which can be calculated directly : $B_{0}\psi_{\frac{2}{3}}\tilde{\psi}^{+}(1)$, $B_{1}\psi_{-\frac{1}{3}}\tilde{\psi}^{+}(1)$,  
and $\psi_{-\frac{4}{3}}B_{2}\tilde{\psi}^{+}(1)$, $\psi_{-\frac{1}{3}}B_{1}\tilde{\psi}^{+}(1)$, $\psi_{\frac{2}{3}}B_{0}\tilde{\psi}^{+}(1)$. 

We introduce the notations $m(n)$, $mc(n)$, $n=0,1,2,...$, 
for the matrix elements, which are explicit in the following. 
With some calculations one obtains :
\beq
m(0)=B_{0}\psi_{\frac{2}{3}}\tilde{\psi}^{+}(1)=\mu B_{0}B(1)\label{eq4.2.21}
\eeq
where
\beq
B_{0}B(1)=\beta^{(112)}_{BB,I}\cdot\partial^{2}T(1)+\beta^{(22)}_{BB,I}\cdot\wedge(1)+b\cdot B(1)\label{eq4.2.22}
\eeq
Next
\beq
m(1)=B_{1}\psi_{-\frac{1}{3}}\tilde{\psi}^{+}(1)=\mu\beta^{(1)}_{\psi\tilde{\psi}^{+},B}\cdot\{4\cdot B_{0}B(1)+\frac{4}{c}\partial^{2}T(1)\}\label{eq4.2.23}
\eeq
And the "conjugate" matrix elements :
\beq
mc(0)=\psi_{-\frac{4}{3}}B_{2}\tilde{\psi}^{+}(1)=\mu\psi_{-\frac{4}{3}}\psi^{+}(1)\label{eq4.2.24}
\eeq
where
\beq
\psi_{-\frac{4}{3}}\psi^{+}(1)=\beta^{(112)}_{\psi\psi^{+},I}\cdot\partial^{2}T(1)+\beta^{(22)}_{\psi\psi^{+},I}\cdot\wedge(1)+\gamma\cdot B(1)\label{eq4.2.25}
\eeq
Next
\beq
mc(1)=\psi_{-\frac{1}{3}}B_{1}\tilde{\psi}^{+}(1)=\mu\beta^{(1)}_{B\tilde{\psi},\psi}\cdot\{(\Delta-\frac{4}{3})\psi_{-\frac{4}{3}}\psi^{+}(1)+\frac{\Delta}{c}\partial^{2}T(1)\} \label{eq4.2.26}
\eeq
\bea
mc(2)=\psi_{\frac{2}{3}}B_{0}\tilde{\psi}^{+}(1)=\mu\beta^{(11)}_{B\tilde{\psi},\psi}\cdot\{(\Delta-\frac{1}{3})(\Delta-\frac{4}{3})\psi_{-\frac{4}{3}}\psi^{+}(1)\nn\\
+2(\Delta-\frac{1}{3})\cdot\frac{\Delta}{c}\partial^{2}T(1)+\frac{2\Delta}{c}\partial^{2}T(1)\}\nn\\
+\mu\beta^{(2)}_{B\tilde{\psi},\psi}\cdot\{(2\Delta-\frac{4}{3})\psi_{-\frac{4}{3}}\psi^{+}(1)+\frac{2\Delta}{c}\wedge(1)\}+\tilde{\gamma}\cdot\mu\cdot B(1)\label{eq4.2.27}
\eea
We also need to know the actions of $\psi_{\frac{2}{3}}$, $(L_{-2}\psi)_{\frac{2}{3}}$ and $\tilde{\psi}_{\frac{2}{3}}$ on $\tilde{\psi}^{+}(1)$, which appear in $R(n)$, eq.(\ref{eq4.2.10}). One finds :
\beq
\psi_{\frac{2}{3}}\tilde{\psi}^{+}(1)=\mu B(1)\label{eq4.2.28}
\eeq
\beq
(L_{-2}\psi)_{\frac{2}{3}}\tilde{\psi}^{+}(1)=(2\Delta+\frac{8}{3})\mu B(1)\label{eq4.2.29}
\eeq
\beq
\tilde{\psi}_{\frac{2}{3}}\tilde{\psi}^{+}(1)=\beta^{(112)}_{\tilde{\psi}\tilde{\psi}^{+},I}\cdot\partial^{2}T(0)+\beta^{(22)}_{\tilde{\psi}\tilde{\psi}^{+},I}\cdot\wedge(1)+\tilde{\gamma}\cdot B(1)\label{eq4.2.30}
\eeq

With the notations for the matrix elements introduced above, the expansion in eq.(\ref{eq4.2.5}) takes the form :
\bea
<B(\infty)B(0)\psi(z)\tilde{\psi}^{+}(1)>\nn\\
=\frac{1}{(z-1)^{2\Delta-2}}\cdot\{m_{B}(0)-(z-1)m_{B}(1)\nn\\
+(z-1)^{2}\cdot m_{B}(2)-(z-1)^{3}\cdot m_{B}(3)+...\}\label{eq4.2.31}
\eea
Here $m_{B}(n)$, $n=0,1,2,...$, are the projections of the matrix elements in (\ref{eq4.2.21}), (\ref{eq4.2.23}) and the following ones which will be calculated by the commutation relations in (\ref{eq4.2.18}), (\ref{eq4.2.19}), (\ref{eq4.2.20}) and so on. These projections (onto $B(\infty)$) 
are obtained in the way explained in the previous section.

We proceed in the standard way, by now. We assume that we have already calculated the coefficients $m_{B}(n)$, $n=1,2,...,$ in (\ref{eq4.2.31}). 
Next we expand the analytic form of our function in the r.h.s. of (\ref{eq4.2.2}). We expand it in $(z-1)$, and we compare the coefficients of the expansion with those in (\ref{eq4.2.31}). Solving the resulting linear equations 
we get the following values the coefficients $\{b_{n}\}$ :
\beq
b_{0}=b\mu\label{eq4.2.32}
\eeq
\beq
b_{1}=-3b\mu\label{eq4.2.33}
\eeq
\beq
b_{2}=\frac{\mu}{4}[12b+\frac{(9248+177c)\gamma}{784+57c}+4\tilde{\gamma}]\label{eq4.2.34}
\eeq
\beq
b_{3}=-\frac{\mu}{2(784+57c)}\cdot[2b(784+57c)+(9248+177c)\gamma+4(784+57c)\tilde{\gamma}]\label{eq4.2.35}
\eeq
\beq
b_{4}=\frac{\mu[(20224+975c)\gamma+4(784+57c)\tilde{\gamma}]}{4(784+57c)}\label{eq4.2.36}
\eeq
\beq
b_{5}=-\frac{7}{2}\gamma\mu\label{eq4.2.37}
\eeq
\beq
b_{6}=\gamma\mu\label{eq4.2.38}
\eeq
Going back to the initial configuration (positioning) of the operators, we 
could state that we have found two expressions for the correlation function $<B(\infty)\tilde{\psi}^{+}(1)\psi(z)B(0)>$ :
\beq
\frac{P_{6}(z)}{(1-z)^{2\Delta-2}(z)^{4}},\quad\mbox{and}\quad\frac{Q_{6}(1-z)}{(1-z)^{2\Delta-2}(z)^{4}}\label{eq4.2.39}
\eeq
The coefficients of the polynomial $P_{6}(z)$, $\{a_{n}\}$, have been defined in the previous section, eq.(\ref{eq4.45}), and the coefficients of $Q_{6}(1-z)$, $\{b_{n}\}$, have been just given above. Now it is simple to verify, with the values 
of the polynomial coefficients found, that
\beq
P_{6}(z)=Q_{6}(1-z)\label{eq4.2.40}
\eeq
$P_{6}(z)$ is defined in (\ref{eq4.28}) and $Q_{6}(1-z)$ is given by (\ref{eq4.2.3}).

The equality of the two polynomials signifies consistency of the  two calculations, obtained for the \underline{unconstrained values} of the operator algebra constants.

\subsection{Correlation function $<B(\infty)\psi(1)\tilde{\psi}(z)\tilde{\psi}(0)>$.}

The analytic form of this function is the following :
\beq
<B(\infty)\psi(1)\tilde{\psi}(z)\tilde{\psi}(0)>=\frac{P_{n}(z)}{(z)^{\Delta+4}\cdot(1-z)^{\Delta+2}}\label{eq4.3.1}
\eeq
In fact, for $z\rightarrow 0$,
\bea
<B(\infty)\psi(1)\tilde{\psi}(z)\tilde{\psi}(0)>\simeq<B(\infty)\psi(1)\frac{\eta}{(z)^{2\tilde{\Delta}-\Delta}}\psi^{+}(0)>\nn\\
=\frac{\eta}{(z)^{\Delta+4}}\cdot<B(\infty)\psi(1)\psi^{+}(0)>=\frac{\eta}{(z)^{\Delta+4}}\cdot\gamma\label{eq4.3.2}
\eea
In particular, it is required that : 
\beq
a_{0}=\eta\cdot\gamma\label{eq4.3.3}
\eeq

$z\rightarrow 1$,
\bea
<B(\infty)\psi(1)\tilde{\psi}(z)\tilde{\psi}(0)>\simeq<B(\infty)\frac{\zeta}{(1-z)^{\tilde{\Delta}}}\psi^{+}(1)\tilde{\psi}(0)\nn\\
=\frac{\zeta}{(1-z)^{\Delta+2}}\cdot<B(\infty)\psi^{+}(1)\tilde{\psi}(0)>=\frac{\zeta}{(1-z)^{\Delta+2}}\cdot\mu\label{eq4.3.4}
\eea
In particular, it should be that :
\beq
\sum^{n}_{k=0}a_{k}=\zeta\mu\label{eq4.3.5}
\eeq

$z\rightarrow\infty$. 

By the analytic continuation of $\tilde{\psi}(z)$ around $\psi(1)$ in the l.h.s. of (\ref{eq4.3.1}), and $z$ around $1$ in the r.h.s., one gets the positioning which is more convenient for analyzing the $z\rightarrow\infty$ limit :
\beq
<B(\infty)\tilde{\psi}(z)\psi(1)\tilde{\psi}(0)> = \frac{P_{n}(z)}{(z)^{\Delta+4}\cdot(z-1)^{\Delta+2}}\label{eq4.3.6}
\eeq
Next, by the operator algebra :
\bea
<B(\infty)\tilde{\psi}(z)\psi(1)\tilde{\psi}(0)>
\simeq<B(\infty)\tilde{\psi}(z)\zeta\cdot\psi^{+}(0)>\nn\\
=\zeta\cdot<B(\infty)\tilde{\psi}(z)\psi^{+}(0)>
=\zeta\cdot\frac{\mu}{(z)^{\tilde{\Delta}+\Delta-4}}=\frac{\zeta\cdot\mu}{(z)^{2\Delta-2}}\label{eq4.3.7}
\eea
By the analytic form, the r.h.s. of eq.(\ref{eq4.3.6}) :
\beq
\frac{P_{n}(z)}{(z)^{\Delta+4}\cdot(z-1)^{\Delta+2}}\simeq\frac{a_{n}z^{n}}{(z)^{2\Delta+6}}=\frac{a_{n}}{(z)^{2\Delta+6-n}}\label{eq4.3.8}
\eeq
Comparing (\ref{eq4.3.7}) and (\ref{eq4.3.8}) one obtains :
\beq
2\Delta-2=2\Delta+6-n,\quad n=8\label{eq4.3.9}
\eeq
\beq
a_{8}=\zeta\cdot\mu\label{eq4.3.10}
\eeq
so that (\ref{eq4.3.1}) takes the form :
\beq
<B(\infty)\psi(1)\tilde{\psi}(z)\tilde{\psi}(0)>\frac{P_{8}(z)}{(z)^{\Delta+4}\cdot(1-z)^{\Delta+2}}\label{eq4.3.11}
\eeq

The coefficients $\{a_{k}\}$ of the polynomial, which will be calculated in the $z\rightarrow 0$ limit expansion, will have to verify, in particular, 
the conditions (\ref{eq4.3.5}) (with $n=8$) and (\ref{eq4.3.10}), due to the limits $z\rightarrow 1$ and $z\rightarrow\infty$.
\vskip0.5cm
We turn now to the operator algebra calculation of the function $<B(\infty)\psi(1)\tilde{\psi}(z)\tilde{\psi}(0)>$, in its $z\rightarrow 0$ limit expansion.

By the operator algebra, expressed in the modes of the operators, the expansion of our function has the following formal form :
\bea
<B(\infty)\psi(1)\tilde{\psi}(z)\tilde{\psi}(0)>\nn\\
=\frac{1}{(z)^{\Delta+4}}\{<B(\infty)\psi_{-\frac{4}{3}}\tilde{\psi}_{2}\tilde{\psi}(0)>+z<B(\infty)\psi_{-\frac{1}{3}}\tilde{\psi}_{1}\tilde{\psi}(0)>\nn\\
+z^{2}<B(\infty)\psi_{\frac{2}{3}}\tilde{\psi}_{0}\tilde{\psi}(0)>+z^{3}<B(\infty)\psi_{\frac{5}{3}}\tilde{\psi}_{-1}\tilde{\psi}(0)>+...\}\label{eq4.3.12}
\eea
In writing (\ref{eq4.3.12}) we have used the knowledge of the action of modes of $\tilde{\psi}(z)$ on $\tilde{\psi}(0)$ which is as follows :
\beq
\tilde{\psi}_{2+n}\tilde{\psi}(0)=0,\quad n>0\label{eq4.3.13}
\eeq
\beq
\tilde{\psi}_{2}\tilde{\psi}(0)=\eta\psi^{+}(0)\label{eq4.3.14}
\eeq
\beq
\tilde{\psi}_{1}\tilde{\psi}(0)=\eta\cdot\beta^{(1)}_{\tilde{\psi}\tilde{\psi},\psi^{+}}\cdot\partial\psi^{+}(0)\label{eq4.3.15}
\eeq
\beq
\tilde{\psi}_{0}\tilde{\psi}(0)=\eta\beta^{(11)}_{\tilde{\psi}\tilde{\psi},\psi^{+}}\cdot\partial^{2}\psi^{+}(0)+\eta\beta^{(2)}_{\tilde{\psi}\tilde{\psi},\psi^{+}}\cdot L_{-2}\psi^{+}(0)+\tilde{\lambda}\tilde{\psi}^{+}(0)\label{eq4.3.16}
\eeq
In writing (\ref{eq4.3.12}) we had to know that the leading action 
of $\tilde{\psi}(z)$ en $\tilde{\psi}(0)$, in modes, 
is that of $\tilde{\psi}_{2}\psi(0)$, and that the total (global) dimension of the operator $\psi_{-\frac{4}{3}}\tilde{\psi}_{2}\psi(0)$ had to be  equal to the dimension of $B$ :
\beq
L_{0}(\psi_{-\frac{4}{3}}\tilde{\psi}_{2}\tilde{\psi}(0)=\frac{4}{3}-2+\tilde{\Delta}=-\frac{2}{3}+\tilde{\Delta}=4\label{eq4.3.17}
\eeq

To calculate the coefficients of the expansion in (\ref{eq4.3.12}) we need the commutation relation $\{\psi,\tilde{\psi}\}\tilde{\psi}(0)$. Obtained in the standard way (as it has been done, e.g., in the Sections 2.1, 2.2) the commutation relation is found to be of the form :
\beq
\sum^{\infty}_{l=0}D^{l}_{\frac{5}{3}}(\psi_{\frac{5}{3}+n-l}\tilde{\psi}_{m+l}+\tilde{\psi}_{\frac{5}{3}+m-l}\psi_{n+l})\tilde{\psi}(0)=R(n,m)\label{eq4.3.17a}
\eeq
\bea
R(n,m)=\{\frac{1}{2}(\Delta+n-1)(\Delta+n-2)\zeta\cdot\psi^{+}_{\frac{5}{3}+n+m}\nn\\
-(\Delta+n-1)(\Delta+n+m+\frac{5}{3})\zeta\cdot\beta^{(1)}_{\psi\tilde{\psi},\psi^{+}}\cdot\psi^{+}_{\frac{5}{3}+n+m}\nn\\
+(\Delta+n+m+\frac{8}{3})(\Delta+n+m+\frac{5}{3})\cdot\zeta\cdot\beta^{(11)}_{\psi\tilde{\psi},\psi^{+}}\cdot\psi^{+}_{\frac{5}{3}+n+m}\nn\\
+\zeta\beta^{(2)}_{\psi\tilde{\psi},\psi^{+}}\cdot(L_{-2}\psi^{+})_{\frac{5}{3}+n+m}+\eta\tilde{\psi}^{+}_{\frac{5}{3}+n+m}\}\tilde{\psi}(0)\label{eq4.3.18}
\eea
Projecting this commutation relation on the level 4 (level of B), we have 
to require that
\bea
-\frac{5}{3}-n-m+\tilde{\Delta}=4,\nn\\
-n-m+3=4,\nn\\
n+m=-1,\quad m=-1-n\label{eq4.3.19}
\eea
With this restriction one gets :
\beq
\sum^{\infty}_{l=0}D^{l}_{\frac{5}{3}}(\psi_{\frac{5}{3}+n-l}\tilde{\psi}_{-1-n+l}+\tilde{\psi}_{\frac{2}{3}-n-l}\psi_{n+l})\tilde{\psi}(0)=R(n)\label{eq4.3.20}
\eeq
\bea
R(n)=R(n,-1-n)=\{\frac{1}{2}(\Delta+n-1)(\Delta+n-2)\zeta\cdot\psi^{+}_{\frac{2}{3}}\nn\\
-(\Delta+n-1)(\Delta+\frac{2}{3})\zeta\cdot\beta^{(1)}_{\psi\tilde{\psi},\psi^{+}}\cdot\psi^{+}_{\frac{2}{3}}\nn\\
+(\Delta+\frac{5}{3})(\Delta+\frac{2}{3})\zeta\cdot\beta^{(11)}_{\psi\tilde{\psi},\psi^{+}}\cdot\psi^{+}_{\frac{2}{3}}\nn\\
+\zeta\beta^{(2)}_{\psi\tilde{\psi},\psi^{+}}\cdot(L_{-2}\psi^{+})_{\frac{2}{3}}+\eta\tilde{\psi}^{+}_{\frac{2}{3}}\}\tilde{\psi}(0)\label{eq4.3.21}
\eea

To use these commutation relations, we need to know the actions of modes of $\psi$ and $\tilde{\psi}$ on $\tilde{\psi}(0)$. For $\tilde{\psi}$ on $\tilde{\psi}(0)$, they are given in (\ref{eq4.3.13}) - (\ref{eq4.3.16}). For $\psi$ on $\tilde{\psi}(0)$ they are as follows :
\beq
\psi_{2+n}\tilde{\psi}(0)=0,\quad n>0\label{eq4.3.22}
\eeq
\beq
\psi_{2}\tilde{\psi}(0)=\zeta\psi^{+}(0)\label{eq4.3.23}
\eeq
\beq
\psi_{1}\tilde{\psi}(0)=\zeta\beta^{(1)}_{\psi\tilde{\psi},\psi^{+}}\cdot\partial\psi^{+}(0)\label{eq4.3.24}
\eeq
\bea
\psi_{0}\tilde{\psi}(0)=\zeta\beta^{(11)}_{\psi\tilde{\psi},\psi^{+}}\cdot\partial^{2}\psi^{+}(0)
+\zeta\beta^{(2)}_{\psi\tilde{\psi},\psi^{+}}\cdot L_{-2}\psi^{+}(0) + 
\eta\tilde{\psi}^{+}(0)\label{eq4.3.25}
\eea

We can now write down the commutation relations, the particular ones, starting from the lowest and going upwards.

The lowest one is obtained from (\ref{eq4.3.20}) for \underline{$n=0$} :
\bea
\psi_{\frac{5}{3}}\tilde{\psi}_{-1}+D^{1}_{\frac{5}{3}}\psi_{\frac{2}{3}}\tilde{\psi}_{0}+D^{2}_{\frac{5}{3}}\psi_{-\frac{1}{3}}\tilde{\psi}_{1}+D^{3}_{\frac{5}{3}}\psi_{-\frac{4}{3}}\tilde{\psi}_{2}\nn\\
+\tilde{\psi}_{\frac{2}{3}}\psi_{0}+D^{1}_{\frac{5}{3}}\tilde{\psi}_{-\frac{1}{3}}\psi_{1}+D^{2}_{\frac{5}{3}}\tilde{\psi}_{-\frac{4}{3}}\psi_{2}\tilde{\psi}(0)=R(0)\label{eq4.3.26}
\eea

\underline{$n=1$} :

\bea
\psi_{\frac{8}{3}}\tilde{\psi}_{-2}+D^{1}_{\frac{5}{3}}\psi_{\frac{5}{3}}\tilde{\psi}_{-1}+D^{2}_{\frac{5}{3}}\psi_{\frac{2}{3}}\tilde{\psi}_{0}+D^{3}_{\frac{5}{3}}\psi_{-\frac{1}{3}}\tilde{\psi}_{1}\nn\\
+D^{4}_{\frac{5}{3}}\psi_{-\frac{4}{3}}\tilde{\psi}_{2}+\tilde{\psi}_{-\frac{1}{3}}\psi_{1}+D^{1}_{\frac{5}{3}}\tilde{\psi}_{-\frac{4}{3}}\psi_{2})\tilde{\psi}(0)=R(1)\label{eq4.3.27}
\eea

\underline{$n=2$} :

\bea
\psi_{\frac{11}{3}}\tilde{\psi}_{-3}+D^{1}_{\frac{5}{3}}\psi_{\frac{8}{3}}\tilde{\psi}_{-2}+D^{2}_{\frac{5}{3}}\psi_{\frac{5}{3}}\tilde{\psi}_{-1}+D^{3}_{\frac{5}{3}}\psi_{\frac{2}{3}}\tilde{\psi}_{0}\nn\\
+D^{4}_{\frac{5}{3}}\psi_{-\frac{1}{3}}\tilde{\psi}_{1}+D^{5}_{\frac{5}{3}}\psi_{-\frac{4}{3}}\tilde{\psi}_{2}+\tilde{\psi}_{-\frac{4}{3}}\psi_{2})\tilde{\psi}(0)=R(2)\label{eq4.3.28}
\eea

We introduce the notations for the matrix elements :
\bea
\psi_{-\frac{4}{3}}\tilde{\psi}_{2}\tilde{\psi}(0)=m(0),\quad \psi_{-\frac{1}{3}}\tilde{\psi}_{1}\tilde{\psi}(0)=m(1)\nn\\
\psi_{\frac{2}{3}}\tilde{\psi}_{0}\tilde{\psi}(0)=m(2),\quad \psi_{\frac{5}{3}}\tilde{\psi}_{-1}\tilde{\psi}(0)=m(3)\nn\\
\psi_{\frac{8}{3}}\tilde{\psi}_{-2}\tilde{\psi}(0)=m(4),\quad \psi_{\frac{11}{3}}\tilde{\psi}_{-3}\tilde{\psi}(0)=m(5)\label{eq4.3.29}
\eea
and so on. Similarly, for the opposite ("conjugate") matrix elements :
\bea
\tilde{\psi}_{-\frac{4}{3}}\psi_{2}\tilde{\psi}(0)=mc(0),\quad \tilde{\psi}_{-\frac{1}{3}}\psi_{1}\tilde{\psi}(0)=mc(1), \quad
\tilde{\psi}_{\frac{2}{3}}\psi_{0}\tilde{\psi}(0)=mc(2)\label{eq4.3.30}
\eea
and so on.

With these notations, the matrix elements $m(n)$, starting with $n=6$, could be defined by the recurrence relation :
\beq
m(n)=R(n-3)-\sum^{n}_{k=1}D^{k}_{\frac{5}{3}}m(n-k)\label{eq4.3.31}
\eeq
We observe that no equations, on the operator algebra constants, are being produced, by the lowest commutation relations. In fact, the very first one, for $n=0$, eq.(\ref{eq4.3.26}), already contains the matrix element $m(3)=\psi_{\frac{5}{3}}\tilde{\psi}_{-1}\tilde{\psi}(0)$ which could not be defined directly, because $\tilde{\psi}_{-1}\tilde{\psi}(0)$ is not defined directly. The list of actions of the modes of $\tilde{\psi}$ on $\tilde{\psi}(0)$ which are defined directly (by the operator algebra expansion in (\ref{eq1.6})) is given in (\ref{eq4.3.13}) - (\ref{eq4.3.16}). We remind that in the $Z_{3}$ charged sector the expansion is explicit only up to the second level, the way our operator algebra is defined, Section 1.

This means that the commutation relation (\ref{eq4.3.26}) serves to define the matrix element $m(3)$, instead of providing an equation on constants.

We observe also that the commutation relations (\ref{eq4.3.20}) for the negative values of $n$, $n=-1,-2,-3,...,$ they define the matrix elements $mc(n)$, which are not needed for the calculation of the correlation function, eq.(\ref{eq4.3.12}). In particular, the $n=-1$ commutation relation will define the matrix element $mc(3)=\tilde{\psi}_{\frac{5}{3}}\psi_{-1}\tilde{\psi}(0)$, which is not explicit neither, compare the actions of the modes of $\psi$ on $\tilde{\psi}(0)$ in the list (\ref{eq4.3.22}) - (\ref{eq4.3.25}).

On the other hand, all the matrix elements get defined. In fact, $m(0)$, $m(1)$, $m(2)$ and also $mc(0)$, $mc(1)$, $mc(2)$ can all be calculated directly, while $m(3)$, $m(4)$, etc. all get defined by the commutation relations,  eq.(\ref{eq4.3.26}) - (\ref{eq4.3.28}) plus the recursion relation (\ref{eq4.3.31}).

All this is perfectly analogous to the way the calculations has been done 
for the correlation function $<B(\infty)\tilde{\psi}^{+}(1)\psi(z)B(0)>$ in 
the preceding Sections, 4.1, 4.2 .

Next we have to define the explicit matrix elements. With some calculations one finds the following expressions :
\beq
m(0)=\psi_{-\frac{4}{3}}\tilde{\psi}_{2}\tilde{\psi}(0)=\eta\cdot\psi_{-\frac{4}{3}}\psi^{+}(0)\label{eq4.3.32}
\eeq
We remind that 
\beq
\psi_{-\frac{4}{3}}\psi^{+}(0)=\beta^{(112)}_{\psi\psi^{+},I}\cdot\partial^{2}T(0)+\beta^{(22)}_{\psi\psi^{+},I}\cdot\wedge(0)+\gamma B(0)\label{eq4.3.33}
\eeq
Next,
\beq
m(1)=\psi_{-\frac{1}{3}}\tilde{\psi}_{1}\tilde{\psi}(0)=\eta\beta^{(1)}_{\tilde{\psi}\tilde{\psi},\psi^{+}}\cdot\{(\Delta-\frac{4}{3})\psi_{-\frac{4}{3}}\psi^{+}(0)+\frac{\Delta}{c}\partial^{2}T(0)\}\label{eq4.3.34}
\eeq
\bea
m(2)=\psi_{\frac{2}{3}}\tilde{\psi}_{0}\tilde{\psi}(0)=\eta\cdot\beta^{(11)}_{\tilde{\psi}\tilde{\psi},\psi^{+}}\cdot\{(\Delta-\frac{1}{3})(\Delta-\frac{4}{3})\psi_{-\frac{4}{3}}\psi^{+}(0)\nn\\
+2(\Delta-\frac{1}{3})\frac{\Delta}{c}\cdot\partial^{2}T(0)+\frac{2\Delta}{c}\partial^{2}T(0)\}\nn\\
+\eta\beta^{(2)}_{\tilde{\psi}\tilde{\psi},\psi^{+}}\cdot\{(2\Delta-\frac{4}{3})\psi_{-\frac{4}{3}}\psi^{+}(0)+\frac{2\Delta}{c}\cdot\wedge(0)\}+\tilde{\lambda}\cdot\mu\cdot B(0)\label{eq4.3.35}
\eea
and
\beq
mc(0)=\tilde{\psi}_{-\frac{4}{3}}\psi_{2}\tilde{\psi}(0)=\zeta\mu\cdot B(0)\label{eq4.3.36}
\eeq
\beq
mc(1)=\tilde{\psi}_{-\frac{1}{3}}\psi_{1}\tilde{\psi}(0)=\zeta\cdot\beta^{(1)}_{\psi\tilde{\psi},\psi^{+}}\cdot(\tilde{\Delta}-\frac{4}{3})\mu B(0)\label{eq4.3.37}
\eeq
\bea
mc(2)=\tilde{\psi}_{\frac{2}{3}}\psi_{0}\tilde{\psi}(0)=\zeta\cdot\beta^{(11)}_{\psi\tilde{\psi},\psi^{+}}\cdot(\tilde{\Delta}-\frac{1}{3})(\tilde{\Delta}-\frac{4}{3})\cdot\mu\cdot B(0)\nn\\
+\zeta\beta^{(2)}_{\psi\tilde{\psi},\psi^{+}}\cdot(2\tilde{\Delta}-\frac{4}{3})\cdot\mu\cdot B(0)\nn\\
+\eta\cdot\{\beta^{(112)}_{\tilde{\psi}\tilde{\psi}^{+},I}\cdot\partial^{2}T(0)+\beta^{(22)}_{\tilde{\psi}\tilde{\psi}^{+},I}\cdot\wedge(0)+\tilde{\gamma}\cdot B(0)\}\label{eq4.3.38}
\eea

We also need the actions of $\psi^{+}_{\frac{2}{3}}$, $(L_{-2}\psi^{+})_{\frac{2}{3}}$ and $\tilde{\psi}^{+}_{\frac{2}{3}}$ onto $\tilde{\psi}(0)$, which appear in $R(n)$, eq.(\ref{eq4.3.21}).
One finds :
\beq
\psi^{+}_{\frac{2}{3}}\tilde{\psi}(0)=\mu B(0)\label{eq4.3.39}
\eeq
\beq
(L_{-2}\psi^{+})_{\frac{2}{3}}\tilde{\psi}(0)=(\Delta+\frac{2}{3}+\tilde{\Delta})\mu B(0)\label{eq4.3.40}
\eeq
\beq
\tilde{\psi}^{+}_{\frac{2}{3}}\tilde{\psi}(0)=\beta^{(112)}_{\tilde{\psi}\tilde{\psi}^{+},I}\cdot\partial^{2}T(0)+\beta^{(22)}_{\tilde{\psi}\tilde{\psi}^{+},I}\cdot\wedge(0)+\tilde{\gamma}\cdot B(0)\label{eq4.3.41}
\eeq

Now we can calculate all the matrix elements. The explicit ones are given in (\ref{eq4.3.32}) - (\ref{eq4.3.38}). $m(3)$, $m(4)$, $m(5)$ are then defined by the commutation relations in (\ref{eq4.3.26}) - (\ref{eq4.3.28}). $m(n)$, $n=6,7,...$ are obtained by the commutation relations which follow, which take the form of the recurrence relation in (\ref{eq4.3.31}).

The matrix elements obtained in this way have still to be projected onto $B(\infty)$ : $m(n)\rightarrow m_{B}(n)$, where, in $m_{B}(n)$, we keep only the $B(0)$ terms, of $m(n)$, with, finally, $B(0)$ suppressed, leaving just the coefficients of $B(0)$.

In fact, it is simpler to first project the explicit matrix elements, 
and also $R(n)$, and then perform the rest of the calculations for the projected matrix elements only.

With our present notations, the $z$ series expansion in (\ref{eq4.3.12}) takes the form :
\bea
<B(\infty)\psi(1)\tilde{\psi}(z)\tilde{\psi}(0)>\nn
=\frac{1}{(z)^{\Delta+4}}\{m_{B}(0)+z\cdot m_{B}(1)+z^{2}\cdot m_{B}(2)\nn\\
+z^{3}\cdot m_{B}(3)+...\}\label{eq4.3.42}
\eea
We can assume now that all the coefficients of this expansion are known,
have been calculated according to the guidelines given above.

Next we expand, in $z$, the analytic form in (\ref{eq4.3.11}) 
of our correlation function. Then we compare the two expansions and determine the coefficients $\{a_{k}\}$ of the polynomial $P_{8}(z)$. In this way we have found the following values of the coefficients :
\beq
a_{0}=\eta\gamma\label{eq4.3.43}
\eeq
\beq
a_{1}=-4\eta\gamma\label{eq4.3.44}
\eeq
\beq
a_{2}=\frac{8(811+42c)\eta\gamma}{784+57c}+\tilde{\lambda}\mu
\eeq
\beq
a_{3}=-\frac{2(4244+105c)\eta\gamma}{784+57c}-3\tilde{\lambda}\mu\label{eq4.3.46}
\eeq
\bea
a_{4}=\frac{1}{8(784+57c)}\{8\eta[13(472+3c)\gamma+(784+57c)\tilde{\gamma}]\nn\\
+\mu[5(5824+237c)\zeta+24(784+57c)\tilde{\lambda}]\}\label{eq4.3.47}
\eea
\bea
a_{5}=\frac{1}{4(784+57c)}\cdot\{8\eta[(-892+3c)\gamma-(784+57c)\tilde{\gamma}]\nn\\
-\mu[5(5824+237c)\zeta+4(784+57c)\tilde{\lambda}]\}\label{eq4.3.48}
\eea
\beq
a_{6}=\eta\tilde{\gamma}+\frac{3(18592+1041c)\zeta\mu}{8(784+57c)}\label{eq4.3.49}
\eeq
\beq
a_{7}=-\frac{17}{4}\zeta\mu\label{eq4.3.50}
\eeq
\beq
a_{8}=\zeta\mu\label{eq4.3.51}
\eeq

One can check that the $z\rightarrow 1$ limit condition (\ref{eq4.3.5}) is verified. The $z\rightarrow\infty$ limit condition (\ref{eq4.3.10}) is also verified. As an additional check, we did also the operator algebra derivation for the next to leading coefficient of the $z\rightarrow\infty$ expansion (expansion in $1/z$), and verified in that way the value of the coefficient $a_{7}$, eq.(\ref{eq4.3.50}).

These checks do not constitute the complete verification, evidently. But for the present level of confidence in the operator algebra we take them as being sufficient.

The complete verification of the consistency between the $z\rightarrow 0$ and $z\rightarrow 1$ limit calculations, we did them in Sections 2.1 and 2.2 for the function $<\psi^{+}\psi\psi\psi^{+}>$, and in 4.1 and 4.2 for the function $<B(\infty)\tilde{\psi}^{+}(1)\psi(z)B(0)>$.

We observe finally that we have found the complete expression for the correlation function $<B(\infty)\psi(1)\tilde{\psi}(z)\tilde{\psi}(0)>$, which is consistent with the operator algebra for the unconstrained values of the constants. 
This is like this was the case also for the function 
$<B(\infty)\tilde{\psi}^{+}(1)\psi(z)B(0)>$ calculated in Sections 4.1, 4.2.

\subsection{Correlation function $<\tilde{\psi}^{+}(\infty)\psi(1)\tilde{\psi}(z)\tilde{\psi}^{+}(0)>$.}

Calculations in this Section will be similar to those in the previous Section. 
For this reason our remarks will be short and limited. More detailed arguments on the methods of calculations could be found, if necessary, in the previous Section. We shall follow the same steps, more or less.

The present function has the following analytic form :
\beq
<\tilde{\psi}^{+}(\infty)\psi(1)\tilde{\psi}(z)\tilde{\psi}^{+}(0)>=\frac{P_{6}(z)}{(z)^{2\tilde{\Delta}-4}\cdot(1-z)^{\tilde{\Delta}}}\label{eq4.4.1}
\eeq
The limits, which justify the above expression :

$z\rightarrow 0$.

\bea
<\tilde{\psi}^{+}(\infty)\psi(1)\tilde{\psi}(z)\tilde{\psi}^{+}(0)>\nn\\
=<\tilde{\psi}^{+}(\infty)\psi(1)\times\frac{1}{(z)^{2\tilde{\Delta}}}\{1+z^{2}\frac{2\tilde{\Delta}}{c}T(0)+z^{3}\cdot\frac{\tilde{\Delta}}{c}\partial T(0)\nn\\
+z^{4}[\beta^{(112)}_{\tilde{\psi}\tilde{\psi}^{+},I}\cdot\partial^{2}T(0)+\beta^{(22)}_{\tilde{\psi}\tilde{\psi}^{+},I}\cdot\wedge(0)+\tilde{\gamma}\cdot B(0)]+...\}>\nn\\
\simeq<\tilde{\psi}^{+}(\infty)\psi(1)\frac{1}{(z)^{2\tilde{\Delta}-4}}\cdot\tilde{\gamma} B(0)>=\frac{\tilde{\gamma}}{(z)^{2\tilde{\Delta}-4}}\cdot\mu\label{eq4.4.2}
\eea
\beq
a_{0}=\tilde{\gamma}\cdot\mu\label{eq4.4.3}
\eeq

$z\rightarrow 1$.

\bea
<\tilde{\psi}^{+}(\infty)\psi(1)\tilde{\psi}(z)\tilde{\psi}^{+}(0)>\nn\\
\simeq<\tilde{\psi}^{+}(\infty)\cdot\frac{1}{(1-z)^{\tilde{\Delta}}}\zeta\psi^{+}(1)\tilde{\psi}^{+}(0)>\nn\\
=\frac{\zeta}{(1-z)^{\tilde{\Delta}}}\cdot\eta\label{eq4.4.4}
\eea
\beq
\sum^{6}_{k=0}a_{k}=\zeta\eta\label{eq4.4.5}
\eeq

$z\rightarrow\infty$.

\beq
<\tilde{\psi}^{+}(\infty)\tilde{\psi}(z)\psi(1)\tilde{\psi}^{+}(0)>=\frac{P_{6}(z)}{(z)^{2\tilde{\Delta}-4}\cdot(z-1)^{\tilde{\Delta}}}\label{eq4.4.5a}
\eeq
\bea
<\tilde{\psi}^{+}(\infty)\tilde{\psi}(z)\psi(1)\tilde{\psi}^{+}(0)>\nn\\
\simeq<\tilde{\psi}^{+}(\infty)\tilde{\psi}(z)\cdot\mu B(0)>=\mu\cdot\frac{\tilde{\gamma}}{(z)^{4}}\label{eq4.4.6}
\eea
\beq
\frac{P_{6}(z)}{(z)^{2\tilde{\Delta}-4}\cdot(z-1)^{\tilde{\Delta}}}\simeq\frac{a_{6}z^{6}}{(z)^{3\tilde{\Delta}-4}}=\frac{a_{6}}{(z)^{3\tilde{\Delta}-10}}=\frac{a_{6}}{(z)^{4}}\label{eq4.4.7}
\eeq
\beq
a_{6}=\mu\tilde{\gamma}\label{eq4.4.8}
\eeq

The actions of the modes of $\tilde{\psi}$ and $\psi$ on $\tilde{\psi}^{+}(0)$, which we shall need further down,
are as follows :
\beq
\tilde{\psi}_{\frac{14}{3}+n}\tilde{\psi}^{+}(0)=0,\quad n>0\label{eq4.4.9}
\eeq
\beq
\tilde{\psi}_{\frac{14}{3}}\tilde{\psi}^{+}(0)=1\label{eq4.4.10}
\eeq
\beq
\tilde{\psi}_{\frac{11}{3}}\tilde{\psi}^{+}(0)=0\label{eq4.4.11}
\eeq
\beq
\tilde{\psi}_{\frac{8}{3}}\tilde{\psi}^{+}(0)=\frac{2\tilde{\Delta}}{c}T(0)\label{eq4.4.12}
\eeq
\beq
\tilde{\psi}_{\frac{5}{3}}\tilde{\psi}^{+}(0)=\frac{\tilde{\Delta}}{c}\partial T(0)\label{eq4.4.13}
\eeq
\beq
\tilde{\psi}_{\frac{2}{3}}\tilde{\psi}^{+}(0)=\beta^{(112)}_{\tilde{\psi}\tilde{\psi}^{+},I}\cdot\partial^{2}T(0)+\beta^{(22)}_{\tilde{\psi}\tilde{\psi}^{+},I}\cdot\wedge(0)+\tilde{\gamma}\cdot B(0)\label{eq4.4.14}
\eeq
\beq
\tilde{\psi}_{-\frac{1}{3}}\tilde{\psi}^{+}(0)=\beta^{(1112)}_{\tilde{\psi}\tilde{\psi}^{+},I}\cdot\partial^{3}T(0)+\beta^{(122)}_{\tilde{\psi}\tilde{\psi}^{+},I}\cdot\partial\wedge(0)+\tilde{\gamma}\cdot \beta^{(1)}_{\tilde{\psi}\tilde{\psi}^{+},B}\cdot\partial B(0)\label{eq4.4.15}
\eeq
and
\beq
\psi_{\frac{2}{3}+n}\tilde{\psi}^{+}(0)=0,\quad n>0\label{eq4.4.16}
\eeq
\beq
\psi_{\frac{2}{3}}\tilde{\psi}^{+}(0)=\mu B(0)\label{eq4.4.17}
\eeq
\beq
\psi_{-\frac{1}{3}}\tilde{\psi}^{+}(0)=\mu\beta^{(1)}_{\psi\tilde{\psi}^{+},B}\cdot\partial B(0)\label{eq4.4.18}
\eeq

The formal expansion of our correlation function
(by the operator algebra) has the following form :
\bea
<\tilde{\psi}^{+}(\infty)\psi(1)\tilde{\psi}(z)\tilde{\psi}^{+}(0)>
=\frac{1}{(z)^{\tilde{\Delta}+\frac{2}{3}}}\{<\tilde{\psi}^{+}(\infty)\psi_{-\frac{2}{3}}\tilde{\psi}_{\frac{2}{3}}\tilde{\psi}^{+}(0)>\nn\\
+z<\tilde{\psi}^{+}(\infty)\psi_{\frac{1}{3}}\tilde{\psi}_{-\frac{1}{3}}
\tilde{\psi}^{+}(0)>
+z^{2}<\tilde{\psi}^{+}(\infty)\psi_{\frac{4}{3}}\tilde{\psi}_{-\frac{4}{3}}\tilde{\psi}^{+}(0)>+...\}\label{eq4.4.19}
\eea
We observe that
\beq
\tilde{\Delta}+\frac{2}{3}=2\tilde{\Delta}-4\label{eq4.4.20}
\eeq
so that the powers of $z$, in front, are the same in (\ref{eq4.4.19}) and in the analytic form (\ref{eq4.4.1}).

The commutation relations $\{\psi,\tilde{\psi}\}\tilde{\psi}^{+}(0)$, in their general form :
\beq
\sum^{\infty}_{l=0}D^{l}_{\frac{5}{3}}(\psi_{\frac{4}{3}+n-l}\tilde{\psi}_{-\frac{1}{3}+m+l}+\tilde{\psi}_{\frac{4}{3}+m-l}\psi_{-\frac{1}{3}+n+l})\tilde{\psi}^{+}(0)=R(n,m)\label{eq4.4.21}
\eeq
\bea
R(n,m)=\{\frac{1}{2}(\Delta-\frac{1}{3}+n-1)(\Delta-\frac{1}{3}+n-2)\zeta\psi^{+}_{n+m+1}\nn\\
-(\Delta-\frac{1}{3}+n-1)\cdot(\Delta+n+m+1)\zeta\beta^{(1)}_{\psi\tilde{\psi},\psi^{+}}\cdot\psi^{+}_{n+m+1}\nn\\
+(\Delta+n+m+2)(\Delta+n+m+1)\zeta\beta^{(11)}_{\psi\tilde{\psi},\psi^{+}}\cdot\psi^{+}_{n+m+1}\nn\\
+\zeta\beta^{(2)}_{\psi\tilde{\psi},\psi^{+}}\cdot(L_{-2}\psi^{+})_{n+m+1}+\eta\tilde{\psi}^{+}_{n+m+1}\}\tilde{\psi}^{+}(0)\label{eq4.4.22}
\eea
Their projection on the level of the operator $\tilde{\psi}(0)$ requires :
\beq
n+m+1=0,\quad m=-1-n\label{eq4.4.23}
\eeq
and we obtain :
\beq
\sum^{\infty}_{l=0}D^{l}_{\frac{5}{3}}(\psi_{\frac{4}{3}+n-l}\tilde{\psi}_{-\frac{4}{3}-n+l}+\tilde{\psi}_{\frac{1}{3}-n-l}\psi_{-\frac{1}{3}+n+l})\tilde{\psi}^{+}(0)=R(n)\label{eq4.4.24}
\eeq
\bea
R(n)=R(n,-1-n)=\{\frac{1}{2}(\Delta-\frac{1}{3}+n-1)
(\Delta-\frac{1}{3}+n-2)\zeta\psi^{+}_{0}\nn\\
-(\Delta-\frac{1}{3}+n-1)\cdot(\Delta)\zeta\beta^{(1)}_{\psi\tilde{\psi},\psi^{+}}\cdot\psi^{+}_{0}\nn\\
+(\Delta+1)(\Delta)\cdot\zeta\beta^{(11)}_{\psi\tilde{\psi},\psi^{+}}\cdot\psi^{+}_{0}\nn\\
+\zeta\beta^{(2)}_{\psi\tilde{\psi},\psi^{+}}\cdot(L_{-2}\psi^{+})_{0}+\eta\tilde{\psi}^{+}_{0}\}\tilde{\psi}^{+}(0)\label{eq4.4.25}
\eea
We need the actions of $\psi^{+}_{0}$, $(L_{-2}\psi^{+})_{0}$ and $\tilde{\psi}^{+}_{0}$ onto $\tilde{\psi}^{+}(0)$, because they appear in $R(n)$ above. One finds :
\beq
\psi^{+}_{0}\tilde{\psi}^{+}(0)=\zeta\beta^{(11)}_{\psi\tilde{\psi},\psi^{+}}\cdot\partial^{2}\psi(0)+\zeta\beta^{(2)}_{\psi\tilde{\psi},\psi^{+}}\cdot L_{-2}\psi(0)+\eta\tilde{\psi}(0)\label{eq4.4.26}
\eeq
\bea
(L_{-2}\psi^{+})_{0}\tilde{\psi}^{+}(0)=\zeta L_{-2}\psi(0)+\zeta\beta^{(1)}_{\psi\tilde{\psi},\psi^{+}}\cdot\partial^{2}\psi(0)\nn\\
+(\Delta+\tilde{\Delta})\cdot\psi^{+}_{0}\tilde{\psi}^{+}(0)\label{eq4.4.27}
\eea
where $\psi^{+}_{0}\tilde{\psi}^{+}(0)$ is given above, eq.(\ref{eq4.4.26}); 

\noindent and
\beq
\tilde{\psi}^{+}_{0}\tilde{\psi}^{+}(0)=\eta\beta^{(11)}_{\tilde{\psi}\tilde{\psi},\psi^{+}}\cdot\partial^{2}\psi(0)+\eta\beta^{(2)}_{\tilde{\psi}\tilde{\psi},\psi^{+}}\cdot L_{-2}\psi(0)+\tilde{\lambda}\cdot\tilde{\psi}(0)\label{eq4.4.28}
\eeq
The low commutation relations coming from (\ref{eq4.4.24}) are the following ones :

\underline{$n=0$} :
\bea
\psi_{\frac{4}{3}}\tilde{\psi}_{-\frac{4}{3}}\tilde{\psi}^{+}(0)+D^{1}_{\frac{5}{3}}\psi_{\frac{1}{3}}\tilde{\psi}_{-\frac{1}{3}}\tilde{\psi}^{+}(0)+D^{2}_{\frac{5}{3}}\psi_{-\frac{2}{3}}\tilde{\psi}_{\frac{2}{3}}\tilde{\psi}^{+}(0)\nn\\
+D^{3}_{\frac{5}{3}}\psi_{-\frac{5}{3}}\tilde{\psi}_{\frac{5}{3}}\tilde{\psi}^{+}(0)
+D^{4}_{\frac{5}{3}}\psi_{-\frac{8}{3}}\tilde{\psi}_{\frac{8}{3}}\tilde{\psi}^{+}(0)+D^{5}_{\frac{5}{3}}\psi_{-\frac{11}{3}}\tilde{\psi}_{\frac{11}{3}}\tilde{\psi}^{+}(0)\nn\\
+D^{6}_{\frac{5}{3}}\psi_{-\frac{14}{3}}\tilde{\psi}_{\frac{14}{3}}\tilde{\psi}^{+}(0)+\tilde{\psi}_{\frac{1}{3}}\psi_{-\frac{1}{3}}\tilde{\psi}^{+}(0)+D^{1}_{\frac{5}{3}}\tilde{\psi}_{-\frac{2}{3}}\psi_{\frac{2}{3}}\tilde{\psi}^{+}(0)=R(0)\label{eq4.4.29}
\eea
We should observe that the matrix elements $\psi_{-\frac{5}{3}}\tilde{\psi}_{\frac{5}{3}}\tilde{\psi}^{+}(0)$, $\tilde{\psi}_{-\frac{8}{3}}\tilde{\psi}_{\frac{8}{3}}\tilde{\psi}^{+}(0)$, $\psi_{-\frac{11}{3}}\tilde{\psi}_{\frac{11}{3}}\tilde{\psi}^{+}(0)$, $\psi_{-\frac{14}{3}}\tilde{\psi}_{\frac{14}{3}}\tilde{\psi}^{+}(0)$, they will not contain the operator $\tilde{\psi}(0)$, as is seen from the actions of modes in (\ref{eq4.4.10}) - (\ref{eq4.4.13}). As a consequence, they will disappear under the projection onto $\tilde{\psi}^{+}(\infty)$, when calculating the correlation function $<\tilde{\psi}^{+}(\infty)\psi(1)\tilde{\psi}(z)\tilde{\psi}^{+}(0)>$ by the eq.(\ref{eq4.4.19}). For this reason we shall not show them explicitly in the following.

\underline{$n=1$} :
\bea
\psi_{\frac{7}{3}}\tilde{\psi}_{-\frac{7}{3}}\tilde{\psi}^{+}(0)+D^{1}_{\frac{5}{3}}\psi_{\frac{4}{3}}\tilde{\psi}_{-\frac{4}{3}}\tilde{\psi}^{+}(0)+D^{2}_{\frac{5}{3}}\psi_{\frac{1}{3}}\tilde{\psi}_{-\frac{1}{3}}\tilde{\psi}^{+}(0)\nn\\
+D^{3}_{\frac{5}{3}}\psi_{-\frac{2}{3}}\tilde{\psi}_{\frac{2}{3}}\tilde{\psi}^{+}(0)
+...+\tilde{\psi}_{-\frac{2}{3}}\psi_{\frac{2}{3}}\tilde{\psi}^{+}(0)=R(1)\label{4.4.30}
\eea

\underline{$n=2$} :
\bea
\psi_{\frac{10}{3}}\tilde{\psi}_{-\frac{10}{3}}\tilde{\psi}^{+}(0)+D^{1}_{\frac{5}{3}}\psi_{\frac{7}{3}}\tilde{\psi}_{-\frac{7}{3}}\tilde{\psi}^{+}(0)+D^{2}_{\frac{5}{3}}\psi_{\frac{4}{3}}\tilde{\psi}_{-\frac{4}{3}}\tilde{\psi}^{+}(0)\nn\\
+D^{3}_{\frac{5}{3}}\psi_{\frac{1}{3}}\tilde{\psi}_{-\frac{1}{3}}\tilde{\psi}^{+}(0)
+D^{4}_{\frac{5}{3}}\psi_{-\frac{2}{3}}\tilde{\psi}_{\frac{2}{3}}\tilde{\psi}^{+}(0)+...=R(2)\label{eq4.4.31}
\eea

Below we given the expression for the matrix elements and we introduce, at the same time, the simplified notations for them. In these simplified notations we choose to note as $m(0)$, the matrix element $\psi_{-\frac{2}{3}}\tilde{\psi}_{\frac{2}{3}}\tilde{\psi}^{+}(0)$, instead of $\psi_{-\frac{14}{3}}\tilde{\psi}_{\frac{14}{3}}\tilde{\psi}^{+}(0)$. 
This is for the same reason as above :  $\psi_{-\frac{2}{3}}\tilde{\psi}_{\frac{2}{3}}\tilde{\psi}^{+}(0)$ is the first matrix element which remains after the projection onto $\tilde{\psi}^{+}(\infty)$. 
(We could observe that with this choice for notations 
we would have $\psi_{-\frac{5}{3}}\tilde{\psi}_{\frac{5}{3}}\tilde{\psi}^{+}(0)=m(-1)$, $\psi_{-\frac{8}{3}}\tilde{\psi}_{\frac{8}{3}}\tilde{\psi}^{+}(0)=m(-2)$, etc.).

The last particular point of the simplifications we did in our present calculations, we shall give below the expressions for the matrix elements which are already projected onto $\tilde{\psi}^{+}(\infty)$ : so to say, we shall present, in the expressions below, only the terms $\sim\tilde{\psi}(0)$, while also suppressing $\tilde{\psi}(0)$ but keeping its coefficients.

One finds, after some calculations :
\beq
m_{\tilde{\psi}}(0)=
(\psi_{-\frac{2}{3}}\tilde{\psi}_{\frac{2}{3}}\tilde{\psi}^{+}(0))_{\tilde{\psi}}=
\tilde{\gamma}\mu\label{eq4.4.32}
\eeq
\beq
m_{\tilde{\psi}}(1)=
(\psi_{\frac{1}{3}}\tilde{\psi}_{-\frac{1}{3}}\tilde{\psi}^{+}(0))_{\tilde{\psi}}=\tilde{\gamma}\beta^{(1)}_{\tilde{\psi}\tilde{\psi}^{+},B}\cdot(\Delta-\frac{2}{3})\cdot\mu \label{eq4.4.33}
\eeq
and
\beq
mc_{\tilde{\psi}}(0)=
(\tilde{\psi}_{-\frac{2}{3}}\psi_{\frac{2}{3}}\tilde{\psi}^{+}(0))_{\tilde{\psi}}=
\mu\tilde{\gamma}\label{eq4.4.33a}
\eeq
\beq
mc_{\tilde{\psi}}(1)=
(\tilde{\psi}_{\frac{1}{3}}\psi_{-\frac{1}{3}}\tilde{\psi}^{+}(0))_{\tilde{\psi}}=
\mu\beta^{(1)}_{\psi\tilde{\psi}^{+}B}\cdot(\tilde{\Delta}-\frac{2}{3})\tilde{\gamma}\label{eq4.4.34}
\eeq

Like for the other two correlation functions, analyzed in Sections 4.1, 4.2, 4.3, the low commutation relations, eq.(\ref{eq4.4.29}) - (\ref{eq4.4.31}), do not provide constraints on the constants, but they allow to define all the matrix elements, provided the explicit matrix elements $m_{\tilde{\psi}}(0)$, $m_{\tilde{\psi}}(1)$, $mc_{\tilde{\psi}}(0)$, $mc_{\tilde{\psi}}(1)$ had been calculated directly. In fact, the commutation relation $n=0$, eq.(\ref{eq4.4.29}), defines the matrix element $m_{\tilde{\psi}}(2)=\psi_{\frac{4}{3}}\tilde{\psi}_{-\frac{4}{3}}\tilde{\psi}^{+}(0)$, and so on. Starting 
with $m_{\tilde{\psi}}(4)$, the matrix elements can be defined by the recurrence relation :
\beq
m_{\tilde{\psi}}(n)=R_{\tilde{\psi}}(n-2)-\sum^{n}_{k=1}D^{k}_{\frac{5}{3}}m_{\tilde{\psi}}(n-k)\label{eq4.4.35}
\eeq
We remark that $R(n)$, in eq.(\ref{eq4.4.25}), with the actions of modes in (\ref{eq4.4.26}) - (\ref{eq4.4.28}), it also has to be taken in the projected form, onto $\tilde{\psi}^{+}(\infty)$.

The correlation function in (\ref{eq4.4.19}) could be rewritten as :
\bea
<\tilde{\psi}^{+}(\infty)\psi(1)\tilde{\psi}(z)\tilde{\psi}(0)>\nn\\
=\frac{1}{(z)^{2\tilde{\Delta}-4}}\cdot\{m_{\tilde{\psi}}(0)+z\cdot m_{\tilde{\psi}}(1)+z^{2}\cdot m_{\tilde{\psi}}(2)+...\}\label{eq4.4.36}
\eea
with all the coefficients $\{m_{\tilde{\psi}}(n), n=0,1,2...\}$ supposed to be known. We remind that $\tilde{\Delta}+\frac{2}{3}=2\tilde{\Delta}-4$.

Next we expand the analytic form (\ref{eq4.4.1}), of our correlation function, and we compare the expansion with the series in (\ref{eq4.4.36}). In this way we get the following values for the coefficients $\{a_{n}, n=0,1,...,6\}$ of the polynomial $P_{6}(z)$ in (\ref{eq4.4.1}) :
\bea
a_{0}=\tilde{\gamma}\cdot\mu, \quad
a_{1}=-\frac{11}{3}\tilde{\gamma}\cdot\mu, \quad
a_{2}=-\frac{7(-488+c)\zeta\eta}{2(784+57c)}+\eta \tilde{\lambda}+\frac{17}{3}\cdot\tilde{\gamma}\mu, \nn\\
a_{3}=\frac{8(-329+8c)\zeta\eta-2(784+57c)(\eta\tilde{\lambda}+3\tilde{\gamma}\mu)}{784+57c}\label{eq4.4.40}
\eea
Next we have found that $a_{4}=a_{2}$, $a_{5}=a_{1}$, $a_{6}=a_{0}$.

We observe that the value of $a_{6}$ is in agreement with the expectation (\ref{eq4.4.8}), of the limit $z\rightarrow\infty$. Also, it is not difficult to verify, with the values of $\{a_{n}\}$ above, that the expectation (\ref{eq4.4.5}), of the limit $z\rightarrow 1$, is also verified. We have also verified the value of $a_{5}$, by calculating, with the operator algebra, the next to leading term of the expansion in $1/z$ (limit $z\rightarrow\infty$).

\numberwithin{equation}{section}

\section{Correlation function $<B(\infty)\tilde{\psi}(1)\tilde{\psi}^{+}(z)B(0)>$}

In this Section we shall calculate the function 
$<B\tilde{\psi}\tilde{\psi}^{+}B>$, to illustrate the techniques to be used in analyzing the triple products of the third group, eq.(\ref{eq3.14})-(\ref{eq3.19}), and the corresponding correlation functions, eq.(\ref{eq3.26})-(\ref{eq3.29}).

First we define the analytic form of our function. It is as follows :
\beq
<B(\infty)\tilde{\psi}(1)\tilde{\psi}^{+}(z)B(0)>=\frac{P_{12}(z)}{(z)^{6}(1-z)^{2\tilde{\Delta}}}\label{eq5.1}
\eeq
In fact :

\underline{$z\rightarrow 0$},
\bea
<B(\infty)\tilde{\psi}(1)\tilde{\psi}^{+}(z)B(0)>\simeq<B(\infty)\tilde{\psi}(1)\frac{\mu}{(z)^{6}}\cdot\psi^{+}(0)>\nn\\
=\frac{\mu}{(z)^{6}}\cdot<B(\infty)\tilde{\psi}(1)\psi^{+}(0)>=\frac{\mu}{(z)^{6}}\cdot\mu\label{eq5.2}
\eea
\beq
a_{0}=\mu^2 \label{eq5.3}
\eeq

\underline{$z\rightarrow 1$},
\bea
<B(\infty)\tilde{\psi}(1)\tilde{\psi}^{+}(z)B(0)>\nn\\
\simeq<B(\infty)\frac{1}{(1-z)^{2\tilde{\Delta}}}B(0)>=\frac{1}{(1-z)^{2\tilde{\Delta}}}\label{eq5.4}
\eea
\beq
\sum^{12}_{k=0}a_{k}=1\label{eq5.5}
\eeq

\underline{$z\rightarrow\infty$},
\beq
<B(\infty)\tilde{\psi}^{+}(z)\tilde{\psi}(1)B(0)>=\frac{P_{12}(z)}{(z)^{6}(z-1)^{2\tilde{\Delta}}}\label{eq5.6}
\eeq
\bea
<B(\infty)\tilde{\psi}^{+}(z)\tilde{\psi}(1)B(0)>\simeq
<B(\infty)\tilde{\psi}^{+}(z)\mu\cdot\psi(0)>\nn\\
=\mu<B(\infty)\tilde{\psi}^{+}(z)\psi(0)>=\mu\cdot\frac{\mu}{(z)^{\tilde{\Delta}+\Delta-4}}=\frac{\mu^{2}}{(z)^{2\Delta-2}}\label{eq5.7}
\eea
\beq
\frac{P_{12}(z)}{(z)^{6}(z-1)^{2\tilde{\Delta}}}\simeq\frac{a_{12}z^{12}}{(z)^{6+2\Delta+4}}=\frac{a_{12}}{(z)^{2\Delta-2}}\label{eq5.8}
\eeq
The order 12 of the polynomial is confirmed. In addition, it should be checked later that
\beq
a_{12}=\mu^{2}\label{eq5.9}
\eeq
More generally, due to the symmetric form of the function $<B(\infty)\tilde{\psi}(1)\tilde{\psi}^{+}(z)B(0)>$, with respect to the limits $z\rightarrow 0$ and $z\rightarrow\infty$, the coefficients $\{a_{k}, k=1,2,...,12\}$ should be symmetric :
\beq
a_{0}=a_{12},\,\, a_{1}=a_{11},\,\, a_{2}=a_{10},\,\,a_{3}=a_{9},\,\,a_{4}=a_{8},\,\,a_{5}=a_{7}\label{eq5.10}
\eeq
-- to be checked later with our calculations.

Now we shall start calculating the function $<B\tilde{\psi}\tilde{\psi}^{+}B>$, in the l.h.s. of (\ref{eq5.1}), by developing it in $z$, with the use of the operator algebra.

The formal expansion is of the form :
\bea
<B(\infty)\tilde{\psi}(1)\tilde{\psi}^{+}(z)B(0)>
=\frac{1}{(z)^{\tilde{\Delta}+\frac{4}{3}}}\{
<B(\infty)\tilde{\psi}_{-\frac{4}{3}}\tilde{\psi}^{+}_{\frac{4}{3}}B(0)>\nn\\
+z<B(\infty)\tilde{\psi}_{-\frac{1}{3}}\tilde{\psi}^{+}_{\frac{1}{3}}(z)B(0)>+
z^{2}<B(\infty)\tilde{\psi}_{\frac{2}{3}}\tilde{\psi}^{+}_{-\frac{2}{3}}B(0)>\nn\\
+z^{3}<B(\infty)\tilde{\psi}_{\frac{5}{3}}\tilde{\psi}^{+}_{-\frac{5}{3}}B(0)>...\}\label{eq5.11}
\eea
We have used the fact that the modes of $\tilde{\psi}^{+}$ (and of $\tilde{\psi}$) act on $B(0)$ as follows :
\beq
\tilde{\psi}^{+}_{\frac{4}{3}+n}B(0)=0, \quad n>0\label{eq5.12}
\eeq
\beq
\tilde{\psi}^{+}_{\frac{4}{3}}B(0)=\mu\psi^{+}(0)\label{eq5.13}
\eeq
\beq
\tilde{\psi}^{+}_{\frac{1}{3}}B(0)=\mu\beta^{(1)}_{\tilde{\psi}B,\psi}\partial\psi^{+}(0)\label{eq5.14}
\eeq
\beq
\tilde{\psi}^{+}_{-\frac{2}{3}}B(0)=\mu\beta^{(11)}_{\tilde{\psi}B,\psi}\cdot\partial^{2}\psi^{+}(0)+\mu\beta^{(2)}_{\tilde{\psi}B,\psi}\cdot L_{-2}\psi^{+}(0)+\tilde{\gamma}\tilde{\psi}^{+}(0)\label{eq5.15}
\eeq
In writing the expansion (\ref{eq5.11}) we have actually used just the fact 
that the non-vanishing mode actions, on $B(0)$, start with $\tilde{\psi}^{+}_{\frac{4}{3}}B(0)$, eq.(\ref{eq5.12}), (\ref{eq5.13}).

To calculate the coefficients of the $z$ expansion in (\ref{eq5.11}) we 
need the commutation relations $\{\tilde{\psi},\tilde{\psi}^{+}\}B(0)$. 
They are found to be of the following general form :
\beq
\sum^{\infty}_{l=0}D^{l}_{\frac{10}{3}}(\tilde{\psi}_{\frac{8}{3}+n-l}\tilde{\psi}_{-\frac{2}{3}+m+l}^{+}-\tilde{\psi}_{\frac{8}{3}+m-l}^{+}\tilde{\psi}_{-\frac{2}{3}+n+l})B(0)=R(n,m)\label{eq5.16}
\eeq
\bea
R(n,m)=\{\frac{1}{5!}(\tilde{\Delta}-\frac{2}{3}+n-1)(\tilde{\Delta}-\frac{2}{3}+n-2)(\tilde{\Delta}-\frac{2}{3}+n-3)\nn\\
\cdot(\tilde{\Delta}-\frac{2}{3}+n-4) (\tilde{\Delta}-\frac{2}{3}+n-5)\cdot\delta_{n+m+2,0}\nn\\
+\frac{1}{3!}(\tilde{\Delta}-\frac{2}{3}+n-1)(\tilde{\Delta}-\frac{2}{3}+n-2)
(\tilde{\Delta}-\frac{2}{3}+n-3)\cdot\frac{2\tilde{\Delta}}{c}L_{n+m+2}\nn\\
-\frac{1}{2}(\tilde{\Delta}-\frac{2}{3}+n-1)(\tilde{\Delta}-\frac{2}{3}+n-2)\cdot(n+m+4)\cdot\frac{\tilde{\Delta}}{c}L_{n+m+2}\nn\\
+(\tilde{\Delta}-\frac{2}{3}+n-1)\cdot[(n+m+5)(n+m+4)\beta^{(112)}_{\tilde{\psi}\tilde{\psi}^{+},I}\cdot L_{n+m+2}\nn\\
+\beta^{(22)}_{\tilde{\psi}\tilde{\psi}^{+},I}\cdot\wedge_{n+m+2}+\tilde{\gamma}\cdot B_{n+m+2}]\nn\\
-[(n+m+6)(n+m+5)(n+m+4)\beta^{(1112)}_{\tilde{\psi}\tilde{\psi}^{+},I}\cdot L_{n+m+2}\nn\\
+(n+m+6)\cdot\beta^{(122)}_{\tilde{\psi}\tilde{\psi}^{+},I}\cdot\wedge_{n+m+2}\nn\\
+(n+m+6)\tilde{\gamma}\beta^{(1)}_{\tilde{\psi}\tilde{\psi}^{+},B}\cdot B_{n+m+2}]\}B(0)\label{eq5.17}
\eea
(\ref{eq5.16}), (\ref{eq5.17}) just translate, into the commutation relations of modes, the operator algebra expansion (\ref{eq1.7}), which, more explicitly, has the form :
\bea
\tilde{\psi}(z')\tilde{\psi}^{+}(z)=\frac{1}{(z'-z)^{2\tilde{\Delta}}}\{1+(z'-z)^{2}\frac{2\tilde{\Delta}}{c}T(z)\nn\\
+(z'-z)^{3}\cdot\frac{\tilde{\Delta}}{c}\partial T(z)+(z'-z)^{4}[\beta^{(112)}_{\tilde{\psi}\tilde{\psi}^{+},I}\cdot\partial^{2}T(z)+\beta^{(22)}_{\tilde{\psi}\tilde{\psi}^{+},I}\cdot\wedge(z)\nn\\
+\tilde{\gamma}\cdot B(z)]+(z'-z)^{5}\cdot[\beta^{(1112)}_{\tilde{\psi}\tilde{\psi}^{+},I}\cdot\partial^{3}T(z)+\beta^{(122)}_{\tilde{\psi}\tilde{\psi},I}\cdot\partial\wedge(z)\nn\\
+\tilde{\gamma}\beta^{(1)}_{\tilde{\psi}\tilde{\psi}^{+},B}\cdot\partial B(z)]+...\}\label{eq5.18}
\eea
The "translation" is being done in a standard way, via the commutation of the contour integrals which define the modes of $\tilde{\psi}$ and $\tilde{\psi}^{+}$, the way this has been done in Sections (\ref{eq2.1}), (\ref{eq2.2}).

It has been explained in the Section 1, and we remind it again, that in the $Z_{3}$ neutral sector we can make explicit the expansion terms up to the 5th level, not further. This is what we did in (\ref{eq5.18}). The higher terms have to be treated as descendants of the operator algebra itself.

We have calculated the commutation relations (\ref{eq5.16}), (\ref{eq5.17}) accordingly, defining the contour integrals the way that the terms in (\ref{eq5.18}) contribute up to the 5th level, not further.

In particular, the $I_{3}$ integral has been taken in the form :
\bea
I_{3}=\frac{1}{(2\pi)^{2}}\oint_{C_{0}}dz(z)^{\tilde{\Delta}-\frac{2}{3}+m-1}\oint_{C_{z}}dz'(z')^{\tilde{\Delta}-\frac{2}{3}+n-1}\nn\\
\times(z'-z)^{2\tilde{\Delta}-6} \tilde{\psi}(z')\tilde{\psi}^{+}(z)B(0)\label{eq5.19}
\eea
The extra power $-6$ of the factor $(z'-z)^{2\tilde{\Delta}-6}$ has been chosen accordingly, so that when the product $\tilde{\psi}(z')\tilde{\psi}^{+}(z)$ is expanded, eq.(\ref{eq5.18}), the terms up to the 5th level only contribute, and not those further down in the expansion.

Due to the fact that not all the terms which are singular, in the expansion (\ref{eq5.18}), contribute to the integral (\ref{eq5.19}) (to the integral over $z'$), the commutation relations (\ref{eq5.16}), (\ref{eq5.17}) will not define all the matrix elements of the descendants of the algebra, some will be missed. But we shall define them differently, whose which are missed, by factorizing the modes of $\tilde{\psi}^{+}$ into the modes 
of principal fields, which are $\psi,\psi^{+}$. 
The commutation relations of the modes of principal fields do define all their matrix elements.

The purpose of this Section is to show the way the calculations for the secondary fields can be realized, so that all their matrix elements get defined, correlation functions calculated and verified to be consistent with the operator algebra.

\vskip0.4cm

After these remarks we go back to the commutation relations (\ref{eq5.16}), (\ref{eq5.17}). If we want to define the matrix elements in the expansion (\ref{eq5.11}), we have to impose the constraint, in (\ref{eq5.16}), (\ref{eq5.17}),
\beq
n+m+2=0,\quad m=-2-n\label{eq5.20}
\eeq
The commutation relations take the form :
\beq
\sum_{l=0}^{\infty}D^{l}_{\frac{10}{3}}(\tilde{\psi}_{\frac{8}{3}+n-l}\tilde{\psi}^{+}_{-\frac{8}{3}-n+l}-\tilde{\psi}^{+}_{\frac{2}{3}-n-l}\tilde{\psi}_{-\frac{2}{3}+n+l})B(0)=R(n)\label{eq5.21}
\eeq
\bea
R(n)=R(n,-2-n) = 
\{\frac{1}{5!}(\tilde{\Delta}-\frac{2}{3}+n-1)(\tilde{\Delta}-\frac{2}{3}+n-2)\nn\\\times(\tilde{\Delta}-\frac{2}{3}+n-3)
(\tilde{\Delta}-\frac{2}{3}+n-4)(\tilde{\Delta}-\frac{2}{3}+n-5)\nn\\
+\frac{1}{3!}(\tilde{\Delta}-\frac{2}{3}+n-1)(\tilde{\Delta}-\frac{2}{3}+n-2)(\tilde{\Delta}-\frac{2}{3}+n-3)\cdot\frac{2\tilde{\Delta}}{c}L_{0}\nn\\
-\frac{1}{2}(\tilde{\Delta}-\frac{2}{3}+n-1)(\tilde{\Delta}-\frac{2}{3}+n-2)\cdot 2\cdot \frac{\tilde{\Delta}}{c}L_{0}\nn\\
+(\tilde{\Delta}-\frac{2}{3}+n-1)\cdot[3\cdot 2\cdot \beta^{(112)}_{\tilde{\psi}\tilde{\psi}^{+},I}\cdot L_{0}
+\beta^{(22)}_{\tilde{\psi}\tilde{\psi}^{+},I}\cdot\wedge_{0}+\tilde{\gamma}\cdot B_{0}]\nn\\
-[4\cdot 3\cdot 2\cdot\beta^{(1112)}_{\tilde{\psi}\tilde{\psi}^{+},I}\cdot L_{0}
+4\cdot\beta^{(122)}_{\tilde{\psi}\tilde{\psi}^{+},I}\cdot\wedge_{0}\nn\\
+4\cdot\tilde{\gamma}\beta^{(1)}_{\tilde{\psi}\tilde{\psi}^{+},B}\cdot B_{0}]\}B(0)\label{eq5.22}
\eea
From (\ref{eq5.21}), the low commutation relations
take the forms :

\underline{$n=-1$}.
\bea
(\tilde{\psi}_{\frac{5}{3}}\tilde{\psi}^{+}_{-\frac{5}{3}}+D^{1}_{\frac{10}{3}}\tilde{\psi}_{\frac{2}{3}}\tilde{\psi}^{+}_{-\frac{2}{3}}+D^{2}_{\frac{10}{3}}\tilde{\psi}_{-\frac{1}{3}}\tilde{\psi}^{+}_{\frac{1}{3}}+D^{3}_{\frac{10}{3}}\tilde{\psi}_{-\frac{4}{3}}\tilde{\psi}^{+}_{\frac{4}{3}}\nn\\
-\tilde{\psi}^{+}_{\frac{5}{3}}\tilde{\psi}_{-\frac{5}{3}}-D^{1}_{\frac{10}{3}}\tilde{\psi}^{+}_{\frac{2}{3}}\tilde{\psi}_{-\frac{2}{3}}-D^{2}_{\frac{10}{3}}\tilde{\psi}^{+}_{-\frac{1}{3}}\tilde{\psi}_{\frac{1}{3}}-D^{3}_{\frac{10}{3}}\tilde{\psi}^{+}_{-\frac{4}{3}}\tilde{\psi}_{\frac{4}{3}})B(0)=R(-1)\label{eq5.23}
\eea
Further below the matrix elements
\bea
\tilde{\psi}_{\frac{2}{3}}\tilde{\psi}^{+}_{-\frac{2}{3}}B(0),\quad
\tilde{\psi}_{-\frac{1}{3}}\tilde{\psi}^{+}_{\frac{1}{3}}B(0),\quad
\tilde{\psi}_{-\frac{4}{3}}\tilde{\psi}^{+}_{\frac{4}{3}}B(0)\nn\\
\tilde{\psi}^{+}_{\frac{2}{3}}\tilde{\psi}_{-\frac{2}{3}}B(0),\quad
\tilde{\psi}^{+}_{-\frac{1}{3}}\tilde{\psi}_{\frac{1}{3}}B(0),\quad
\tilde{\psi}^{+}_{-\frac{4}{3}}\tilde{\psi}_{\frac{4}{3}}B(0)\label{eq5.24}
\eea
will be calculated directly, as well as the actions of $L_{0},\wedge_{0},B_{0}$ onto $B(0)$, in $R(n)$, eq(\ref{eq5.22}). Using those expressions, to be found below, one can check that the equation (\ref{eq5.23}) reduces to
\beq
\tilde{\psi}_{\frac{5}{3}}\tilde{\psi}^{+}_{-\frac{5}{3}}B(0)-\tilde{\psi}^{+}_{\frac{5}{3}}\tilde{\psi}_{-\frac{5}{3}}B(0)=0\label{eq5.25}
\eeq
So to say, it reduces to the statement that the matrix element $\tilde{\psi}_{\frac{5}{3}}\tilde{\psi}^{+}_{-\frac{5}{3}}B(0)$ is $Z_{3}$ conjugation symmetric, like all the lower matrix elements in (\ref{eq5.24}), to be defined directly. Otherwise (\ref{eq5.23}) does not provide any information on the value of the first non-explicit matrix element
\beq
\tilde{\psi}_{\frac{5}{3}}\tilde{\psi}^{+}_{-\frac{5}{3}}B(0)\label{eq5.26}
\eeq

\underline{$n=0$}.
\bea
(\tilde{\psi}_{\frac{8}{3}}\tilde{\psi}^{+}_{-\frac{8}{3}}+D^{1}_{\frac{10}{3}}\tilde{\psi}_{\frac{5}{3}}\tilde{\psi}^{+}_{-\frac{5}{3}}+D^{2}_{\frac{10}{3}}\tilde{\psi}_{\frac{2}{3}}\tilde{\psi}^{+}_{-\frac{2}{3}}+D^{3}_{\frac{10}{3}}\tilde{\psi}_{-\frac{1}{3}}\tilde{\psi}^{+}_{\frac{1}{3}}+D^{4}_{\frac{10}{3}}\tilde{\psi}_{-\frac{4}{3}}\tilde{\psi}^{+}_{\frac{4}{3}}\nn\\
-\tilde{\psi}^{+}_{\frac{2}{3}}\tilde{\psi}_{-\frac{2}{5}}-D^{1}_{\frac{10}{3}}\tilde{\psi}^{+}_{-\frac{1}{3}}\tilde{\psi}_{\frac{1}{3}}-D^{2}_{\frac{10}{3}}\tilde{\psi}^{+}_{-\frac{4}{3}}\tilde{\psi}_{\frac{4}{3}})B(0)=R(0)\label{eq5.27}
\eea

\underline{$n=1$}
\bea
(\tilde{\psi}_{\frac{11}{3}}\tilde{\psi}^{+}_{-\frac{11}{3}}+D^{1}_{\frac{10}{3}}\tilde{\psi}_{\frac{8}{3}}\tilde{\psi}^{+}_{-\frac{8}{3}}+...+D^{5}_{\frac{10}{3}}\cdot\tilde{\psi}_{-\frac{4}{3}}\tilde{\psi}^{+}_{\frac{4}{3}}\nn\\
-\tilde{\psi}_{-\frac{1}{3}}^{+}\tilde{\psi}_{\frac{1}{3}}-D^{1}_{\frac{10}{3}}\tilde{\psi}^{+}_{-\frac{4}{3}}\tilde{\psi}_{\frac{4}{3}})B(0)=R(1)\label{eq5.28}
\eea

\underline{$n=2$}
\beq
(\tilde{\psi}_{\frac{14}{3}}\tilde{\psi}^{+}_{-\frac{14}{3}}+...+D^{6}_{\frac{10}{3}}\tilde{\psi}_{-\frac{4}{3}}\tilde{\psi}^{+}_{\frac{4}{3}}-\tilde{\psi}^{+}_{-\frac{4}{3}}\tilde{\psi}_{\frac{4}{3}})B(0)=R(2)\label{eq5.29}
\eeq
If we denote :
\bea
m(0)=\tilde{\psi}_{-\frac{4}{3}}\tilde{\psi}^{+}_{\frac{4}{3}}B(0)\nn\\
m(1)=\tilde{\psi}_{-\frac{1}{3}}\tilde{\psi}^{+}_{\frac{1}{3}}B(0)\nn\\
m(2)=\tilde{\psi}_{\frac{2}{3}}\tilde{\psi}^{+}_{-\frac{2}{3}}B(0)\nn\\
m(3)=\tilde{\psi}_{\frac{5}{3}}\tilde{\psi}^{+}_{-\frac{5}{3}}B(0)\nn\\
m(4)=\tilde{\psi}_{\frac{8}{3}}\tilde{\psi}^{+}_{-\frac{8}{3}}B(0)\nn\\
m(5)=\tilde{\psi}_{\frac{11}{3}}\tilde{\psi}^{+}_{-\frac{11}{3}}B(0)\nn\\
m(6)=\tilde{\psi}_{\frac{14}{3}}\tilde{\psi}^{+}_{-\frac{14}{3}}B(0)\label{eq5.30}
\eea
and so on, then the higher matrix elements, starting from $m(7)$, 
could be calculated by the recurrence relation :
\beq
m(n)=R(n-4)-\sum_{k=1}^{n}D^{k}_{\frac{10}{3}}\cdot m(n-k)\label{eq5.31}
\eeq

The problem, which is contained in the first now-trivial commutation relation, $n=0$, eq.(\ref{eq5.27}), -- the problem is that (\ref{eq5.27}) contains two unknown matrix elements, $m(3)$ and $m(4)$. We could take this commutation relation as defining $m(4)$, but in terms of $m(3)$ which remains undefined. And then all the higher matrix elements will be defined, by (\ref{eq5.28}), (\ref{eq5.29}), (\ref{eq5.31}), -- all in terms of the unknown matrix element (\ref{eq5.26}), $m(3)$.

We have to define the matrix element (\ref{eq5.26}) independently, by different means. This is done in the Appendix D, its expression is given by
the eq. (\ref{eqD.70}).

Assuming that the matrix element (\ref{eq5.26}), $m(3)$, is known, we still need expressions for the low (explicit) matrix elements, $m(0)$, $m(1)$, $m(2)$ in (\ref{eq5.30}), those which could be calculated directly.

Still one remark is in order. From the expressions given below for $m(0)$, $m(1)$, $m(2)$ one observes that they are explicitly $Z_{3}$ conjugation invariant. Which means that the $Z_{3}$ conjugate matrix elements, which appear in (\ref{eq5.27}), (\ref{eq5.28}), (\ref{eq5.29}), they have the same values :
\bea
\tilde{\psi}_{-\frac{4}{3}}^{+}\tilde{\psi}_{\frac{4}{3}}B(0)=
\tilde{\psi}_{-\frac{4}{3}}\tilde{\psi}^{+}_{\frac{4}{3}}B(0)=m(0)\nn\\
\tilde{\psi}^{+}_{-\frac{1}{3}}\tilde{\psi}_{\frac{1}{3}}B(0)=
\tilde{\psi}_{-\frac{1}{3}}\tilde{\psi}^{+}_{\frac{1}{3}}B(0)=m(1)\nn\\
\tilde{\psi}^{+}_{\frac{2}{3}}\tilde{\psi}_{-\frac{2}{3}}B(0)=
\tilde{\psi}_{\frac{2}{3}}\tilde{\psi}^{+}_{-\frac{2}{3}}B(0)=m(2)\label{eq5.32}
\eea

With some calculations, which use, in particular, the actions of modes in (\ref{eq5.12})-(\ref{eq5.15}) and some commutations, we have  found 
the following expressions for the explicit matrix elements :
\beq
m(0)=\tilde{\psi}_{-\frac{4}{3}}\tilde{\psi}^{+}_{\frac{4}{3}}B(0)=\mu^{2}B(0)\label{eq5.33}
\eeq
\beq
m(1)=\tilde{\psi}_{-\frac{1}{3}}\tilde{\psi}^{+}_{\frac{1}{3}}B(0)=\mu\beta^{(1)}_{\tilde{\psi}B,\psi}\cdot(\tilde{\Delta}-\frac{4}{3})\cdot\mu\cdot B(0)\label{eq5.34}
\eeq
\bea
m(2)=\tilde{\psi}_{\frac{2}{3}}\tilde{\psi}^{+}_{-\frac{2}{3}}B(0)=\mu\cdot\beta^{(11)}_{\tilde{\psi}B,\psi}\cdot(\tilde{\Delta}-\frac{1}{3})(\tilde{\Delta}-\frac{4}{3})\cdot\mu\cdot B(0)\nn\\
+\mu\beta^{(2)}_{\tilde{\psi}B,\psi}\cdot(2\tilde{\Delta}-\frac{4}{3})\cdot\mu\cdot B(0)\nn\\
+\tilde{\gamma}\cdot(\beta^{(112)}_{\tilde{\psi}\tilde{\psi}^{+},I}\cdot\partial^{2}T(0)+\beta^{(22)}_{\tilde{\psi}\tilde{\psi}^{+},I}\cdot\wedge(0)+\tilde{\gamma}\cdot B(0))\label{eq5.35}
\eea

In addition, we need the actions of $L_{0}$, $\wedge_{0}$, $B_{0}$ onto $B(0)$, which appear in $R(n)$, eq.(\ref{eq5.22}). They are as follows :
\beq
L_{0}B(0)=4\cdot B(0)\label{eq5.36}
\eeq
\beq
\wedge_{0}B(0)=L_{0}(L_{0}+2)B(0)=24\cdot B(0)\label{eq5.37}
\eeq
\beq
B_{0}B(0)=\beta^{(112)}_{BB,I}\cdot\partial^{2}T(0)+\beta^{(22)}_{BB,I}\cdot\wedge(0)+b\cdot B(0)\label{eq5.38}
\eeq

Now we can consider all the matrix elements of the type
\bea
\tilde{\psi}_{\mu}\tilde{\psi}_{-\mu}^{+}B(0), \quad
\mu=-\frac{4}{3},-\frac{1}{3},+\frac{2}{3},+\frac{5}{3},... \label{eq5.39}
\eea
to be known, being defined by the commutation relations 
(\ref{eq5.27})--(\ref{eq5.29}), (\ref{eq5.31}).

The expansion (\ref{eq5.11}), for our correlation function, can be rewritten in the form :
\bea
<B(\infty)\tilde{\psi}(1)\tilde{\psi}^{+}(z)B(0)>\nn\\
=\frac{1}{(z)^6}\{m_{B}(0)+z\cdot m_{B}(1)+z^{2}\cdot m_{B}(2)\nn\\
+z^{3}\cdot m_{B}(3)+...\}\label{eq5.40}
\eea
where $m_{B}(n)$ are the projections of the matrix elements $m(n)$ onto $B(\infty)$ : they are the full coefficients of $B(0)$ in the expressions for $m(n)$.

We can assume that we know all the coefficients in the expansion (\ref{eq5.40}).
We can compare them with the corresponding coefficients of the expansion for the analytic form in the r.h.s. of (\ref{eq5.1}). In this way we determine the values of the coefficients $\{a_{n}\}$ of the polynomial $P_{12}(z)$ in (\ref{eq5.1}).
We have found the following values for them :
\beq
a_{0}=\mu^{2}\label{eq5.41}
\eeq
\beq
a_{1}=-\frac{29\mu^{2}}{4}\label{eq5.42}
\eeq
\beq
a_{2}=\tilde{\gamma}^{2}+\frac{(160448+10275c)\mu^{2}}{8(784+57c)}\label{eq5.43}
\eeq
\bea
a_{3}=\frac{1}{108 (784 + 57 c)^2 \zeta} 
\{  9\zeta [63 (688 + 7 c) (1024 + 27 c) \gamma^2\nn\\ + 
      4 (784 + 57 c) (10192 + 369 c) \gamma \tilde{\gamma} - 
      112 (784 + 57 c)^2 \tilde{\gamma}^2]\nn\\ + 
   6 [6 (784 + 57 c) \eta ((-4512 + 99 c) \gamma + 
         16 (784 + 57 c) \tilde{\gamma})\nn\\ + (216 b (-116 + c) (784 + 
            57 c) + (37473152 + c (14142088 + 451911 c)) \gamma\nn\\ + 
         96 (-692 + 3 c) (784 + 57 c) \tilde{\gamma}) \lambda] \mu\nn\\ - 
   2 (784 + 57 c) ((2562112 + 132249 c) \zeta + 
      108 (784 + 57 c) \tilde{\lambda}) \mu^2  \}     \label{eq5.44}
\eea

\bea
a_{4}= \frac{1}{18 c (22 + 5 c) (784 + 57 c)^2 \zeta}   
  \{3 \zeta [-112 (-200 + c) (784 + 57 c)^2\nn\\ - 
      189 c (22 + 5 c) (688 + 7 c) (1024 + 27 c) \gamma^2\nn\\ + 
      6 c (22 + 5 c) (784 + 57 c) (b (784 + 57 c)\nn\\ - 
         2 (10192 + 369 c) \gamma) \tilde{\gamma} + 
      216 c (22 + 5 c) (784 + 57 c)^2 \tilde{\gamma}^2]\nn\\ - 
6 c (22 + 5 c) [6 (784 + 57 c) \eta ((-4512 + 99 c) \gamma + 
         16 (784 + 57 c) \tilde{\gamma})\nn\\ + (216 b (-116 + c) (784 + 
            57 c) + (37473152 + c (14142088 + 451911 c)) \gamma\nn\\ + 
         96 (-692 + 3 c) (784 + 57 c) \tilde{\gamma}) \lambda] \mu\nn\\ + 
2 c (22 + 5 c) (784 + 57 c) [(768952 + 33159 c) \zeta + 
      108 (784 + 57 c) \tilde{\lambda}] \mu^2\}      \label{eq5.45}
\eea

\bea
a_{5}=  \frac{1}{36 c (22 + 5 c) (784 + 57 c)^2 \zeta}
\{   3 \zeta [3136 (-54 + c) (784 + 57 c)^2\nn\\ + 
    945 c (22 + 5 c) (688 + 7 c) (1024 + 27 c) \gamma^2\nn\\ + 
    12 c (22 + 5 c) (784 + 57 c) (-4 b (784 + 57 c) + 
       5 (10192 + 369 c) \gamma) \tilde{\gamma}\nn\\ - 
    912 c (22 + 5 c) (784 + 57 c)^2 \tilde{\gamma}^2]\nn\\ - 
 c (22 + 5 c) (784 + 57 c) (4383280 + 178797 c) \zeta \mu^2\nn\\ + 
 30 c (22 + 5 c) \mu [6 (784 + 57 c) \eta ((-4512 + 99 c) \gamma + 
       16 (784 + 57 c) \tilde{\gamma})\nn\\ + (216 b (-116 + c) (784 + 
          57 c) + (37473152 + c (14142088 + 451911 c)) \gamma\nn\\ + 
       96 (-692 + 3 c) (784 + 57 c) \tilde{\gamma}) \lambda - 
    36 (784 + 57 c)^2 \tilde{\lambda} \mu]   \}   
    \label{eq5.46}
\eea

\bea
a_{6}=   \frac{1}{108 c (22 + 5 c) (784 + 57 c)^2 \zeta}
\{   36 \zeta [38275858432\nn\\ + 
    c (4711250432 + 87308928 c - 3175470 c^2 + 48735 c^3\nn\\ - 
       4882268160 \gamma^2 - 1288012320 c \gamma^2 - 
       41856570 c^2 \gamma^2 - 297675 c^3 \gamma^2\nn\\ + 
       2 (22 + 5 c) (784 + 57 c) (9 b (784 + 57 c) - 
          10 (10192 + 369 c) \gamma) \tilde{\gamma}\nn\\ + 
       290 (22 + 5 c) (784 + 57 c)^2 \tilde{\gamma}^2)]\nn\\ - 
 120 c (22 + 5 c) [6 (784 + 57 c) \eta ((-4512 + 99 c) \gamma + 
       16 (784 + 57 c) \tilde{\gamma})\nn\\ + (216 b (-116 + c) (784 + 
          57 c) + (37473152 + c (14142088 + 451911 c)) \gamma\nn\\ + 
       96 (-692 + 3 c) (784 + 57 c) \tilde{\gamma}) \lambda] \mu\nn\\ + 
 c (22 + 5 c) (784 + 57 c) ((14819584 + 605487 c) \zeta + 
    4320 (784 + 57 c) \tilde{\lambda}) \mu^2   \}
    \label{eq5.47}
\eea
Next we have found that $a_{7}=a_{5}$, $a_{8}=a_{4}$, $a_{9}=a_{3}$,
$a_{10}=a_{2}$, $a_{11}=a_{1}$, $a_{12}=a_{0}$.

With the values of the coefficients found, one concludes that the conditions (\ref{eq5.9}), (\ref{eq5.10}), of the limit $z\rightarrow\infty$, all get verified.

One could also check, with the values of the coefficients given above,
that the condition (\ref{eq5.5}), of the limit $z\rightarrow 1$, is also verified. With the present level of confidence in the operator algebra we consider this verification, of the limit $z\rightarrow 1$, to be sufficient. So to say, we skip the complete calculation of the present function by the $z\rightarrow 1$ limit expansion (expansion in $1-z$), the way we did it for different functions before, starting with the function $<\psi^{+}(\infty)\psi(1)\psi(z)\psi^{+}(0)>$ in Section 2.2.

The verification of the limit $z\rightarrow\infty$, on the other hand, is complete. Due to the symmetry of our function with respect to the limits $z\rightarrow 0$ and $z\rightarrow\infty$, the verification of the conditions in eq.(\ref{eq5.10}) is sufficient.

We observe it again that all the consistency conditions have been verified 
for the \underline{unconstrained values of the constants}, in accordance with our classifications in Sec. 3.

\section{Conclusions.}

In this paper we have given the proof, which we consider to be
complete, of the associativity of the chiral algebra $Z^{(3)}_{3}$
generated by the principal parafermionic fields $\psi(z)$,
$\psi^{+}(z)$ with the conformal dimensions
$\Delta_{\psi}=\Delta_{\psi^{+}}=8/3$.

The principal guidelines of the methods and classifications are given
in Sections 1 and 3, with some additional precisions in Section
4.1. In the rest of the paper the methods of the associativity
calculations, suggested in Section 1 and 3, are just realized, in much
detail, in the context of the chiral algebra $Z^{(3)}_{3}$.

This algebra closes by the fields :
\beq
\psi(z),\,\,\psi^{+}(z),\,\,\tilde{\psi}(z),\,\,\tilde{\psi}^{+}(z),\,\,B(z)\label{eq6.1}
\eeq
having dimensions :
\beq
\Delta_{\psi}=\Delta_{\psi^{+}}=8/3,\quad\Delta_{\tilde{\psi}}=\Delta_{\tilde{\psi}^{+}}=\Delta_{\psi}+2=14/3,\quad\Delta_{B}=4\label{eq6.2}
\eeq 

 It has been stated in Section 1 that the associativity solution with the fields in (\ref{eq6.1}) is a minimal one, with respect to the number of fields. In addition to this minimal solution, we have found two non-minimal chiral
algebras, we call them solutions 2 and 3 (mutually different in the values of the operator algebra constants) in which the operator algebra closes by the fields :
 \bea
 \psi(z),\,\,\psi^{+}(z),\,\,\tilde{\psi}(z),\,\,\tilde{\psi}^{+}(z),\nn\\
U(z),\,\,B(z),\,\,W(z)\label{eq6.3}
\eea
with dimensions :
\bea
\Delta_{\psi}=\Delta_{\psi^{+}}=8/3,\quad\Delta_{\tilde{\psi}}=\Delta_{\tilde{\psi}^{+}}=14/3,\nn\\
\Delta_{U}=3,\,\,\,\Delta_{B}=4,\,\,\,\Delta_{W}=5\label{eq6.4}
\eea
These two other solutions will be presented in the next publication [6].

We would like to stress, in conclusion, that the solution 1, presented here, and the solutions 2 and 3 in [6], they make a full set. This is in the sense 
that there could be no other associative parafermionic chiral algebras, realizing the $Z_{3}$ symmetry, with dimensions $\Delta_{\psi}=\Delta_{\psi^{+}}=8/3$ of the principal fields $\psi,\psi^{+}$, and having the central charge unconstrained, as a free parameter.

\vskip1cm

{\bf Acknowledgments.}

I am particularly grateful to Raoul Santachiara. In the present paper
I have developed somewhat our initial methods which led us to the algebra (\ref{eq1.1})--(\ref{eq1.9}), [1]. 

I am also grateful to Benoit Estienne and Raoul Santachiara for
numerous discussions on the approaches to associativity calculations
presented in this paper.

All kinds of help and encouragement from Marco Picco,
in the process of this work, were extremely helpful.

\appendix
\section{ Values of the $\beta$ coefficients.}

In this Appendix we shall provide the values of the $\beta$ coefficients
which are needed for the calculations of the main text. 
They are as follows:
\bea
\beta^{(2)}_{\psi \psi^{+},I} = \frac{16}{3 c}, \quad
\beta^{(12)}_{\psi \psi^{+},I} = \frac{8}{3 c}, \quad
\beta^{(112)}_{\psi \psi^{+},I} = \frac{4 (-4 + 3 c)}{3 c (22 + 5 c)}, \nn\\
\beta^{(22)}_{\psi \psi^{+},I} = \frac{688}{9 c (22 + 5 c)}, \quad
\beta^{(1112)}_{\psi \psi^{+},I} = \frac{4 (-17 + 2 c)}{9 c (22 + 5 c)}, \quad
\beta^{(122)}_{\psi \psi^{+},I} = \frac{344}{9 c (22 + 5 c)}
\eea
\beq
\beta^{(1)}_{\psi \psi, \psi^{+}} = \frac{1}{2}, \quad
\beta^{(11)}_{\psi \psi, \psi^{+}} = \frac{128 + 33 c}{4 (784 + 57 c)}, \quad
\beta^{(2)}_{\psi \psi, \psi^{+}} = \frac{344}{784 + 57 c}
\eeq
\beq
\beta^{(1)}_{\psi \psi, \tilde{\psi}^{+}} = \frac{1}{2}, \quad
\beta^{(11)}_{\psi \psi, \tilde{\psi}^{+}} = \frac{1112 + 51 c}{4 (2716 + 93 c)}, 
\quad
\beta^{(2)}_{\psi \psi, \tilde{\psi}^{+}} = \frac{650}{2716 + 93 c}
\eeq
\beq
\beta^{(1)}_{\psi \tilde{\psi}, \psi^{+}} = \frac{1}{8}, \quad
\beta^{(11)}_{\psi \tilde{\psi}, \psi^{+}} = \frac{5 (-224 + 3 c)}{16 (784 + 57 c)},
\quad
\beta^{(2)}_{\psi \tilde{\psi}, \psi^{+}} = \frac{350}{784 + 57 c}
\eeq
\beq
\beta^{(1)}_{\psi \tilde{\psi}, \tilde{\psi}^{+}} = \frac{2}{7}, \quad
\beta^{(11)}_{\psi \tilde{\psi}, \tilde{\psi}^{+}} = \frac{224 + 33 c}{7 (2716 + 93 c)}, \quad
\beta^{(2)}_{\psi \tilde{\psi}, \tilde{\psi}^{+}} = \frac{728}{2716 + 93 c}
\eeq
\beq
\beta^{(1)}_{\tilde{\psi} \psi, \psi^{+}} = \frac{7}{8}, \quad
\beta^{(11)}_{\tilde{\psi} \psi, \psi^{+}} = \frac{7 (512 + 51 c)}{16 (784 + 57 c)},
\quad
\beta^{(2)}_{\tilde{\psi} \psi, \psi^{+}} = \frac{350}{784 + 57 c}
\eeq
\beq
\beta^{(1)}_{\tilde{\psi} \tilde{\psi}, \psi^{+}} = \frac{1}{2}, \quad
\beta^{(11)}_{\tilde{\psi} \tilde{\psi}, \psi^{+}} = \frac{11 (-8 + 3 c)}{4 (784 + 57 c)}, \quad
\beta^{(2)}_{\tilde{\psi} \tilde{\psi}, \psi^{+}} = \frac{572}{784 + 57 c}
\eeq
\bea
\beta^{(2)}_{\tilde{\psi} \tilde{\psi}^{+}, I} = \frac{28}{3 c}, \quad
\beta^{(12)}_{\tilde{\psi} \tilde{\psi}^{+}, I} = \frac{14}{3 c}, \quad
\beta^{(112)}_{\tilde{\psi} \tilde{\psi}^{+}, I} = \frac{7 (-16 + 3 c)}{3 c (22 + 5 c)}, \nn\\
\beta^{(22)}_{\tilde{\psi} \tilde{\psi}^{+}, I} = \frac{2044}{9 c (22 + 5 c)}, \quad
\beta^{(1112)}_{\tilde{\psi} \tilde{\psi}^{+}, I} = \frac{7 (-35 + 2 c)}{9 c (22 + 5 c)}, \quad
\beta^{(122)}_{\tilde{\psi} \tilde{\psi}^{+}, I} = \frac{1022}{9 c (22 + 5 c)}
\eea
\beq
\beta^{(1)}_{\psi B, \psi} = \frac{1}{4}, \quad
\beta^{(11)}_{\psi B, \psi} = \frac{-416 + 21 c}{8 (784 + 57 c)}, \quad
\beta^{(2)}_{\psi B, \psi} = \frac{372}{784 + 57 c}
\eeq
\beq
\beta^{(1)}_{B \psi, \psi} = \frac{3}{4}, \quad
\beta^{(11)}_{B \psi, \psi} = \frac{9 (128 + 15 c)}{8 (784 + 57 c)}, \quad
\beta^{(2)}_{B \psi, \psi} = \frac{372}{784 + 57 c}
\eeq
\beq
\beta^{(1)}_{B \tilde{\psi}, \psi} = \frac{3}{8}, \quad
\beta^{(11)}_{B \tilde{\psi}, \psi} = \frac{27 (-32 + 3 c)}{16 (784 + 57 c)}, \quad
\beta^{(2)}_{B \tilde{\psi}, \psi} = \frac{522}{784 + 57 c}
\eeq
\beq
\beta^{(1)}_{\tilde{\psi} B, \psi} = \frac{5}{8}, \quad
\beta^{(11)}_{\tilde{\psi} B, \psi} = \frac{704 + 195 c}{16 (784 + 57 c)}, \quad
\beta^{(2)}_{\tilde{\psi} B, \psi} = \frac{522}{784 + 57 c}
\eeq
\bea
\beta^{(2)}_{B B, I} = \frac{8}{c}, \quad
\beta^{(12)}_{B B, I} = \frac{4}{c}, \quad
\beta^{(112)}_{B B, I} = \frac{6 (-4 + c)}{c (22 + 5 c)}, \nn\\
\beta^{(22)}_{B B, I} = \frac{168}{c (22 + 5 c)}, \quad
\beta^{(1112)}_{B B, I} = \frac{-58 + 4 c}{c(66 + 15 c)}, \quad
\beta^{(122)}_{B B, I} = \frac{84}{c (22 + 5 c)}
\eea

Finally we shall put into a single line several coefficients belonging 
to diferent OPEs, the expansions for which we need only one (first) coefficient:
\beq
\beta^{(1)}_{\psi \psi^{+}, B} =\frac {1}{2}, \quad
\beta^{(1)}_{\psi \tilde{\psi}^{+}, B} = \frac{1}{4}, \quad
\beta^{(1)}_{\tilde{\psi} \psi^{+}, B} = \frac{3}{4}, \quad
\beta^{(1)}_{\tilde{\psi} \tilde{\psi}^{+},B} = \frac{1}{2}, \quad
\beta^{(1)}_{B B, B} = \frac{1}{2}
\eeq

\section{Calculation of the matrix elements (\ref{eq2.69}), (\ref{eq2.70}).}

In the calculation which follow will be useful the following relations :

\beq
[\psi_{\mu},L_{n}]=(\mu-n(\Delta-1))\psi_{\mu+n}\label{eqB.1}
\eeq
More generally :
\beq
[\Phi_{m},L_{n}]=(m-n(\Delta_{\Phi}-1))\Phi_{m+n}\label{eqB.2}
\eeq
Next,
\beq
\wedge_{n}=\sum^{\infty}_{l=0}(L_{-2-l}L_{n+l+2}+L_{n-l+1}L_{l-1})\label{eqB.3}
\eeq
In particular :
\beq
\wedge_{0} = L_{0}(L_{0}+2)+2\sum^{\infty}_{l=1}L_{-l}L_{l}\label{eqB.4}
\eeq
Next,
\beq
(L_{-2}\psi^{+})_{n}\psi^{+}(0)=\sum^{\infty}_{l=0}(L_{-2-l}\psi^{+}_{n+l+2}+\psi^{+}_{n-l+1}L_{l-1})\psi^{+}(0)\label{eqB.5}
\eeq
\beq
\psi_{-\frac{8}{3}}I=\psi(0),\quad\psi_{-\frac{11}{3}}I=\partial\psi(0),\quad\psi_{-\frac{14}{3}}I=\frac{1}{2}\partial^{2}\psi(0)\label{B.6}
\eeq
\bea
\psi_{-\frac{2}{3}}T(0)=\Delta\psi(0),\nn\\
\psi_{-\frac{5}{3}}T(0)=(\Delta-1)\partial\psi(0),\nn\\
\psi_{-\frac{8}{3}}T(0)=(\frac{\Delta}{2}-1)\partial^{2}\psi(0)+L_{-2}\Psi(0)\label{eqB.7}
\eea

Now we turn to the calculation of the matrix elements (\ref{eq2.69}). We shall be using the action of modes in equations ({\ref{eq2.23})-(\ref{eq2.29}).

One finds :
\beq
\psi_{-\frac{8}{3}}\psi_{\frac{8}{3}}\psi^{+}(0)=\psi_{-\frac{8}{3}}I=\psi(0)\label{eqB.8}
\eeq
\beq
\psi_{-\frac{5}{3}}\psi_{\frac{5}{3}}\psi^{+}(0)=\psi_{-\frac{5}{3}}0=0\label{eqB.8a}
\eeq
\beq
\psi_{-\frac{2}{3}}\psi_{\frac{2}{3}}\psi^{+}(0)=\psi_{-\frac{2}{3}}\frac{2\Delta}{c}T(0)=\frac{2\Delta^{2}}{c}\cdot\psi(0)\label{eqB.9}
\eeq

\bea
\psi_{\frac{1}{3}}\psi_{-\frac{1}{3}}\psi^{+}(0)=\psi_{\frac{1}{3}}\frac{\Delta}{c}\partial T(0)=\frac{\Delta}{c}\psi_{\frac{1}{3}}\partial T(0)\nn\\
=\frac{\Delta}{c}[(\frac{1}{3}+(\Delta-1))\psi_{-\frac{2}{3}}T(0)+\partial\psi_{\frac{1}{3}}T(0)]\nn\\
=\frac{\Delta}{c}[(\Delta-\frac{2}{3})\psi_{-\frac{2}{3}}T(0)+0]\nn\\
=\frac{\Delta^{2}}{c}(\Delta-\frac{2}{3})\psi(0)\label{eqB.10}
\eea

\bea
\psi_{\frac{4}{3}}\psi_{-\frac{4}{3}}\psi^{+}(0)=\psi_{\frac{4}{3}}(\beta^{(112)}_{\psi\psi^{+},I}\cdot\partial^{2}T(0)+\beta^{(22)}_{\psi\psi^{+},I}\cdot\wedge(0)+\gamma\cdot B(0))\nn\\
\psi_{\frac{4}{3}}\partial^{2}T(0)=(\frac{4}{3}+(\Delta-1))\psi_{\frac{1}{3}}\partial T(0)+\partial\psi_{\frac{4}{3}}\partial T(0)\nn\\
=(\Delta+\frac{1}{3})\cdot(\Delta-\frac{2}{3})\cdot\psi_{-\frac{2}{3}}T(0) + (\Delta+\frac{1}{3})\partial\psi_{\frac{1}{3}}T(0)+\partial\psi_{\frac{4}{3}}\partial T(0)\nn\\
=(\Delta+\frac{1}{3})(\Delta-\frac{2}{3})\Delta\psi(0)+0+0\nn\\
\psi_{\frac{4}{3}}\wedge(0)=\psi_{\frac{4}{3}}L_{-2}T(0)=(\frac{4}{3}+2\cdot(\Delta-1))\psi_{-\frac{2}{3}}T(0)+L_{-2}\psi_{\frac{4}{3}}T(0)\nn\\
=(2\Delta-\frac{2}{3})\Delta\psi(0)+0\nn\\
\psi_{\frac{4}{3}}B(0)=\gamma\cdot\psi(0)\nn\\
\psi_{\frac{4}{3}}\psi_{-\frac{4}{3}}\psi^{+}(0)=\beta^{(112)}_{\psi\psi^{+},I}\cdot(\Delta+\frac{1}{3})(\Delta-\frac{2}{3})\Delta\psi(0)\nn\\
+\beta^{(22)}_{\psi\psi^{+},I}\cdot(2\Delta-\frac{2}{3})\Delta\psi(0)+\gamma^{2}\cdot\psi(0)\label{eqB.11}
\eea

\bea
\psi_{\frac{7}{3}}\psi_{-\frac{7}{3}}\psi^{+}(0)=\psi_{\frac{7}{3}}(\beta^{(1112)}_{\psi\psi^{+},I}\cdot\partial^{3}T(0)+\beta^{(122)}_{\psi\psi^{+},I}\cdot\partial\wedge(0)+\gamma\beta^{(1)}_{\psi\psi^{+},B}\cdot\partial B(0))\nn\\
\psi_{\frac{7}{3}}\partial^{3}T(0)=(\frac{7}{3}+(\Delta-1))\psi_{\frac{4}{3}}\partial^{2} T(0)+0\nn\\
=(\Delta+\frac{4}{3})\cdot(\Delta+\frac{1}{3})\cdot\psi_{\frac{1}{3}}\partial T(0)+0\nn\\
=(\Delta+\frac{4}{3})(\Delta+\frac{1}{3})(\Delta-\frac{2}{3})\psi_{-\frac{2}{3}}T(0)+0\nn\\
=(\Delta+\frac{4}{3})(\Delta+\frac{1}{3})(\Delta-\frac{2}{3})\Delta\psi(0)\nn\\
\psi_{\frac{7}{3}}\partial\wedge(0)=(\frac{7}{3}+(\Delta-1))\psi_{\frac{4}{3}}\wedge(0)+\partial\psi_{\frac{7}{3}}\wedge(0)\nn\\
=(\Delta+\frac{4}{3})\cdot(2\Delta-\frac{2}{3})\Delta\psi(0)+0\nn\\
\psi_{\frac{7}{3}}\partial B(0)=(\frac{7}{3}+(\Delta-1))\psi_{\frac{4}{3}}B(0)+\partial\psi_{\frac{7}{3}}B(0)\nn\\
=(\Delta+\frac{4}{3})\gamma\psi(0)+0\nn\\
\psi_{\frac{7}{3}}\psi_{-\frac{7}{3}}\psi^{+}(0)=\beta^{(1112)}_{\psi\psi^{+},I}\cdot(\Delta+\frac{4}{3})(\Delta+\frac{1}{3})(\Delta-\frac{2}{3})\Delta\psi(0)\nn\\
+\beta^{(122)}_{\psi\psi^{+},I}\cdot(\Delta+\frac{4}{3})\cdot(2\Delta-\frac{2}{3})\cdot\Delta\psi(0)\nn\\
+\gamma^{2}\beta^{(1)}_{\psi\psi^{+},I}\cdot(\Delta+\frac{4}{3})\cdot\psi(0)\label{eqB.12}
\eea

Next, the matrix elements in (\ref{eq2.70}).
\beq
\psi^{+}_{0}\psi^{+}(0)=\lambda\psi(0)\label{eqB.13}
\eeq
\bea
(L_{-2}\psi^{+})_{0}\psi^{+}(0)=\sum^{\infty}_{l=0}(L_{-2-l}\psi^{+}_{l+2}+\psi^{+}_{-l+1}L_{l-1})\psi^{+}(0)\nn\\
=(L_{-2}\psi^{+}_{2}+...+\psi^{+}_{1}L_{-1}+\psi^{+}_{0}L_{0})\psi^{+}(0)\nn\\
=0+\psi^{+}_{1}L_{-1}\psi^{+}(0)+\psi^{+}_{0}\Delta\psi^{+}(0)\nn\\
=(1+(\Delta-1))\psi^{+}_{0}\psi^{+}(0)+L_{-1}\psi^{+}_{1}\psi^{+}(0)+\Delta\psi^{+}_{0}\psi^{+}(0)\nn\\
=\Delta\cdot\lambda\psi(0)+0+\Delta\cdot\lambda\psi(0)\nn\\
=2\Delta\lambda\psi(0)\label{eqB.14}
\eea
\beq
\tilde{\psi}^{+}_{0}\psi^{+}(0)=\zeta\cdot\psi(0)\label{eqB.15}
\eeq

\section{Calculation of the matrix elements (\ref{eq2.3.13}) and (\ref{eq2.3.16}).}

We calculate first the matrix elements (\ref{eq2.3.13}). We shall be using the actions of modes in (\ref{eq2.3.8})-(\ref{eq2.3.10}) and the commutations (\ref{eqB.1}).
\bea
\psi_{-\frac{2}{3}}\psi_{\frac{2}{3}}\tilde{\psi}^{+}(0)=\psi_{-\frac{2}{3}}\mu B(0)\nn\\
=\mu(\gamma\beta^{(11)}_{\psi B,\psi}\cdot\partial^{2}\psi(0)+\gamma\beta^{(2)}_{\psi B,\psi}\cdot L_{-2}\psi(0)+\mu\tilde{\psi}(0))\label{eqC.1}
\eea
\bea
\psi_{\frac{1}{3}}\psi_{-\frac{1}{3}}\tilde{\psi}^{+}(0)=\psi_{\frac{1}{3}}\mu\beta^{(1)}_{\psi\tilde{\psi}^{+},B}\cdot\partial B(0)\nn\\
=\mu\beta^{(1)}_{\psi\tilde{\psi}^{+},B}\cdot[(\frac{1}{3}+(\Delta-1))\psi_{-\frac{2}{3}}B(0)+\partial\psi_{\frac{1}{3}}B(0)]\nn\\
=\mu\beta^{(1)}_{\psi\tilde{\psi}^{+},B}\cdot[(\Delta-\frac{2}{3})(\gamma\beta^{(11)}_{\psi B,\psi}\cdot\partial^{2}\psi(0)+\gamma\beta^{(2)}_{\psi B,\psi}L_{-2}\psi(0)\nn\\
+\mu\tilde{\psi}(0)+\gamma\beta^{(1)}_{\psi B,\psi}\cdot\partial^{2}\psi(0)]\label{eqC.2}
\eea
We have used in addition the action of modes $\psi_{-\frac{2}{3}}B(0)$ and $\psi_{\frac{1}{3}}B(0)$, which should be evident from the operator product expansion of $\psi(z)B(0)$ in (\ref{eq1.5}).

Next we calculate the matrix elements (\ref{eq2.3.16}).
\beq
\psi^{+}_{0}\tilde{\psi}^{+}(0)=\zeta\beta^{(11)}_{\psi\tilde{\psi},\psi^{+}}\cdot\partial^{2}\psi(0)+\zeta\beta^{(2)}_{\psi\tilde{\psi},\psi^{+}}L_{-2}\psi(0)+\eta\tilde{\psi}(0)\label{eqC.3}
\eeq
-- cf. operator product expansion (\ref{eq1.3});
\bea
(L_{-2}\psi^{+})_{0}\tilde{\psi}^{+}(0)=\sum^{\infty}_{l=0}(L_{-2-l}\psi^{+}_{2+l}+\psi^{+}_{-l+1}L_{l-1})\tilde{\psi}^{+}(0)\nn\\
=(L_{-2}\psi^{+}_{2}+L_{-3}\psi^{+}_{3}+...+\psi^{+}_{1}L_{-1}+\psi^{+}_{0}L_{0})\tilde{\psi}^{+}(0)\nn\\
=L_{-2}\zeta\psi(0)+0+(1+(\Delta-1))\psi^{+}_{0}\tilde{\psi}^{+}(0)+L_{-1}\psi^{+}_{1}\tilde{\psi}^{+}(0)+\psi^{+}_{0}\tilde{\Delta}\tilde{\psi}^{+}(0)\nn\\
=\zeta\cdot L_{-2}\psi(0)+(\Delta+\tilde{\Delta})\cdot(\zeta\beta^{(11)}_{\psi\tilde{\psi},\psi^{+}}\partial^{2}\psi(0)+\zeta\beta^{(2)}_{\psi\tilde{\psi},\psi^{+}}L_{-2}\psi(0)\nn\\
+\eta\tilde{\psi}(0))+\zeta\beta^{(1)}_{\psi\tilde{\psi},\psi^{+}}\partial^{2}\psi(0)\label{eqC.4}
\eea
-- we have used the relation (\ref{eqB.5}), with $\psi^{+}(0)$ replaced by $\tilde{\psi}^{+}(0)$;
\beq
\tilde{\psi}^{+}_{0}\tilde{\psi}^{+}(0)=\eta\beta^{(11)}_{\tilde{\psi}\tilde{\psi},\psi^{+}}\partial^{2}\psi(0)+\eta\beta^{(2)}_{\tilde{\psi}\tilde{\psi},\psi^{+}}L_{-2}\psi(0)+\tilde{\lambda}\tilde{\psi}(0)\label{eqC.5}
\eeq

\section{Calculation of the matrix element $\tilde{\psi}_{\frac{5}{3}}\tilde{\psi}^{+}_{-\frac{5}{3}}B(0)$.}

We start by factorizing $\tilde{\psi}^{+}_{-\frac{5}{3}}B(0)$, in the matrix element $\tilde{\psi}_{\frac{5}{3}}\tilde{\psi}^{+}_{-\frac{5}{3}}B(0)$, by using the commutation relation $\{\psi,\psi\}B(0)$. Its general form :
\beq
\sum^{\infty}_{l=0}D^{l}_{-\frac{1}{3}}(\psi_{n-1-l}\psi_{-\frac{2}{3}+m+l}+\psi_{m-1-l}\psi_{-\frac{2}{3}+n+l})B(0)=R(n,m)\label{eqD.1}
\eeq
\bea
R(n,m)=\{\frac{1}{2}(\Delta-\frac{2}{3}+n-1)(\Delta-\frac{2}{3}+n-2)\cdot\lambda\psi^{+}_{-\frac{5}{3}+n+m}\nn\\
-(\Delta-\frac{2}{3}+n-1)(n+m+1)\lambda\beta^{(1)}_{\psi\psi,\psi^{+}}\cdot\psi^{+}_{-\frac{5}{3}+n+m}\nn\\
+(n+m+2)(n+m+1)\lambda\beta^{(11)}_{\psi\psi,\psi^{+}}\cdot\psi^{+}_{-\frac{5}{3}+n+m}\nn\\
+\lambda\beta^{(2)}_{\psi\psi,\psi^{+}}\cdot(L_{-2}\psi^{+})_{-\frac{5}{3}+n+m}+\zeta\tilde{\psi}^{+}_{-\frac{5}{3}+n+m}\}B(0)\label{eqD.2}
\eea
From this commutation relation we want to express $\tilde{\psi}^{+}_{-\frac{5}{3}}B(0)$. So we put
\beq
n+m=0,\quad m=-n\label{eqD.3}
\eeq
The expressions in (\ref{eqD.1}), (\ref{eqD.2}) take the form :
\beq
\sum^{\infty}_{l=0}D^{l}_{-\frac{1}{3}}(\psi_{n-1-l}\psi_{-\frac{2}{3}-n+l}+\psi_{-n-1-l}\psi_{-\frac{2}{3}+n+l})B(0)=R(n)\label{eqD.4}
\eeq
\bea
R(n)=R(n,-n)=\{\frac{1}{2}(\Delta-\frac{2}{3}+n-1)(\Delta-\frac{2}{3}+n-2)\cdot\lambda\psi^{+}_{-\frac{5}{3}}\nn\\
-(\Delta-\frac{2}{3}+n-1)\lambda\beta^{(1)}_{\psi\psi,\psi^{+}}\cdot\psi^{+}_{-\frac{5}{3}}\nn\\
+2\lambda\beta^{(11)}_{\psi\psi,\psi^{+}}\cdot\psi^{+}_{-\frac{5}{3}}+\lambda\beta^{(2)}_{\psi\psi,\psi^{+}}\cdot(L_{-2}\psi^{+})_{-\frac{5}{3}}+\zeta\tilde{\psi}^{+}_{-\frac{5}{3}}\}B(0)\label{eqD.5}
\eea
We choose, in addition, $n=0$, because the l.h.s. of (\ref{eqD.4}) is simpler with this choice, and we express $\tilde{\psi}^{+}_{-\frac{5}{3}}B(0)$ in terms of the rest. We obtain :
\bea
\zeta\tilde{\psi}^{+}_{-\frac{5}{3}}B(0)=\lambda\beta^{(1)}_{\psi\psi,\psi^{+}}\cdot\psi^{+}_{-\frac{5}{3}}B(0)\nn\\
-2\lambda\beta^{(11)}_{\psi\psi,\psi^{+}}\cdot\psi^{+}_{-\frac{5}{3}}B(0)-\lambda\beta^{(2)}_{\psi\psi,\psi^{+}}\cdot(L_{-2}\psi^{+})_{-\frac{5}{3}}B(0)\nn\\
+2\cdot\psi_{-1}\psi_{-\frac{2}{3}}B(0)+2D^{1}_{-\frac{1}{3}}\cdot\psi_{-2}\psi_{\frac{1}{3}}B(0)+2D^{2}_{-\frac{1}{3}}\cdot\psi_{-3}\psi_{\frac{4}{3}}B(0)\label{eqD.6}
\eea
Multiplying this equation by $\tilde{\psi}_{\frac{5}{3}}$, on the left, one obtains :
\bea
\zeta\tilde{\psi}_{\frac{5}{3}}\tilde{\psi}^{+}_{-\frac{5}{3}}B(0)=\lambda\beta^{(1)}_{\psi\psi,\psi^{+}}\cdot\tilde{\psi}_{\frac{5}{3}}\psi^{+}_{-\frac{5}{3}}B(0)\nn\\
-2\lambda\beta^{(11)}_{\psi\psi,\psi^{+}}\cdot\tilde{\psi}_{\frac{5}{3}}\psi^{+}_{-\frac{5}{3}}B(0)-\lambda\beta^{(2)}_{\psi\psi,\psi^{+}}\cdot\tilde{\psi}_{\frac{5}{3}}(L_{-2}\psi^{+})_{-\frac{5}{3}}B(0)\nn\\
+2\cdot\tilde{\psi}_{\frac{5}{3}}\psi_{-1}\psi_{-\frac{2}{3}}B(0)+2D^{1}_{-\frac{1}{3}}\cdot\tilde{\psi}_{\frac{5}{3}}\psi_{-2}\psi_{\frac{1}{3}}B(0)+2D^{2}_{-\frac{1}{3}}\cdot\tilde{\psi}_{\frac{5}{3}}\psi_{-3}\psi_{\frac{4}{3}}B(0)\label{eqD.7}
\eea
We have expressed our matrix element in terms of simpler (lower) ones. We have the following list of lower matrix elements to be calculated :
\beq
M1=\tilde{\psi}_{\frac{5}{3}}\psi^{+}_{-\frac{5}{3}}B(0)\label{eqD.8}
\eeq
\beq
M2=\tilde{\psi}_{\frac{5}{3}}(L_{-2}\psi^{+})_{-\frac{5}{3}}B(0)\label{eqD.9}
\eeq
\beq
M3=\tilde{\psi}_{\frac{5}{3}}\psi_{-3}\psi_{\frac{4}{3}}B(0)\label{eqD.10}
\eeq
\beq
M4=\tilde{\psi}_{\frac{5}{3}}\psi_{-2}\psi_{\frac{1}{3}}B(0)\label{eqD.11}
\eeq
\beq
M5=\tilde{\psi}_{\frac{5}{3}}\psi_{-1}\psi_{-\frac{2}{3}}B(0)\label{eqD.12}
\eeq

\underline{$M1=\tilde{\psi}_{\frac{5}{3}}\psi^{+}_{-\frac{5}{3}}B(0)$}.

This matrix element is defined by the eq(\ref{eq4.14}).
After some calculation, which uses the expressions obtained in Section 4.1 
and the values of $\beta$ coefficients in Appendix A, one finds :
\bea
M1=\tilde{\psi}_{\frac{5}{3}}\psi^{+}_{-\frac{5}{3}}B(0)
=\frac{\mu}{162}  \{\frac{90 (-62 + 39 c) \partial^2 T(0) 
+ 70080 \wedge(0)}{c (22 + 5 c)} \nn\\
+ \frac{162 b (784 + 57 c) + (387280 + 1407 c) \gamma + 216 (784 + 57 c) \tilde{\gamma}}{784 + 57 c}B(0)\}
\label{eqD.13}
\eea

As we shall use all these matrix elements to calculate, eventually, the correlation function $<B\tilde{\psi}\tilde{\psi}^{+}B>$, we need only the expressions projected onto $B(\infty)$, which are simpler. The projected form of the matrix element $M1$ above is as follows :
\bea
M1_{B}=(\tilde{\psi}_{\frac{5}{3}}\psi^{+}_{-\frac{5}{3}}B(0))_{B}\nn\\
= \frac{\mu}{162} \frac{162 b (784 + 57 c) + (387280 + 1407 c) \gamma + 216 (784 + 57 c) \tilde{\gamma}}{784 + 57 c}
\label{eqD.14}
\eea
\vskip3cm
\underline{$M2=\tilde{\psi}_{\frac{5}{3}}(L_{2}\psi^{+})_{-\frac{5}{3}}B(0)$}.

\bea
(L_{-2}\psi^{+})_{-\frac{5}{3}}B(0)=\sum^{\infty}_{l=0}(L_{-2-l}\psi^{+}_{-\frac{5}{3}+l+2}+\psi^{+}_{-\frac{5}{3}-l+1}L_{l-1})B(0)\nn\\
=(L_{-2}\psi^{+}_{\frac{1}{3}}+L_{-3}\psi^{+}_{\frac{4}{3}}+\psi^{+}_{-\frac{2}{3}}L_{-1}+\psi^{+}_{-\frac{5}{3}}L_{0})B(0)\nn\\
=L_{-2}\psi^{+}_{\frac{1}{3}}B(0)+L_{-3}\psi^{+}_{\frac{4}{3}}B(0)+(\Delta-\frac{5}{3})\psi^{+}_{-\frac{5}{3}}B(0)\nn\\
+L_{-1}\psi^{+}_{-\frac{2}{3}}B(0)+4\cdot\psi^{+}_{-\frac{5}{3}}B(0)\label{eqD.15}
\eea
We have used the relation (\ref{eqB.5}), generalized.
\bea
M2=\tilde{\psi}_{\frac{5}{3}}(L_{-2}\psi^{+})_{-\frac{5}{3}}B(0)=\tilde{\psi}_{\frac{5}{3}}L_{-2}\psi^{+}_{\frac{1}{3}}B(0)+\tilde{\psi}_{\frac{5}{3}}L_{-3}\psi^{+}_{\frac{4}{3}}B(0)\nn\\
+\tilde{\psi}_{\frac{5}{3}}\psi^{+}_{-\frac{5}{3}}B(0)+\tilde{\psi}_{\frac{5}{3}}L_{-1}\psi^{+}_{-\frac{2}{3}}B(0)+4\tilde{\psi}_{\frac{5}{3}}\psi^{+}_{-\frac{5}{3}}B(0)\label{eqD.16}
\eea
To define $M2$, we have to calculate the matrix elements :
\beq
M2.1=\tilde{\psi}_{\frac{5}{3}}L_{-3}\psi^{+}_{\frac{4}{3}}B(0)\label{eqD.17}
\eeq
\beq
M2.2=\tilde{\psi}_{\frac{5}{3}}L_{-2}\psi^{+}_{\frac{1}{3}}B(0)\label{eqD.18}
\eeq
\beq
M2.3=\tilde{\psi}_{\frac{5}{3}}L_{-1}\psi^{+}_{-\frac{2}{3}}B(0)\label{eqD.19}
\eeq
\underline{$M2.1$}.
\bea
\tilde{\psi}_{\frac{5}{3}}L_{-3}\psi^{+}_{\frac{4}{3}}B(0)=\tilde{\psi}_{\frac{5}{3}}L_{-3}\gamma\psi^{+}(0)\nn\\
=\gamma\cdot(\frac{5}{3}+3(\tilde{\Delta}-1))\tilde{\psi}_{-\frac{4}{3}}\psi^{+}(0)+\gamma L_{-3}\tilde{\psi}_{\frac{5}{3}}\psi^{+}(0)\nn\\
=\gamma\cdot(3\tilde{\Delta}-\frac{4}{3})\cdot\mu B(0)+0\label{eqD.20}
\eea
\underline{$M2.2$}
\bea
\tilde{\psi}_{\frac{5}{3}}L_{-2}\psi^{+}_{\frac{1}{3}}B(0)=\tilde{\psi}_{\frac{5}{3}}L_{-2}\gamma\beta^{(1)}_{\psi B,\psi}\cdot\partial\psi^{+}(0)\nn\\
=\gamma\beta^{(1)}_{\psi B,\psi}\cdot(\frac{5}{3}+2(\tilde{\Delta}-1))\tilde{\psi}_{-\frac{1}{3}}\partial\psi^{+}(0)+\gamma\beta^{(1)}_{\psi B\psi} L_{-2}\tilde{\psi}_{\frac{5}{3}}\partial\psi^{+}(0)\nn\\
=\gamma\beta^{(1)}_{\psi B,\psi}\cdot(2\tilde{\Delta}-\frac{1}{3})
\cdot[(-\frac{1}{3}+(\tilde{\Delta}-1))\tilde{\psi}_{-\frac{4}{3}}\psi^{+}(0)+\partial\tilde{\psi}_{-\frac{1}{3}}\psi^{+}(0)] + 0\nn\\
=\gamma\beta^{(1)}_{\psi B,\psi}\cdot[(2\tilde{\Delta}-\frac{1}{3})(\tilde{\Delta}-\frac{4}{3})\cdot\mu B(0)+0]\label{eqD.21}
\eea
\underline{$M2.3$}
\beq
\tilde{\psi}_{\frac{5}{3}}L_{-1}\psi^{+}_{-\frac{2}{3}}B(0)=(\tilde{\Delta}+\frac{2}{3})\tilde{\psi}_{\frac{2}{3}}\psi^{+}_{-\frac{2}{3}}B(0)+L_{-1}\tilde{\psi}_{\frac{5}{3}}\psi^{+}_{-\frac{2}{3}}B(0)
\label{eqD.22}
\eeq
The matrix element $\tilde{\psi}_{\frac{2}{3}}\psi^{+}_{-\frac{2}{3}}B(0)$ has been defined in Section 4.1, eq.(\ref{eq4.21}). Next,
\bea
L_{-1}\tilde{\psi}_{\frac{5}{3}}\psi^{+}_{-\frac{2}{3}}B(0)=L_{-1}\tilde{\psi}_{\frac{5}{3}}(\gamma\beta^{(11)}_{\psi B,\psi}\cdot\partial^{2}\psi^{+}(0)+\gamma\beta^{(2)}_{\psi B,\psi}\cdot L_{-2}\psi^{+}(0)+\mu\tilde{\psi}^{+}(0))\nn\\
=L_{-1}(\gamma\beta^{(11)}_{\psi B,\psi}\cdot\tilde{\psi}_{\frac{5}{3}}\partial^{2}\psi^{+}(0)+\gamma\beta^{(2)}_{\psi B,\psi}\cdot\tilde{\psi}_{\frac{5}{3}}L_{-2}\psi^{+}(0)+\mu\tilde{\psi}_{\frac{5}{3}}\tilde{\psi}^{+}(0))\nn\\
=0+0+\mu\cdot\frac{\tilde{\Delta}}{c}\partial^{2}T(0)\label{eqD.23}
\eea
Putting together eq.(\ref{eq4.21}) for $\tilde{\psi}_{\frac{2}{3}}\psi^{+}_{-\frac{2}{3}}B(0)$ and eq.(\ref{eqD.23}) for $L_{-1}\tilde{\psi}_{\frac{5}{3}}\psi^{+}_{-\frac{2}{3}}B(0)$, one obtains, from (\ref{eqD.22}) :
\bea
\tilde{\psi}_{\frac{5}{3}}L_{-1}\psi^{+}_{-\frac{2}{3}}B(0)=
(\tilde{\Delta}+\frac{2}{3}) [\gamma\beta^{(11)}_{\psi B,\psi}\cdot(\tilde{\Delta}-\frac{1}{3})(\tilde{\Delta}-\frac{4}{3})\mu B(0)\nn\\
+\gamma\beta^{(2)}_{\psi B,\psi}\cdot(2\tilde{\Delta}-\frac{4}{3})\mu B(0)\nn\\
+\mu\cdot(\beta^{(112)}_{\tilde{\psi}\tilde{\psi}^{+},I}\cdot\partial^{2}T(0)+\beta^{(22)}_{\tilde{\psi}\tilde{\psi}^{+},I}\cdot\wedge(0)+\tilde{\gamma}\cdot B(0))]\nn\\  
+ \mu\cdot\frac{\tilde{\Delta}}{c}\partial^{2}T(0)
\label{eqD.24}
\eea

Finally, putting (\ref{eqD.20}), (\ref{eqD.21}), (\ref{eqD.24}), plus (\ref{eqD.13}), into (\ref{eqD.16}) one gets the following expression for the matrix element $M2$ : 
\bea
M2=\tilde{\psi}_{\frac{5}{3}}(L_{-2}\psi^{+})_{-\frac{5}{3}}B(0)=
\frac{\mu}{81}  
\{ \frac{117 (3 (-62 + 39 c) \partial^2 T(0) + 2336 \wedge(0))}{c (22 + 5 c)}\nn\\
+ \frac{405 b (784 + 57 c) + (3210016 + 113007 c) \gamma + 
       972 (784 + 57 c) \tilde{\gamma}}{784 + 57 c} B(0)   \}
\label{eqD.25}
\eea
In getting this expression for $M2$ we have substituted the values 
of the $\beta$ coefficients and the values of $\Delta$ and $\tilde{\Delta}$.

The projected form (onto $B(\infty)$) is as follows :
\bea
M2_{B}=(\tilde{\psi}_{\frac{5}{3}}(L_{-2}\psi^{+})_{-\frac{5}{3}}B(0))_{B}\nn\\
=\frac{\mu}{81} \frac{405 b (784 + 57 c) + (3210016 + 113007 c) \gamma + 
       972 (784 + 57 c) \tilde{\gamma}}{784 + 57 c}
\label{eqD.26}
\eea

\vskip3cm
\underline{$M3=\tilde{\psi}_{\frac{5}{3}}\psi_{-3}\psi^{+}_{\frac{4}{3}}B(0)$}.

\bea
\tilde{\psi}_{\frac{5}{3}}\psi_{-3}\psi_{\frac{4}{3}}B(0)=\tilde{\psi}_{\frac{5}{3}}\psi_{-3}\gamma\psi(0)\nn\\
M3=\tilde{\psi}_{\frac{5}{3}}\psi_{-3}\psi_{\frac{4}{3}}B(0)=\gamma\tilde{\psi}_{\frac{5}{3}}\psi_{-3}\psi(0)\label{eqD.27}
\eea
The matrix element $\tilde{\psi}_{\frac{5}{3}}\psi_{-3}\psi(0)$ cannot be calculated directly. This is because $\psi_{-3}\psi(0)$ is a descendant of the third level, in the charged sector (sector of $\psi^{+}(0)$), while only the levels 0, 1, and 2 are explicites (in the charged sector), cf. Section 1.

The matrix element $\tilde{\psi}_{\frac{5}{3}}\psi_{-3}\psi(0)$ has to be defined by the commutation relations $\{\tilde{\psi},\psi\}\psi(0)$. They have the following form:
\beq
\sum^{\infty}_{l=0}D^{l}_{\frac{5}{3}}(\tilde{\psi}_{\frac{5}{3}+n-l}\psi_{m+l}+\psi_{\frac{5}{3}+m-l}\tilde{\psi}_{n+l})\psi(0)=R(n,m)\label{eqD.28}
\eeq
\bea
R(n,m)=\{\frac{1}{2}(\tilde{\Delta}+n-1)(\tilde{\Delta}+n-2)\cdot\zeta\psi^{+}_{\frac{5}{3}+n+m}\nn\\
-(\tilde{\Delta}+n-1)(\Delta+n+m+\frac{5}{3})\cdot\zeta\beta^{(1)}_{\tilde{\psi}\psi,\psi^{+}}\cdot\psi^{+}_{\frac{5}{3}+n+m}\nn\\
+(\Delta+n+m+\frac{8}{3})(\Delta+n+m+\frac{5}{3})\cdot\zeta\beta^{(11)}_{\tilde{\psi}\psi,\psi^{+}}\cdot\psi^{+}_{\frac{5}{3}+n+m}\nn\\
+\zeta\beta^{(2)}_{\tilde{\psi}\psi,\psi^{+}}\cdot(L_{-2}\psi^{+})_{\frac{5}{3}+n+m}+\eta\tilde{\psi}^{+}_{\frac{5}{3}+n+m}\}\psi(0)\label{eqD.29}
\eea
We need to have the matrix element $\tilde{\psi}_{\frac{5}{3}}\psi_{-3}\psi(0)$ in the l.h.s. of (\ref{eqD.28}). So we take $n=0$, $m=-3$. With this choice we obtain the equation :
\bea
(\tilde{\psi}_{\frac{5}{3}}\psi_{-3}+D^{1}_{\frac{5}{3}}\tilde{\psi}_{\frac{2}{3}}\psi_{-2}+D^{2}_{\frac{5}{3}}\tilde{\psi}_{-\frac{1}{3}}\psi_{-1}+D^{3}_{\frac{5}{3}}\tilde{\psi}_{-\frac{4}{3}}\psi_{0}\nn\\
+\psi_{-\frac{4}{3}}\tilde{\psi}_{0})\psi(0)=R(0,-3)\label{eqD.30}
\eea
\bea
R(0,-3)=\{\frac{1}{2}(\tilde{\Delta}-1)(\tilde{\Delta}-2)\cdot\zeta\psi^{+}_{-\frac{4}{3}}\nn\\
-(\tilde{\Delta}-1)(\Delta-\frac{4}{3})\cdot\zeta\beta^{(1)}_{\tilde{\psi}\psi,\psi^{+}}\cdot\psi^{+}_{-\frac{4}{3}}\nn\\
+(\Delta-\frac{1}{3})(\Delta-\frac{4}{3})\cdot\zeta\beta^{(11)}_{\tilde{\psi}\psi,\psi^{+}}\cdot\psi^{+}_{-\frac{4}{3}}\nn\\
+\zeta\cdot\beta^{(2)}_{\tilde{\psi}\psi,\psi^{+}}\cdot(L_{-2}\psi^{+})_{-\frac{4}{3}}+\eta\cdot\tilde{\psi}^{+}_{-\frac{4}{3}}\}\psi(0)\label{eqD.31}
\eea
The matrix element $\tilde{\psi}_{\frac{5}{3}}\psi_{-3}\psi(0)$ can be defined by the equation (\ref{eqD.30}). But first we have to calculate several matrix elements which enter (\ref{eqD.30}), (\ref{eqD.31}), all of which are explicit (could be calculated directly).

\underline{$\tilde{\psi}_{-\frac{4}{3}}\psi_{0}\psi(0)$}.
\beq
\tilde{\psi}_{-\frac{4}{3}}\psi_{0}\psi(0)=\tilde{\psi}_{-\frac{4}{3}}\lambda\psi^{+}(0)=\lambda\cdot\mu B(0)\label{eqD.32}
\eeq

\underline{$\tilde{\psi}_{-\frac{1}{3}}\psi_{-1}\psi(0)$}.
\bea
\tilde{\psi}_{-\frac{1}{3}}\psi_{-1}\psi(0)=\tilde{\psi}_{-\frac{1}{3}}\lambda\beta^{(1)}_{\psi\psi,\psi^{+}}\cdot\partial\psi^{+}(0)\nn\\
=\lambda\beta^{(1)}_{\psi\psi,\psi^{+}}\cdot(\tilde{\Delta}-\frac{4}{3})\cdot\tilde{\psi}_{-\frac{4}{3}}\psi^{+}(0)+\lambda\beta^{(1)}_{\psi\psi,\psi^{+}}\cdot\partial\tilde{\psi}_{-\frac{1}{3}}\psi^{+}(0)\nn\\
=\lambda\beta^{(1)}_{\psi\psi,\psi^{+}}\cdot(\tilde{\Delta}-\frac{4}{3})\cdot\mu B(0)+0\label{eqD.33}
\eea

\underline{$\tilde{\psi}_{\frac{2}{3}}\psi_{-2}\psi(0)$}.
\bea
\tilde{\psi}_{\frac{2}{3}}\psi_{-2}\psi(0)=\tilde{\psi}_{\frac{2}{3}}(\lambda\beta^{(11)}_{\psi\psi,\psi^{+}}\cdot\partial^{2}\psi^{+}(0)+\lambda\beta^{(2)}_{\psi\psi,\psi^{+}}\cdot L_{-2}\psi^{+}(0)+\zeta\cdot\tilde{\psi}^{+}(0))\nn\\
=\lambda\beta^{(11)}_{\psi\psi,\psi^{+}}\cdot\tilde{\psi}_{\frac{2}{3}}\partial^{2}\psi^{+}(0)+\lambda\beta^{(2)}_{\psi\psi,\psi^{+}}\cdot\tilde{\psi}_{\frac{2}{3}}L_{-2}\psi^{+}(0)+\zeta\cdot\tilde{\psi}_{\frac{2}{3}}\tilde{\psi}^{+}(0)\nn\\
\tilde{\psi}_{\frac{2}{3}}\partial^{2}\psi(0)=(\tilde{\Delta}-\frac{1}{3})\tilde{\psi}_{-\frac{1}{3}}\partial\psi^{+}(0)+(\tilde{\Delta}-\frac{1}{3})\partial\tilde{\psi}_{-\frac{1}{3}}\psi^{+}(0)\nn\\
+\partial^{2}\tilde{\psi}_{\frac{2}{3}}\psi^{+}(0)\nn\\
=(\tilde{\Delta}-\frac{1}{3})(\tilde{\Delta}-\frac{4}{3})\tilde{\psi}_{-\frac{4}{3}}\psi^{+}(0)+0+0\nn\\
=(\tilde{\Delta}-\frac{1}{3})(\tilde{\Delta}-\frac{4}{3})\cdot\mu B(0)\nn\\
\tilde{\psi}_{\frac{2}{3}}L_{-2}\psi^{+}(0)=(\frac{2}{3}+2(\tilde{\Delta}-1))\tilde{\psi}_{-\frac{4}{3}}\psi^{+}(0)+L_{-2}\tilde{\psi}_{\frac{2}{3}}\psi^{+}(0)\nn\\
=(2\tilde{\Delta}-\frac{4}{3})\cdot\mu B(0)+0\nn\\
\tilde{\psi}_{\frac{2}{3}}\tilde{\psi}^{+}(0)=\beta^{(112)}_{\tilde{\psi}\tilde{\psi}^{+},I}\cdot\partial^{2}T(0)+\beta^{(22)}_{\tilde{\psi}\tilde{\psi}^{+},I}\cdot\wedge(0)+\tilde{\gamma}\cdot B(0)\nn\\
\tilde{\psi}_{\frac{2}{3}}\psi_{-2}\psi(0)=\lambda\beta^{(11)}_{\psi\psi,\psi^{+}}\cdot(\tilde{\Delta}-\frac{1}{3})(\tilde{\Delta}-\frac{4}{3})\cdot\mu B(0)\nn\\
+\lambda\beta^{(2)}_{\psi\psi,\psi^{+}}\cdot(2\tilde{\Delta}-\frac{4}{3})\mu B(0)\nn\\
+\zeta\cdot(\beta^{(112)}_{\tilde{\psi}\tilde{\psi}^{+},I}\cdot\partial^{2}T(0)+\beta^{(22)}_{\tilde{\psi}\tilde{\psi}^{+},I}\cdot\wedge(0)+\tilde{\gamma}\cdot B(0))\label{eqD.34}
\eea

\underline{${\psi}_{-\frac{4}{3}}\tilde{\psi}_{0}\psi(0)$}.
\bea
{\psi}_{-\frac{4}{3}}\tilde{\psi}_{0}\psi(0)=\psi_{-\frac{4}{3}}\cdot\zeta\psi^{+}(0)\nn\\
=\zeta\cdot(\beta^{(112)}_{\psi\psi^{+},I}\cdot\partial^{2}T(0)+\beta^{(22)}_{\psi\psi^{+},I}\cdot\wedge(0)+\gamma\cdot B(0)\label{eqD.35}
\eea

\underline{$\psi^{+}_{-\frac{4}{3}}\psi(0)$}.
\bea
\psi^{+}_{-\frac{4}{3}}\psi(0)=\beta^{(112)}_{\psi\psi^{+},I}\cdot\partial^{2}T(0)+\beta^{(22)}_{\psi\psi^{+},I}\cdot\wedge(0)+\gamma\cdot B(0)\label{eqD.36}
\eea

\underline{$(L_{-2}\psi^{+})_{-\frac{4}{3}}\psi(0)$}.
\bea
(L_{-2}\psi^{+})_{-\frac{4}{3}}\psi(0)=\sum^{\infty}_{l=0}(L_{-2-l}\psi^{+}_{-\frac{4}{3}+l+2}+\psi^{+}_{-\frac{4}{3}-l+1}L_{l-1})\psi(0)\nn\\
=(L_{-2}\psi^{+}_{\frac{2}{3}}+L_{-3}\psi^{+}_{\frac{5}{3}}+L_{-4}\psi^{+}_{\frac{8}{3}}+\psi^{+}_{-\frac{1}{3}}L_{-1}+\psi^{+}_{-\frac{4}{3}}L_{0})\psi(0)\nn\\
=L_{-2}\frac{2\Delta}{c}\cdot T(0)+0+L_{-4}I+(\Delta-\frac{4}{3})\psi^{+}_{-\frac{4}{3}}\psi(0)+L_{-1}\psi^{+}_{-\frac{1}{3}}\psi(0)
+\psi^{+}_{-\frac{4}{3}}\cdot\Delta\psi(0)\nn\\
=\frac{2\Delta}{c}\cdot\wedge(0)+\frac{1}{2}\partial^{2}T(0)\nn\\
+(2\Delta-\frac{3}{4})\cdot(\beta^{(112)}_{\psi\psi^{+},I}\cdot\partial^{2}T(0)+\beta^{(22)}_{\psi\psi^{+},I}\cdot\wedge(0)+\gamma\cdot B(0))
+ \frac{\Delta}{c}\cdot \partial^2 T(0)
\label{eqD.37}
\eea

\underline{$\tilde{\psi}_{-\frac{4}{3}}^{+}\psi(0)$}.
\beq
\tilde{\psi}^{+}_{-\frac{4}{3}}\psi(0)=\mu B(0)\label{eqD.38}
\eeq
Now, putting (\ref{eqD.32})-(\ref{eqD.38}) into (\ref{eqD.30}), (\ref{eqD.31}), we can define the matrix element $\tilde{\psi}_{\frac{5}{3}}\psi_{-3}\psi(0)$, and then the matrix element $M3$ in eq.(\ref{eqD.27}). One gets :
\bea
\tilde{\psi}_{\frac{5}{3}}\psi_{-3}\psi(0)
= \frac{1}{324} \{      \frac{84 \zeta (3 (-62 + 39 c) \partial^2 T(0) 
+ 2336 \wedge(0))}{c (22 + 5 c)} \nn\\
+  \frac{1}{784 + 57 c}[567 (1024 + 27 c) \zeta \gamma + 
      540 (784 + 57 c) \zeta \tilde{\gamma}\nn\\
      +  2 (162 (784 + 57 c) \eta 
      + 145 (5120 + 159 c) \lambda) \mu] B(0)     \}
\label{eqD.39}
\eea
In getting the expression above we have substituted the values 
of the $\beta$ coefficients and also the values of $\Delta$, $\tilde{\Delta}$.
Finally one obtains :
\bea
M3=\tilde{\psi}_{\frac{5}{3}}\psi_{-3}\psi_{\frac{4}{3}}B(0)
= \gamma \frac{1}{324} \{      \frac{84 \zeta (3 (-62 + 39 c) \partial^2 T(0) 
+ 2336 \wedge(0))}{c (22 + 5 c)} \nn\\
+  \frac{1}{784 + 57 c}[567 (1024 + 27 c) \zeta \gamma + 
      540 (784 + 57 c) \zeta \tilde{\gamma}\nn\\
      +  2 (162 (784 + 57 c) \eta 
      + 145 (5120 + 159 c) \lambda) \mu] B(0)     \}
\label{eqD.40}
\eea
The projected form :
\bea
M3_{B}=(\tilde{\psi}_{\frac{5}{3}}\psi_{-3}\psi_{\frac{4}{3}}B(0))_{B}
=  \frac{\gamma}{324(784 + 57 c)}[567 (1024 + 27 c) \zeta \gamma\nn\\ + 
      540 (784 + 57 c) \zeta \tilde{\gamma}
      +  2 (162 (784 + 57 c) \eta 
      + 145 (5120 + 159 c) \lambda) \mu]
\label{eqD.41}
\eea

\vskip2cm
\underline{$M4=\tilde{\psi}_{\frac{5}{3}}\psi_{-2}\psi_{\frac{1}{3}}B(0)$}.
\bea
\tilde{\psi}_{\frac{5}{3}}\psi_{-2}\psi_{\frac{1}{3}}B(0)=\tilde{\psi}_{\frac{5}{3}}\psi_{-2}\gamma\beta^{(1)}_{\psi B,\psi}\partial\psi(0)\nn\\
M4=\tilde{\psi}_{\frac{5}{3}}\psi_{-2}\psi_{\frac{1}{3}}B(0)=\gamma\beta^{(1)}_{\psi B,\psi}\cdot\tilde{\psi}_{\frac{5}{3}}\psi_{-2}\partial\psi(0)\label{eqD.42}
\eea
\bea
\tilde{\psi}_{\frac{5}{2}}\psi_{-2}\partial\psi(0)=\tilde{\psi}_{\frac{5}{3}}(-2+(\Delta-1))\psi_{-3}\psi(0)+\tilde{\psi}_{\frac{5}{3}}\partial\psi_{-2}\psi(0)\nn\\
=(\Delta-3)\tilde{\psi}_{\frac{5}{3}}\psi_{-3}\psi(0)+(\tilde{\Delta}+\frac{2}{3}) \tilde{\psi}_{\frac{2}{3}}\psi_{-2}\psi(0)\nn\\
+\partial\tilde{\psi}_{\frac{5}{3}}\psi_{-2}\psi(0)\label{eqD.43}
\eea
The matrix elements $\tilde{\psi}_{\frac{5}{3}}\psi_{-3}\psi(0)$ and also 
$\tilde{\psi}_{\frac{2}{3}}\psi_{-2}\psi(0)$ have been defined just before,
when calculating the matrix element M3,  eq.(\ref{eqD.39}) 
and (\ref{eqD.34}). Next,
\bea
\partial\tilde{\psi}_{\frac{5}{3}}\psi_{-2}\psi(0)=\partial\tilde{\psi}_{\frac{5}{3}}(\lambda\beta^{(11)}_{\psi\psi,\psi^{+}}\partial^{2}\psi^{+}(0)+\lambda\beta^{(2)}_{\psi\psi,\psi^{+}}L_{-2}\psi^{+}(0)\nn\\
+\zeta\tilde{\psi}^{+}(0))\nn\\
=\lambda\beta^{(11)}_{\psi\psi,\psi^{+}}\cdot\partial\tilde{\psi}_{\frac{5}{3}}\partial^{2}\psi^{+}(0)+\lambda\beta^{(2)}_{\psi\psi,\psi^{+}}\cdot\partial\tilde{\psi}_{\frac{5}{6}}L_{-2}\psi^{+}(0)\nn\\
+\zeta\cdot\partial\tilde{\psi}_{\frac{5}{3}}\tilde{\psi}^{+}(0)\nn\\
=0+0+\zeta\cdot\partial\frac{\tilde{\Delta}}{c}\partial T(0)=\zeta\cdot\frac{\tilde{\Delta}}{c}\partial^{2}T(0)\label{eqD.44}
\eea
Putting(\ref{eqD.39}), (\ref{eqD.34}), (\ref{eqD.44}) into (\ref{eqD.43}) one defines $\tilde{\psi}_{\frac{5}{3}}\psi_{-2}\partial\psi(0)$, and then, by (\ref{eqD.42}), one gets the expression for $M4$ :
\bea
\tilde{\psi}_{\frac{5}{3}}\psi_{-2}\partial\psi(0)=
\frac{420 \zeta (3 (-62 + 39 c) \partial^2 T(0) 
+ 2336 \wedge(0))}{972 c (22 + 5 c)} \nn\\
+  \frac{1}{972 (784 + 57 c)} [-567 (1024 + 27 c) \zeta \gamma 
+  4644 (784 + 57 c) \zeta \tilde{\gamma} \nn\\
+  2 (-162 (784 + 57 c) \eta + (7588864 + 285825 c) \lambda) \mu] B(0)
\label{eqD.45}
\eea
\bea
M4=\tilde{\psi}_{\frac{5}{3}}\psi_{-2}\psi_{\frac{1}{3}}B(0)=
\frac{420 \gamma\zeta (3 (-62 + 39 c) \partial^2 T(0) 
+ 2336 \wedge(0))}{3888 c (22 + 5 c)} \nn\\
+  \frac{\gamma}{3888 (784 + 57 c)} [-567 (1024 + 27 c) \zeta \gamma 
+  4644 (784 + 57 c) \zeta \tilde{\gamma} \nn\\
+  2 (-162 (784 + 57 c) \eta + (7588864 + 285825 c) \lambda) \mu] B(0)
\label{eqD.46}
\eea
The projected form :
\bea
M4_{B}=(\tilde{\psi}_{\frac{5}{3}}\psi_{-2}\psi_{\frac{1}{3}}B(0))_{B}=
 \frac{\gamma}{3888 (784 + 57 c)} [-567 (1024 + 27 c) \zeta \gamma \nn\\
+  4644 (784 + 57 c) \zeta \tilde{\gamma} 
+  2 (-162 (784 + 57 c) \eta + (7588864 + 285825 c) \lambda) \mu]
\label{eqD.46a}
\eea

\vskip1cm

\underline{$M5=\tilde{\psi}_{\frac{5}{3}}\psi_{-1}\psi_{-\frac{2}{3}}B(0)$}.
\bea
\tilde{\psi}_{\frac{5}{3}}\psi_{-1}\psi_{-\frac{2}{3}}B(0)=\tilde{\psi}_{\frac{5}{3}}\psi_{-1}(\gamma\beta^{(11)}_{\psi B,\psi}\cdot\partial^{2}\psi(0)+\gamma\beta^{(2)}_{\psi B,\psi}\cdot L_{-2}\psi(0)+\mu\cdot\tilde{\psi}(0))\nn\\
M5=\tilde{\psi}_{\frac{5}{3}}\psi_{-1}\psi_{-\frac{2}{3}}B(0)
=\gamma\beta^{(11)}_{\psi B\psi}
\cdot\tilde{\psi}_{\frac{5}{3}}\psi_{-1}\partial^{2}\psi(0)
+\gamma\beta^{(2)}_{\psi B,\psi}
\cdot\tilde{\psi}_{\frac{5}{3}}\psi_{-1}L_{-2}\psi(0)\nn\\
+\mu
\cdot\tilde{\psi}_{\frac{5}{3}}\psi_{-1}\tilde{\psi}(0)\nn\\
=\gamma\beta^{(11)}_{\psi B\psi}\cdot M5.1 
+ \gamma\beta^{(2)}_{\psi B,\psi}\cdot M5.2 + \mu\cdot M5.3
\label{eqD.47}
\eea
To define $M5$, we have to calculate the matrix elements :
\beq
M5.1=\tilde{\psi}_{\frac{5}{3}}\psi_{-1}\partial^{2}\psi(0)\label{eqD.48}
\eeq
\beq
M5.2=\tilde{\psi}_{\frac{5}{3}}\psi_{-1}L_{-2}\psi(0)\label{eqD.49}
\eeq
\beq
M5.3=\tilde{\psi}_{\frac{5}{3}}\psi_{-1}\tilde{\psi}(0)\label{eqD.50}
\eeq

\underline{$M5.1$}
\bea
\tilde{\psi}_{\frac{5}{3}}\psi_{-1}\partial^{2}\psi(0)=(\Delta-2)\tilde{\psi}_{\frac{5}{3}}\psi_{-2}\partial\psi(0)
+(\Delta-2)\tilde{\psi}_{\frac{5}{3}}\partial\psi_{-2}\psi(0)\nn\\
+\tilde{\psi}_{\frac{5}{3}}\partial^{2}\psi_{-1}\psi(0)\nn\\
=(\Delta-2)\tilde{\psi}_{\frac{5}{3}}\psi_{-2}\partial\psi(0)
+(\Delta-2)(\tilde{\Delta}
+\frac{2}{3})\tilde{\psi}_{\frac{2}{3}}\psi_{-2}\psi(0)\nn\\
+(\Delta-2)\partial \tilde{\psi}_{\frac{5}{3}}\psi_{-2} \psi(0)\nn\\
+(\tilde{\Delta}+\frac{2}{3})\tilde{\psi}_{\frac{2}{3}}\partial\psi_{-1}\psi(0)
+(\tilde{\Delta}+\frac{2}{3})\partial\tilde{\psi}_{\frac{2}{3}}\psi_{-1}\psi(0)\nn\\
+\partial^{2}\tilde{\psi}_{\frac{5}{3}}\psi_{-1}\psi(0)\nn\\
=(\Delta-2)\tilde{\psi}_{\frac{5}{3}}\psi_{-2}\partial\psi(0)
+(\Delta-2)(\tilde{\Delta}+\frac{2}{3})\tilde{\psi}_{\frac{2}{3}}\psi_{-2}\psi(0)\nn\\
+(\Delta-2)\partial\tilde{\psi}_{\frac{5}{3}}\psi_{-2}\psi(0)\nn\\
+(\tilde{\Delta}+\frac{2}{3})(\tilde{\Delta}-\frac{1}{3})\tilde{\psi}_{-\frac{1}{3}}\psi_{-1}\psi(0)+2(\tilde{\Delta}+\frac{2}{3})\partial\tilde{\psi}_{\frac{2}{3}}\psi_{-1}\psi(0)\nn\\
+\partial^{2}\tilde{\psi}_{\frac{5}{3}}\psi_{-1}\psi(0)\label{eqD.51}
\eea

Because $\psi_{-1}\psi(0)=\lambda\beta^{(1)}_{\psi\psi,\psi^{+}}\partial\psi^{+}(0)$, and $\tilde{\psi}_{\frac{2}{3}}\partial\psi^{+}(0)=0$, $\tilde{\psi}_{\frac{5}{3}}\partial\psi^{+}(0)=0$, the last two terms in (\ref{eqD.51}) vanish.

$\tilde{\psi}_{\frac{5}{3}}\psi_{-2}\partial\psi(0)$ is given by (\ref{eqD.45}),
$\tilde{\psi}_{\frac{2}{3}}\psi_{-2}\psi(0)$ is in (\ref{eqD.34}), $\partial\tilde{\psi}_{\frac{5}{3}}\psi_{-2}\psi(0)$ is in (\ref{eqD.44}), $\tilde{\psi}_{-\frac{1}{3}}\psi_{-1}\psi(0)$ is in (\ref{eqD.33}). Substituting these expressions into (\ref{eqD.51}), one obtains the matrix element M5.1 :
\bea
M5.1=\tilde{\psi}_{\frac{5}{3}}\psi_{-1}\partial^{2}\psi(0)
=\frac{154 \zeta (3 (-62 + 39 c) \partial^2 T(0) 
+ 2336 \wedge(0))}{243 c (22 + 5 c)}\nn\\
- \frac{1}{1458 (784+57c)} [189 \zeta ((3072 + 81 c) \gamma
- 52 (784 + 57 c) \tilde{\gamma})\nn\\
+ 2 (162 (784 + 57 c) \eta - (37934848 + 2195265 c) \lambda) \mu] B(0)
\label{eqD.52}
\eea

\underline{M5.2.}
\bea
\tilde{\psi}_{\frac{5}{3}}\psi_{-1}L_{-2}\psi(0)=(-1+2(\Delta-1))\tilde{\psi}_{\frac{5}{3}}\psi_{-3}\psi(0)+(\frac{5}{3}+2(\tilde{\Delta}-1))\tilde{\psi}_{-\frac{1}{3}}\psi_{-1}\psi(0)\nn\\
+L_{-2}\tilde{\psi}_{\frac{5}{3}}\psi_{-1}\psi(0)\nn\\
=(2\Delta-3)\tilde{\psi}_{\frac{5}{3}}\psi_{-3}\psi(0)+(2\tilde{\Delta}-\frac{1}{3})\tilde{\psi}_{-\frac{1}{3}}\psi_{-1}\psi(0) + 0
\label{eqD.53}
\eea
Substituting $\tilde{\psi}_{\frac{5}{3}}\psi_{-3}\psi(0)$ from (\ref{eqD.39}) 
and  $\tilde{\psi}_{-\frac{1}{3}}\psi_{-1}\psi(0)$  from (\ref{eqD.33})  one obtains :
\bea
M5.2=\tilde{\psi}_{\frac{5}{3}}\psi_{-1}L_{-2}\psi(0)=
15 \lambda \mu B(0) + 
 \frac{49 \zeta (3 (-62 + 39 c) \partial^2 T(0) 
 + 2336 \wedge(0))}{81 c (22 + 5 c)} \nn\\
 + \frac{7}{792 (784 + 57 c)} [567 (1024 + 27 c) \zeta \gamma 
 + 540 (784 + 57 c) \zeta \tilde{\gamma} \nn\\
 + 2 (162 (784 + 57 c) \eta + 145 (5120 + 159 c) \lambda) \mu] B(0)
\label{eqD.54}
\eea

\underline{M5.3}

$\tilde{\psi}_{\frac{5}{3}}\psi_{-1}\tilde{\psi}(0)$, this matrix element cannot be calculated directly. This is because the operator $\psi_{-1}\tilde{\psi}(0)$ belong to the third level in the charged sector (the sector of $\psi^{+}(0)$), which is not explicit. To calculate $M5.3$ we have to use the commutation relation $\{\tilde{\psi},\psi\}\tilde{\psi}(0)$. Its general form is the same as in (\ref{eqD.28}), (\ref{eqD.29}), with just $\psi(0)$ replaced by $\tilde{\psi}(0)$. As we need the appearance of the matrix element $\tilde{\psi}_{\frac{5}{3}}\psi_{-1}\tilde{\psi}(0)$, we choose $n=0$, $m=-1$. One obtains, from (\ref{eqD.28}), (\ref{eqD.29}), with $\psi(0)\rightarrow\tilde{\psi}(0)$ :
\bea
(\tilde{\psi}_{\frac{5}{3}}\psi_{-1}+D^{1}_{\frac{5}{3}}\tilde{\psi}_{\frac{2}{3}}\psi_{0}+D^{2}_{\frac{5}{3}}\tilde{\psi}_{-\frac{1}{3}}\psi_{1}+D^{3}_{\frac{5}{3}}\tilde{\psi}_{-\frac{4}{3}}\psi_{2}\nn\\
+\psi_{\frac{2}{3}}\tilde{\psi}_{0}+D^{1}_{\frac{5}{3}}\psi_{-\frac{1}{3}}\tilde{\psi}_{1}+D^{2}_{\frac{5}{3}}\psi_{-\frac{4}{3}}\tilde{\psi}_{2})\tilde{\psi}(0)=R(0,-1)\label{eqD.55}
\eea
\bea
R(0,-1)=\{\frac{1}{2}(\tilde{\Delta}-1)(\tilde{\Delta}-2)\zeta\psi^{+}_{\frac{2}{3}}\nn\\
-(\tilde{\Delta}-1)(\Delta+\frac{2}{3})\zeta\beta^{(1)}_{\tilde{\psi}\psi,\psi^{+}}\cdot\psi^{+}_{\frac{2}{3}}\nn\\
+(\Delta+\frac{5}{3})(\Delta+\frac{2}{3})\zeta\beta^{(11)}_{\tilde{\psi}\psi,\psi^{+}}\cdot\psi^{+}_{\frac{2}{3}}\nn\\
+\zeta\beta^{(2)}_{\tilde{\psi}\psi,\psi^{+}}\cdot(L_{-2}\psi^{+})_{\frac{2}{3}}+\eta\tilde{\psi}^{+}_{\frac{2}{3}}\}\tilde{\psi}(0)\label{eqD.56}
\eea

We have to calculate several (explicit) matrix elements in (\ref{eqD.55}), (\ref{eqD.56}).

\underline{$\tilde{\psi}_{-\frac{4}{3}}\psi_{2}\tilde{\psi}(0).$}
\beq
\tilde{\psi}_{-\frac{4}{3}}\psi_{2}\tilde{\psi}(0)=\tilde{\psi}_{-\frac{4}{3}}\zeta\psi^{+}(0)=\zeta\cdot\mu B(0)\label{eqD.57}
\eeq

\underline{$\tilde{\psi}_{-\frac{1}{3}}\psi_{1}\tilde{\psi}(0).$}
\bea
\tilde{\psi}_{-\frac{1}{3}}\psi_{1}\tilde{\psi}(0)=\tilde{\psi}_{-\frac{1}{3}}\zeta\beta^{(1)}_{\psi\tilde{\psi},\psi^{+}}\cdot\partial\psi^{+}(0)\nn\\
=\zeta\beta^{(1)}_{\psi\tilde{\psi},\psi^{+}}\cdot(\tilde{\Delta}-\frac{4}{3})\tilde{\psi}_{-\frac{4}{3}}\psi^{+}(0)+\zeta\beta^{(1)}_{\psi\tilde{\psi},\psi^{+}}\cdot\partial\tilde{\psi}_{-\frac{1}{3}}\psi^{+}(0)\nn\\=
\zeta\beta^{(1)}_{\psi\tilde{\psi},\psi^{+}}\cdot(\tilde{\Delta}-\frac{4}{3})\mu B(0)+0\label{eqD.58}
\eea

\underline{$\tilde{\psi}_{\frac{2}{3}}\psi_{0}\tilde{\psi}(0).$}
\bea
\tilde{\psi}_{\frac{2}{3}}\psi_{0}\tilde{\psi}(0)=\tilde{\psi}_{\frac{2}{3}}(\zeta\beta^{(11)}_{\psi\tilde{\psi},\psi^{+}}\cdot\partial^{2}\psi^{+}(0)+\zeta\beta^{(2)}_{\psi\tilde{\psi},\psi^{+}}\cdot L_{-2}\psi^{+}(0)+\eta\tilde{\psi}^{+}(0))\nn\\
=\zeta\beta^{(11)}_{\psi\tilde{\psi},\psi^{+}}\cdot\tilde{\psi}_{\frac{2}{3}}\partial^{2}\psi^{+}(0)+\zeta\beta^{(2)}_{\psi\tilde{\psi},\psi^{+}}\cdot\tilde{\psi}_{\frac{2}{3}}L_{-2}\psi^{+}(0)+\eta\tilde{\psi}_{\frac{2}{3}}\tilde{\psi}^{+}(0)\nn\\
\tilde{\psi}_{\frac{2}{3}}\partial^{2}\psi^{+}(0)=(\tilde{\Delta}-\frac{1}{3})\tilde{\psi}_{-\frac{1}{3}}\partial\psi^{+}(0)+(\tilde{\Delta}-\frac{1}{3})\partial\tilde{\psi}_{-\frac{1}{3}}\psi^{+}(0)+\partial^{2}\tilde{\psi}_{\frac{2}{3}}\psi^{+}(0)\nn\\
=(\tilde{\Delta}-\frac{1}{3})\tilde{\psi}_{-\frac{1}{3}}\partial\psi^{+}(0)+0+0\nn\\
=(\tilde{\Delta}-\frac{1}{3})(\tilde{\Delta}-\frac{4}{3})\tilde{\psi}_{-\frac{4}{3}}\psi^{+}(0)+(\tilde{\Delta}-\frac{1}{3})\partial\tilde{\psi}_{-\frac{1}{3}}\psi^{+}(0)\nn\\
=(\tilde{\Delta}-\frac{1}{3})(\tilde{\Delta}-\frac{4}{3})\mu B(0)+0\nn\\
\tilde{\psi}_{\frac{2}{3}}L_{-2}\psi^{+}(0)=(\frac{2}{3}+2(\tilde{\Delta}-1))\tilde{\psi}_{-\frac{4}{3}}\psi^{+}(0)+L_{-2}\tilde{\psi}_{\frac{2}{3}}\psi^{+}(0)\nn\\
=(2\tilde{\Delta}-\frac{4}{3})\cdot\mu B(0)+0\nn\\
\tilde{\psi}_{\frac{2}{3}}\tilde{\psi}^{+}(0)=\beta^{(112)}_{\tilde{\psi}\tilde{\psi}^{+},I}\cdot\partial^{2}T(0)+\beta^{(22)}_{\tilde{\psi}\tilde{\psi}^{+},I}\cdot\wedge(0)+\tilde{\gamma}\cdot B(0)\nn\\
\tilde{\psi}_{\frac{2}{3}}\psi_{0}\tilde{\psi}(0)=\zeta\beta^{(11)}_{\psi\tilde{\psi},\psi^{+}}\cdot(\tilde{\Delta}-\frac{1}{3})(\tilde{\Delta}-\frac{4}{3})\mu B(0)\nn\\
+\zeta\beta^{(2)}_{\psi\tilde{\psi},\psi^{+}}\cdot(2\tilde{\Delta}-\frac{4}{3})\cdot\mu B(0)\nn\\
+\eta\cdot(\beta^{(112)}_{\tilde{\psi}\tilde{\psi}^{+},I}\cdot\partial^{2}T(0)+\beta^{(22)}_{\tilde{\psi}\tilde{\psi}^{+},I}\cdot\wedge(0)
+\tilde{\gamma}\cdot B(0))\label{eqD.59}
\eea

\underline{$\psi_{-\frac{4}{3}}\tilde{\psi}_{2}\tilde{\psi}(0).$}
\bea
\psi_{-\frac{4}{3}}\tilde{\psi}_{2}\tilde{\psi}(0)=\psi_{-\frac{4}{3}}\cdot\eta\psi^{+}(0)\nn\\
=\eta\cdot(\beta^{(112)}_{\psi\psi^{+},I}\cdot\partial^{2}T(0)+\beta^{(22)}_{\psi\psi^{+},I}\cdot\wedge(0)+\gamma\cdot B(0))\label{eqD.60}
\eea

\underline{$\psi_{-\frac{1}{3}}\tilde{\psi}_{1}\tilde{\psi}(0).$}
\bea
\psi_{-\frac{1}{3}}\tilde{\psi}_{1}\tilde{\psi}(0)=\psi_{-\frac{1}{3}}\eta\beta^{(1)}_{\tilde{\psi}\tilde{\psi},\psi^{+}}\cdot\partial\psi^{+}(0)\nn\\
=\eta\beta^{(1)}_{\tilde{\psi}\tilde{\psi},\psi^{+}}\cdot(\Delta-\frac{4}{3})\psi_{-\frac{4}{3}}\psi^{+}(0)+\eta\beta^{(1)}_{\tilde{\psi}\tilde{\psi},\psi^{+}}\cdot\partial\psi_{-\frac{1}{3}}\psi^{+}(0)\nn\\
=\eta\beta^{(1)}_{\tilde{\psi}\tilde{\psi},\psi^{+}}\cdot(\Delta-\frac{4}{3})\cdot(\beta^{(112)}_{\psi\psi^{+},I}\cdot\partial^{2}T(0)+\beta^{(22)}_{\psi\psi^{+},I}\cdot\wedge(0)+\gamma\cdot B(0))\nn\\
+\eta\cdot\beta^{(1)}_{\tilde{\psi}\tilde{\psi},\psi^{+}}\cdot\frac{\Delta}{c}\partial^{2}T(0)\label{eqD.61}
\eea

\underline{$\psi_{\frac{2}{3}}\tilde{\psi}_{0}\tilde{\psi}(0).$}
\bea
\psi_{\frac{2}{3}}\tilde{\psi}_{0}\tilde{\psi}(0)=\psi_{\frac{2}{3}}(\eta\beta^{(11)}_{\tilde{\psi}\tilde{\psi},\psi^{+}}\cdot\partial^{2}\psi^{+}(0)+\eta\beta^{(2)}_{\tilde{\psi}\tilde{\psi},\psi^{+}}\cdot L_{-2}\psi^{+}(0)+\tilde{\lambda}\tilde{\psi}^{+}(0))\nn\\
=\eta\beta^{(11)}_{\tilde{\psi}\tilde{\psi},\psi^{+}}\cdot\psi_{\frac{2}{3}}\partial^{2}\psi^{+}(0)+\eta\beta^{(2)}_{\tilde{\psi}\tilde{\psi},\psi^{+}}\cdot\psi_{\frac{2}{3}}L_{-2}\psi^{+}(0)+\tilde{\lambda}\cdot\psi_{\frac{2}{3}}\tilde{\psi}^{+}(0)\nn\\
\psi_{\frac{2}{3}}\partial^{2}\psi^{+}(0)=(\Delta-\frac{1}{3})\psi_{-\frac{1}{3}}\partial\psi^{+}(0)+(\Delta-\frac{1}{3})\partial\psi_{-\frac{1}{3}}\psi^{+}(0)\nn\\
+\partial^{2}\psi_{\frac{2}{3}}\psi^{+}(0)\nn\\
=(\Delta-\frac{1}{3})(\Delta-\frac{4}{3})\psi_{-\frac{4}{3}}\psi^{+}(0)+2(\Delta-\frac{1}{3})\partial\psi_{-\frac{1}{3}}\psi^{+}(0)\nn\\
+\partial^{2}\psi_{\frac{2}{3}}\psi^{+}(0)\nn\\
=(\Delta-\frac{1}{3})(\Delta-\frac{4}{3})\cdot(\beta^{(112)}_{\psi\psi^{+},I}\cdot\partial^{2}T(0)+\beta^{(22)}_{\psi\psi^{+},I}\cdot\wedge(0)+\gamma B(0))\nn\\
+2(\Delta-\frac{1}{3})\frac{\Delta}{c}\cdot\partial^{2}T(0)+\frac{2\Delta}{c}\partial^{2}T(0)\nn\\
\psi_{\frac{2}{3}}L_{-2}\psi^{+}(0)=(2\Delta-\frac{4}{3})\psi_{-\frac{4}{3}}\psi^{+}(0)+L_{-2}\psi_{\frac{2}{3}}\psi^{+}(0)\nn\\
=(2\Delta-\frac{4}{3})(\beta^{(112)}_{\psi\psi^{+},I}\cdot\partial^{2}T(0)+\beta^{(22)}_{\psi\psi^{+},I}\cdot\wedge(0)+\gamma\cdot B(0))+\frac{2{\Delta}}{c}\cdot\wedge(0)\nn\\
\psi_{\frac{2}{3}}\tilde{\psi^{+}}(0)=\mu\cdot B(0)\nn\\
\psi_{\frac{2}{3}}\tilde{\psi}_{0}\tilde{\psi}(0)=\eta\beta^{(11)}_{\tilde{\psi}\tilde{\psi},\psi^{+}}\cdot[(\Delta-\frac{1}{3})(\Delta-\frac{4}{3})(\beta^{(112)}_{\psi\psi^{+},I}\partial^{2}T(0)+\beta^{(22)}_{\psi\psi^{+},I}\wedge(0)+\gamma B(0))\nn\\
+2(\Delta-\frac{1}{3})\cdot\frac{\Delta}{c}\partial^{2}T(0)+\frac{2\Delta}{c}\partial^{2}T(0)]\nn\\
+\eta\beta^{(2)}_{\tilde{\psi}\tilde{\psi},\psi^{+}}\cdot[(2\Delta-\frac{4}{3}) (\beta^{(112)}_{\psi\psi^{+},I}\partial^{2}T(0)+\beta^{(22)}_{\psi\psi^{+},I}\wedge(0)+\gamma B(0))
+\frac{2\Delta}{c}\wedge(0)]\nn\\
+\tilde{\lambda}\cdot\mu\cdot B(0)\label{eqD.62}
\eea

\underline{$\psi_{\frac{2}{3}}^{+}\tilde{\psi}(0).$}
\beq
\psi^{+}_{\frac{2}{3}}\tilde{\psi}(0)=\mu B(0)\label{eqD.63}
\eeq

\underline{$(L_{-2}\psi^{+})_{\frac{2}{3}}\tilde{\psi}(0).$}
\bea
(L_{-2}\psi^{+})_{\frac{2}{3}}\tilde{\psi}(0)=\sum^{\infty}_{l=0}(L_{-2-l}\psi^{+}_{\frac{2}{3}+2+l}+\psi^{+}_{\frac{2}{3}-l+1}L_{l-1})\tilde{\psi}(0)\nn\\
=(\psi^{+}_{\frac{5}{3}}L_{-1}+\psi^{+}_{\frac{2}{3}}L_{0})\tilde{\psi}(0)\nn\\
=(\Delta+\frac{2}{3})\psi^{+}_{\frac{2}{3}}\tilde{\psi}(0)+L_{-1}\psi^{+}_{\frac{5}{3}}\tilde{\psi}(0)+\psi^{+}_{\frac{2}{3}}\tilde{\Delta}\tilde{\psi}(0)\nn\\
=(\Delta+\tilde{\Delta}+\frac{2}{3})\cdot\mu B(0)+0\label{eqD.64}
\eea

\underline{$\tilde{\psi}^{+}_{\frac{2}{3}}\tilde{\psi}(0).$}
\beq
\tilde{\psi}^{+}_{\frac{2}{3}}\tilde{\psi}(0)=\beta^{(112)}_{\tilde{\psi}\tilde{\psi}^{+},I}\cdot\partial^{2}T(0)+\beta^{(22)}_{\tilde{\psi}\tilde{\psi}^{+},I}\cdot\wedge(0)+\tilde{\gamma}\cdot B(0)\label{eqD.65}
\eeq
\vskip1.5cm
Substituting (\ref{eqD.57})-(\ref{eqD.65}) into the equations (\ref{eqD.55}), (\ref{eqD.56}) we define the matrix $\tilde{\psi}_{\frac{5}{3}}\psi_{-1}\tilde{\psi}(0)$. One gets the following expression :
\bea
M5.3=\tilde{\psi}_{\frac{5}{3}}\psi_{-1}\tilde{\psi}(0)
=  \frac{4 \eta (3 (-62 + 39 c) \partial^2 T(0) 
+ 2336 \wedge(0))}{27c (22 + 5 c)}\nn\\
+\frac{1}{324 (784 + 57 c)} [216 \eta (3 (-892 + 3 c) \gamma 
+ 4 (784 + 57 c) \tilde{\gamma})\nn\\
- ((-1372336 + 897 c) \zeta + 
         324 (784 + 57 c) \tilde{\lambda}) \mu] B(0)
\label{eqD.66}
\eea

\vskip3cm

Putting the expressions (\ref{eqD.52}), (\ref{eqD.54}), (\ref{eqD.66}), for M5.1, M5.2, M5.3, into the expression (\ref{eqD.47}) for M5, one gets :
\bea
M5=\tilde{\psi}_{\frac{5}{3}}\psi_{-1}\psi_{-\frac{2}{3}}B(0)\nn\\
=\frac{(7 (26672 + 231 c) \zeta \gamma + 
      144 (784 + 57 c) \eta \mu)
      (3 (-62 + 39 c) \partial^2 T(0) + 2336 \wedge(0))}{972 c (22 + 5 c)(784+57c)} \nn\\
      -  \frac{1}{11664 (784 + 57 c)^2} 
      [189 \zeta \gamma (3 (-31664 + 21 c) (1024 + 27 c) \gamma \nn\\
      - 28 (2416 + 39 c) (784 + 57 c) \tilde{\gamma})\nn\\
             - 2 (162 (784 + 57 c) \eta (5 (-6512 + 39 c) \gamma + 
            96 (784 + 57 c) \tilde{\gamma}) \nn\\
            + (32930985472 + 
            3 c (819583376 + 15366855 c)) \gamma \lambda) \mu \nn\\
            + 36 (784 + 57 c) ((-1372336 + 897 c) \zeta + 
         324 (784 + 57 c) \tilde{\lambda}) \mu^2] B(0)
\label{eqD.67}
\eea
The projected form :
\bea
M5_{B}=(\tilde{\psi}_{\frac{5}{3}}\psi_{-1}\psi_{-\frac{2}{3}}B(0))_{B}\nn\\
= -  \frac{1}{11664 (784 + 57 c)^2} 
      [189 \zeta \gamma (3 (-31664 + 21 c) (1024 + 27 c) \gamma \nn\\
      - 28 (2416 + 39 c) (784 + 57 c) \tilde{\gamma})\nn\\
             - 2 (162 (784 + 57 c) \eta (5 (-6512 + 39 c) \gamma + 
            96 (784 + 57 c) \tilde{\gamma}) \nn\\
            + (32930985472 + 
            3 c (819583376 + 15366855 c)) \gamma \lambda) \mu \nn\\
            + 36 (784 + 57 c) ((-1372336 + 897 c) \zeta + 
         324 (784 + 57 c) \tilde{\lambda}) \mu^2]
\label{eqD.68}
\eea

\vskip3cm

We could summarize that all the matrix elements in the list (\ref{eqD.8})-(\ref{eqD.12}) has been defined. Substituting their projected forms, in (\ref{eqD.14}), (\ref{eqD.26}), (\ref{eqD.41}), (\ref{eqD.46a}), (\ref{eqD.68}), into the equation (\ref{eqD.7}), we get the following expression for the projected matrix element $\tilde{\psi}_{\frac{5}{3}}\tilde{\psi}_{-\frac{5}{3}}B(0)$:
\bea
(\tilde{\psi}_{\frac{5}{3}}\tilde{\psi}_{-\frac{5}{3}}B(0))_{B}
= \frac{1}{\zeta} \{     \lambda \beta^{(1)}_{\psi \psi, \psi^{+}}\cdot M1_B  
-   2 \lambda \beta^{(11)}_{\psi \psi, \psi^{+}}\cdot M1_B   
-   \lambda \beta^{(2)}_{\psi \psi, \psi^{+}}\cdot M2_B   \nn\\
+   2 \cdot M5_B   +   2 D^{1}_{-\frac{1}{3}} \cdot M4_B   
+   2 D^{2}_{-\frac{1}{3}} \cdot M3_B     \}
\label{eqD.69}
\eea
\bea
(\tilde{\psi}_{\frac{5}{3}}\tilde{\psi}_{-\frac{5}{3}}B(0))_{B}
=\frac{1}{324 (784 + 57 c)^2 \zeta}
 [     27 \zeta \gamma (63 (688 + 7 c) (1024 + 27 c) \gamma\nn\\ + 
      4 (784 + 57 c) (10192 + 369 c) \tilde{\gamma})\nn\\ 
      + 18 (6 (784 + 57 c) \eta ((-4512 + 99 c) \gamma + 
         16 (784 + 57 c) \tilde{\gamma})\nn\\ 
         + (216 b (-116 + c) (784 + 57 c) 
         + (37473152 + c (14142088 + 451911 c)) \gamma\nn\\
          + 96 (-692 + 3 c) (784 + 57 c) \tilde{\gamma}) \lambda) \mu \nn\\
         - 2 (784 + 57 c) ((-1372336 + 897 c) \zeta + 
      324 (784 + 57 c) \tilde{\lambda}) \mu^2     ]
      \label{eqD.70}
\eea

\end{document}